\documentclass[a4paper,fleqn]{cas-sc}

\usepackage[authoryear,longnamesfirst]{natbib}
\usepackage{float}
\usepackage[linesnumbered,ruled]{algorithm2e}
\usepackage{caption}
\usepackage[section]{placeins}

\usepackage{times}
\usepackage{soul}
\usepackage{url}
\usepackage{algorithmic}

\usepackage{multirow}
\usepackage{multicol}
\usepackage{rotating}
\usepackage[graphicx]{realboxes}
\usepackage{booktabs}
\usepackage{subfig}
\usepackage{caption}
\usepackage{xspace}
\newcommand{\name}[0]{GBO\xspace}

\usepackage{xcolor}

\def\tsc#1{\csdef{#1}{\textsc{\lowercase{#1}}\xspace}}
\tsc{WGM}
\tsc{QE}
\usepackage[english]{babel}
\usepackage{amsthm}

\theoremstyle{definition}
\newtheorem{definition}{Definition}[section]
\begin{document}
\let\WriteBookmarks\relax
\def\floatpagepagefraction{1}
\def\textpagefraction{.001}

\captionsetup[figure]{labelfont={bf},name={Fig.},labelsep=period}
\shorttitle{Geometric Batch Optimization for the PECC Problem on Large Scale}

\shortauthors{Jianrong Zhou, Kun He, Joingzi Zheng, Chu-Min Li}  

\title [mode = title]{Geometric Batch Optimization for the Packing Equal Circles in a Circle Problem on Large Scale} 

%

\author[1]{Jianrong Zhou}[]
\author[1]{Kun He$^*$}[]
\cortext[1]{Corresponding author. Email: brooklet60@hust.edu.cn}
\author[1]{Jiongzhi Zheng}[]
\author[2]{Chu-Min Li}[]

\affiliation[1]{organization={School of Computer Science and Technology, Huazhong University of Science and Technology},
            city={Wuhan},
            postcode={80039}, 
            country={China}}

\affiliation[2]{organization={MIS, University of Picardie Jules Verne},
            city={Amiens},
            postcode={430074}, 
            country={France}}
















\begin{abstract}
The problem of packing equal circles in a circle is a classic and famous packing problem, which is well-studied in academia and has a variety of applications in industry. This problem is computationally challenging, and researchers mainly focus on small-scale instances with the number of circular items $n$ less than 320 in the literature. In this work, we aim to solve this problem on large scale. Specifically, we propose a novel geometric batch optimization method that not only can significantly speed up the convergence process of continuous optimization but also reduce the memory requirement during the program's runtime. Then we propose a heuristic search method, called solution-space exploring and descent, that can discover a feasible solution efficiently on large scale. Besides, we propose an adaptive neighbor object maintenance method to maintain the neighbor structure applied in the continuous optimization process.
In this way, we can find high-quality solutions on large scale instances within reasonable computational times. Extensive experiments on the benchmark instances sampled from $n = $ 300 to 1,000 show that our proposed algorithm outperforms the state-of-the-art algorithms and performs excellently on large scale instances. In particular, our algorithm found 10 improved solutions out of the 21 well-studied moderate scale instances and 95 improved solutions out of the 101 sampled large scale instances. Furthermore, our geometric batch optimization,  heuristic search, and adaptive maintenance methods are general and can be adapted to other packing and continuous optimization problems.
\end{abstract}


\maketitle

\section{Introduction} \label{sec:01intro}
The packing problems are a typical class of optimization problems that aim to pack a set of geometric objects into one or multiple containers, and the goal is to either find a configuration that is as dense as possible or find a solution with as few containers (bins) as possible. 
The packing problems have a rich research history, with numerous variants being proposed, and a number of studies and methods being published for solving these problems, 
such as circle packing~\citep{huang2011global, he2018efficient, lai2022iterated}, sphere packing~\citep{hartman2019sphere, hifi2019local}, square packing~\citep{leung1990packing, fekete2017online}, cube packing~\citep{miyazawa2003cube, epstein2005online}, irregular packing~\citep{leao2020irregular, zhao2020hybrid, rao2021hybridizing} and bin packing~\citep{baldi2019generalized, zhao2021online,He2021CircleBin}, etc.
 


As one of the most popular packing problems, the Circle Packing Problem (CPP) has been widely studied in mathematics and computer science fields. Given $n$ circular items with fixed radii, the CPP aims to pack all the $n$ items into a container so that all the items are totally contained in the container without overlapping with each other, and the container size is minimized. Specifically, if the container is square, then the goal is to minimize the side length of the square container; and if the container is circular, then the goal is to minimize the radius of the circular container. 
There are also other CPP variants~\citep{lopez2011heuristic} based on different geometric containers. 

CPP has a lot of industrial applications, such as 
facility layout, cylinder packing, circular cutting, container loading, dashboard layout~\citep{castillo2008solving}, structure design~\citep{yanchevskyi2020circular} and satellite packaging~\citep{wang2019stimulus}. Furthermore, the solution of CPP can be applied to data visualization and data analysis~\citep{wang2006visualization, murakami2015clonepacker, gortler2017bubble}. 
On the other hand, CPP is proved to be NP-hard~\citep{demaine2010circle}, therefore solving this problem is computationally challenging. 
In particular, there is a special variant of CPP, called Packing Equal Circles in a Circle (PECC), which aims to pack $n$ unit circles into a circular container with the smallest possible radius. Since PECC is representative and simple in form, it has become the most famous and well-studied 
problem in the CPP family, to which numerous efforts have been devoted.  
Since the CPP is NP-hard, the computational resource and difficulty for obtaining a high-quality configuration grow exponentially with the number of circular items to be packed, and solving the large scale instances of CPP is extremely difficult. Therefore, most works for CPP focus on solving instances on small or moderate scale, and only a few efforts are devoted to solving large scale instances.
In this work, we address the PECC variant of CPP.
Based on the classic elastic model~\citep{huang1999two, he2018efficient} (also known as Quasi-Physical Quasi-Human model, QPQH), we propose a Geometric Batch Optimization (\name) method that can construct a feasible solution or an infeasible solution with the overlaps being as few as possible on large scale instances. Different from existing methods, \name divides the packing circles into several geometric batches. Then, in the non-convex continuous optimization process, \name alternately 
updates each batch of packing circles instead of updating all the circles. In this way, \name 
reduce the time and space complexity, which
can not only significantly speed up the convergence process of the non-convex continuous optimization but also reduce the resident memory requirement during the program's runtime.

In addition, we propose an Adaptive Neighbor object Maintenance (ANM) method to maintain the neighbor structure~\citep{he2018efficient} for the continuous optimization process. ANM uses two variables ``$counter$'' and ``$deferring~length$'' to accomplish the adaptive feature. When the layout is changing significantly, ANM maintains the neighbors in each iteration, otherwise, ANM defers the maintaining process. Our ANM method can handle some issues and disadvantages of the existing methods~\citep{he2018efficient, lai2022iterated} and it can adopt to solve dynamic packing problems or online packing problems.

Finally, we propose an advanced local search heuristic, called Solution-space Exploring and Descent (SED), based on the \name  module to solve the PECC problem. SED perturbs the current solution to obtain several perturbed candidate solutions.
Then, SED employs the \name module to minimize the overlap of each candidate solution and takes a solution with minimal overlap among the candidate solutions to replace the current solution. The algorithm could efficiently find a high-quality solution by iteratively executing the SED heuristic procedure. 

The experimental results show that our \name method dramatically accelerates the convergence speed of the non-convex continuous process and reduces memory consumption. 
Specifically, \name reduces the convergence time by 32.74\% to 54.11\% and the runtime resident memory by 36.00\% to 39.04\% compared with the typical non-batch method on large scale instances for $n =$ 500 to 1,000. 
By combining the \name module into SED to solve 101 large scale instances, including 51 regular numbers and 50 irregular numbers selected from 500 $\leq n \leq$ 1,000, our method improves the best-known solution for 95 instances reported in Packomania website~\citep{Spechtweb}. Besides, our method is also efficient on moderate scale instances, because it improves the best-known solution for 10 instances among the 21 well-studied moderate scale instances, 300 $\leq n \leq$ 320, which were solved by the state-of-the-art algorithm called IDTS~\citep{lai2022iterated}. 
Through these experiments, \name shows clear advantages over existing methods for solving large scale packing instances and SED is an efficient heuristic.

The main contributions of this work are summarized as follows:
\begin{itemize}
    \item We propose a novel method called Geometric Batch Optimization (\name), which divides the circles into several geometric batches and speeds up the computation significantly. 
    \item We propose an Adaptive Neighbor object Maintenance (ANM) method of the neighbor structure for solving the PECC problem.
    \item We propose an efficient Solution-space Exploring and Descent (SED) heuristic with the \name as a sub-module for solving the PECC problem.
    \item Extensive experiments on large scale instances sampled from $n = 500$ to 1,000, as well as moderate instances for $n = 300$ to 320,  demonstrate the excellent performance and efficiency of our proposed algorithm, gaining new best solutions on many instances.
\end{itemize}


The rest of this paper is organized as follows. Section~\ref{sec:02RW} presents the related works of the CPP and PECC problem, including the landmark models, the landmark heuristics, and the recent works for solving the CPP and PECC problems. Section~\ref{sec:03pre} introduces the mathematical formula of the PECC problem and the famous and powerful elastic model (QPQH) that is adopted in this work. Section~\ref{sec:04opti} presents the continuous optimization methods, including our proposed \name method, the container adjusting method, and our ANM method, which are the essential modules adopted in our final heuristic. Section~\ref{sec:05heur} presents the discrete optimization method (i.e., SED heuristic), which is our final algorithm for solving the PECC problem. Section~\ref{sec:06exp} presents the experimental results of our proposed algorithm compared with the best-known results and the state-of-the-art algorithm and parameter study. 
The conclusion is drawn in the end. 

\section{Related Work} \label{sec:02RW}
In this section, We briefly review the works on CPP, including some typical works, landmark models, and landmark heuristics, 
then we review the recent works on the PECC problem. 

Early works for solving CPP focus on solving the PECC variant using mathematical analysis with the goal of finding and proving an optimal solution for small scale instances. Starting from 1967, \citet{kravitz1967packing} first provided the solutions for $2 \leq n \leq 19$. Subsequently, \citet{graham1968e1910} proved the optimality for $2 \leq n \leq 7$. In 1969, \citet{pirl1969mindestabstand} further proved the optimality of the solutions for $2 \leq n \leq 10$ and also provided solutions for $11 \leq n \leq 19$. In 1971, \citet{goldberg1971packing} improved the solutions for $n = 14, 16, 17$ and further provided the solution for $n = 20$. In 1975, \citet{reis1975dense} provided the solutions for $21 \leq n \leq 25$. \citet{melissen1994densest} proved the optimality for $n = 11$ and \citet{fodor1999densest, fodor2000densest, fodor2003densest} proved the optimality for $n = 12, 13, 19$. To summarize, the optimality for $2 \leq n \leq 13$ and $n = 19$ has been proven. For larger instances, it is hard to find and prove the optimality of a solution by exploiting mathematical analysis. Thereafter, most efforts are devoted to designing efficient heuristic algorithms. 

The discrete optimization model is one of the popular techniques for solving CPP, of which the idea is to pack each circle one by one into the container. If there exists a placement order by which all the circles can be packed into the container, then the solution is found. A circle placement heuristic is essential to this model, because it directly impacts the placement strategies and the performance. \citet{huang2003two} and \citet{huang2006new} propose a classic circle placement heuristic, called Corner-Occupying Placement (COP), and a metric function for placement action, named Maximal Hole Degree (MHD), on the unequal circle packing problem. COP requests each current packing circle contact with two packed circles or the boundary of the container and does not overlap with other circles, and the MHD function can measure the quality of the candidate action of the circle placement. There are several follow-up works based on COP heuristic and MHD function to either solve other variants or improve the algorithm efficiency: 
\citet{huang2005greedy} adopt this heuristic for packing circles in a rectangular container; 
\cite{lu2008perm} apply a Pruned–Enriched-Rosenbluth Method (PERM) to improve the performance on the unequal circle packing problem; 
\citet{akeb2009beam, akeb2010adaptive} further employ the beam search and adaptive beam search algorithms to improve the performance on unequal circle packing problems, and 
\citet{chen2018greedy} present a greedy heuristic algorithm for solving the PECC problem.
There also exist other circle placement heuristics for solving the packing problems, such as Bottom-Left-Fill~\citep{martello2003exact, burke2006new} and Best Local Position~\citep{hifi2004approximate, hifi2007dynamic}. 

However, due to the characteristic of the discrete optimization model, it takes massive computational time to obtain a dense solution on moderate or large scale instances. Therefore, most researchers prefer to adopt the non-convex continuous optimization model to solve moderate scale CPP. Inspired by the physical model, some quasi-physical models have been proposed for solving CPP. \citet{graham1998dense} design a molecular repulsion model and a Billiards model for solving the PECC problem, and these models are also applied to solve other packing variants~\citep{nurmela1997packing}. 
Some efforts are devoted to designing an efficient local search heuristic based on the elastic model, termed Quasi-Physical Quasi-Human (QPQH), proposed by \citet{huang1999two} for solving CPP and its variants: 
\citet{huang2011global} propose a basin hopping heuristic of attractive force model to solve the PECC problem;
\citet{he2018efficient} propose a shrinking basin-hopping heuristic to solve the PECC problem;
\citet{lai2022iterated} adopt the iterated dynamic thresholding search to solve the PECC problem;
\citet{he2013coarse} employ the elastic model to solve circle packing problem with equilibrium constraints and 
\citet{liu2016heuristic} employ the elastic model to solve the weighted circle packing problem.

There also exist other methods and heuristics for solving CPP, such as monotonic basin hopping and population basin hopping heuristics~\citep{addis2008disk, grosso2010solving}, simulated annealing approach~\citep{hifi2004simulated}, genetic algorithm-based approach~\citep{hifi2004approximate}, tabu search approach~\citep{carrabs2014tabu}, non-linear programming based approach~\citep{mladenovic2005reformulation, birgin2008minimizing, stoyan2014packing}, mixed-strategy~\citep{stoyan2020optimized}, etc.

The well-known Packomania website~\citep{Spechtweb} maintained by Specht presents many circle packing problems and records their best-known solutions. According to the number of solutions, the scale of solutions, and the recently updated history, we observe that the PECC problem and the problem of circle packing in a rectangle are popular and well-studied, and the recent algorithms for solving the two problems are similar to the methods mentioned above, suggesting that these methods and algorithms have good universality and generality on different circle packing problems and their variants. 

From the recently updated history on PECC with $n \geq 100$ at Packomania, we observe the 
current best-known records as follows: 
\citet{huang2011global} hold several best-known solutions for $100 \leq n \leq 200$; Cantrell holds several best-known solutions for $110 \leq n \leq 1039$ with unpublished methods; Donovan holds three best-known solutions for $n = 85, 109$ and $121$ with unpublished methods; \citet{lai2022iterated} hold a number of best-known solutions for $126 \leq n \leq 319$, \citet{stoyan2020optimized} hold 16 best-known solutions for $1077 \leq n \leq 5000$, and the remaining best-known solutions are held by Specht with unpublished methods. 
In summary, the elastic model (QPQH) based methods~\citep{huang2011global, lai2022iterated} and mixed-strategy method~\citep{stoyan2020optimized} can be regarded as the state-of-the-art algorithms for solving the PECC problem. 

\section{Preliminaries} \label{sec:03pre}
\subsection{Problem Formulation} \label{ssec:03-01formu}

The PECC problem aims to pack $n$ unit circles $\{ c_1, c_2, ..., c_n \}$ into a circular container with the smallest possible radius while subjected to two constraints: (I) Any two circles do not overlap; (II) Any circle does not exceed the container. The problem can be formulated in the Cartesian coordinate system as a non-linear constrained optimization problem:
\begin{align}
    \mathrm{Minimize} \quad & R &\nonumber \\
    \textit{s.t.} \quad & \sqrt{(x_i - x_j)^2 + (y_i - y_j)^2} \geq 2, \quad 1 \leq i, j \leq n, \ i \neq j  \label{eq-1}, \\
    \quad &  \sqrt{x_i^2 + y_i^2} + 1 \leq R, \quad 1 \leq i \leq n \label{eq-2},
\end{align}
where $R$ is the radius of the circular container centered at the origin $(0, 0)$, and the center of unit circle $c_i$ is located at $(x_i, y_i)$. Eqs.~\eqref{eq-1} and \eqref{eq-2} correspond to the two constraints (I) and (II).

\subsection{The Elastic Model (QPQH) for PECC} \label{ssec:03-02QPQH}

The elastic model~\citep{huang1999two, he2018efficient, lai2022iterated} could be regarded as a relaxation of the PECC problem, considering that each circle is elastic. An algorithm following this model first forces all circles to be packed into the container with possible circle-circle and circle-container overlapping, and quantifies the overlapping degree using a metric function. The main work of the algorithm is then to minimize the value of the metric function (i.e., minimize the overlapping area). In this way, the elastic model converts the PECC problem to an unconstrained non-convex continuous optimization problem. 

\begin{figure}[t]
    \centering
    \includegraphics[width=0.7\columnwidth]{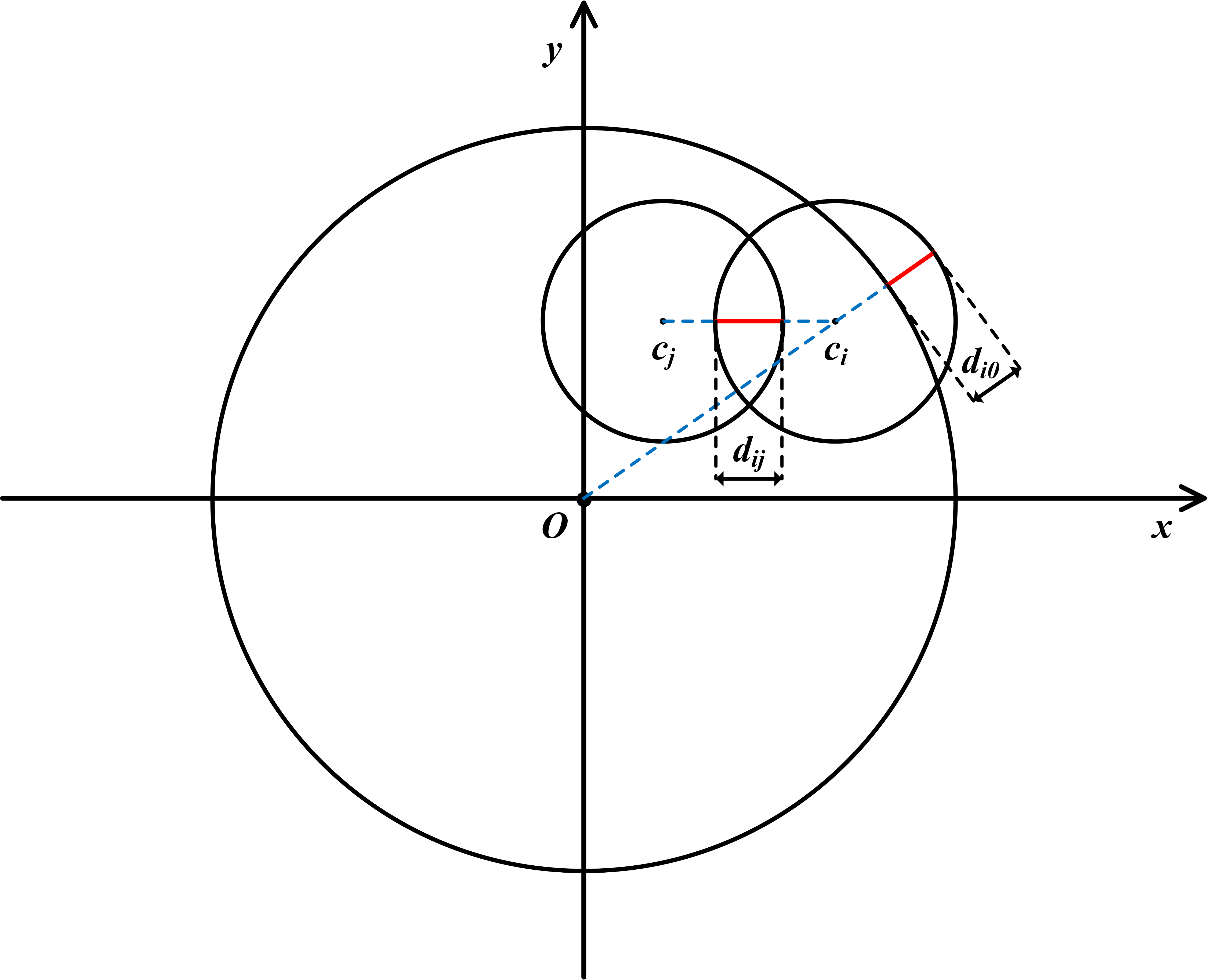}
    \caption{Illustration of a conflicting example with two types of overlaps, where $d_{ij}$ indicates the circle-circle overlapping distance and $d_{i0}$ indicates the circle-container overlapping distance.}
    \label{fig-PECC}
\end{figure}

\begin{definition}[Overlapping distance]
The overlapping distance of two unit circles $c_i$ and $c_j$, denoted as $d_{ij}$, is defined as follows:
\begin{equation} \label{eq-3}
    d_{ij} = \max \left( 0, 2 - \sqrt{(x_i - x_j)^2 + (y_i - y_j)^2} \right), 
\end{equation}
And the overlapping distance of circle $c_i$ to the container, denoted as $d_{i0}$, is defined as follows:
\begin{equation} 
    d_{i0} = \max \left( 0, \sqrt{x_i^2 + y_i^2} + 1 - R \right). \nonumber
\end{equation}
\end{definition}

Figure~\ref{fig-PECC} shows a conflicting example with the two types of overlaps. According to the theory of elasticity, the elastic potential energy between two elastic objects is proportional to the square of the embedded distance, giving the following elastic energy definition.


\begin{definition}[Elastic energy]
The total elastic energy $E$ of the PECC system is defined as follows:
\begin{equation} \label{eq-5}
    E_R(\boldsymbol{x}) = \sum_{i=1}^{n} \sum_{j=i+1}^{n} d_{ij}^2 + \sum_{i=1}^{n} d_{i0}^2,
\end{equation}
where $\boldsymbol{x} = [x_1, y_1, x_2, y_2, ... , x_n, y_n]^{\rm T}$ is a vector, $\boldsymbol{x} \in \mathbb{R}^{2n}$, representing a candidate solution, and $R$ is the container radius.
\end{definition}



Note that, in the elastic model of the PECC problem, the energy quantifies the overlapping degree. When the energy $E$, i.e., Eq.~\eqref{eq-5}, is equal to zero, Eqs.~\eqref{eq-1} and \eqref{eq-2} are satisfied and the solution is feasible for the PECC problem. Since the radius $R$ of the container is fixed in the rest of the paper, we omit the subscript $R$ in $E_R(\boldsymbol{x})$ and denote it simply as $E(\boldsymbol{x})$ in the rest of the paper for readability reason.

\section{Continuous Optimization} \label{sec:04opti}
Since we employ the elastic model for solving the PECC problem, our target switches to minimizing the energy $E$ so as to discover a feasible solution that can be solved using 
continuous optimization algorithms. 
 In this section, we present 1) 
our proposed Geometric Batch Optimization (\name) method for minimizing the energy of a conflicting solution with a fixed container radius, which can be regarded as a batch non-convex continuous optimization algorithm,
2) the optimization method for adjusting container radius to obtain a feasible solution with a minimal radius $R$, 3) our proposed Adaptive Neighbor object Maintenance (ANM) method that can adaptively and efficiently maintain the neighbor structure in the continuous optimization process, and finally, 4) we give a complexity analysis of our proposed \name method. 

\begin{algorithm}[tb]
\caption{\name($\boldsymbol{x}$, $R$)} \label{alg_batch}
\textbf{Input}: A candidate solution $\boldsymbol{x}$; and the fixed container radius $R$\\
\textbf{Output}: A local minimum solution $\boldsymbol{x}^{*}$
\begin{algorithmic}[1] 
\STATE $( \boldsymbol{x}_1, \boldsymbol{x}_2, \ldots, \boldsymbol{x}_k )$ $\leftarrow$ $\mathrm{partition} (\boldsymbol{x})$
\STATE $( \boldsymbol{H}_1, \boldsymbol{H}_2, \ldots, \boldsymbol{H}_k )$ $\leftarrow$ $(\boldsymbol{I}, \boldsymbol{I}, \ldots, \boldsymbol{I})$
\STATE $cnt \leftarrow 0, \ len \leftarrow 1$ 
\STATE construct the current neighbor $\Gamma$ 
\FOR {$t$ for $1$ to $MaxIter$}
\FOR {$p$ from $1$ to $k$} 
\STATE calculate $g(\boldsymbol{x}_p)$
\STATE calculate $\alpha_p$ by using Eq. (\ref{eq-9})
\STATE $\boldsymbol{x}_p$ $\leftarrow$ $\boldsymbol{x}_p - \alpha_p \boldsymbol{H}_p g(\boldsymbol{x}_p)$
\STATE update $\boldsymbol{H}_p$ by using Eqs. (\ref{eq-10}) and (\ref{eq-11})
\ENDFOR
\STATE ($cnt, len, \Gamma$) $\leftarrow$ ANM($cnt, len, \Gamma$) 
\IF {$\sum_{p=1}^{k} {\|g(\boldsymbol{x}_p)\|_2} \leq 10^{-12}$}
\STATE \textbf{break}
\ENDIF
\ENDFOR
\STATE $\boldsymbol{x}^{*}$ $\leftarrow$ $\mathrm{merge} ( \boldsymbol{x}_1, \boldsymbol{x}_2, \ldots, \boldsymbol{x}_k )$
\STATE \textbf{return} $\boldsymbol{x}^{*}$
\end{algorithmic}
\end{algorithm}

 
\subsection{The \name Method for Solving PECC} \label{ssec:04-01GBO}

Let $k$ be a hyperparameter. Given an infeasible solution in which $n$ elastic unit circles are already forced to be packed in the container, \name first evenly partitions these circles into $k$ disjoint sets: $B_1, B_2, ..., B_k$, with the following properties:  
\begin{gather}
    B_p \cap B_q = \emptyset, \quad 1 \leq p, \ q \leq k, \ p \neq q \nonumber \\
    \bigcup_{p=1}^{k} B_p = \{c_1, c_2,  \ldots, c_n\}. \nonumber
\end{gather}
Each subset of circles is regarded as a batch. The circles in batch $B_p$ can be represented as a vector $\boldsymbol{x}_p$, and the energy $E(\boldsymbol{x}_p)$ of the circles in batch $B_p$ can be reformulated as follows:
\begin{equation} \label{eq-7} 
    E(\boldsymbol{x}_p) = \sum_{i \in B_p} \sum_{j=1}^{n} d_{ij}^2 [j \notin B_p \vee j < i] + \sum_{i \in B_p} d_{i0}^2,
\end{equation}
where $[]$ is the Iverson bracket that $[P]=1$ if statement $P$ is true, and otherwise $[P]=0$. The statement ``$j \notin B_p \vee j < I$'' guarantees the circle-circle overlaps in batch $B_p$ only be calculated once. 
We employ the Broyden–Fletcher–Goldfarb–Shanno (BFGS) algorithm~\citep{ren1983convergence} as our basic optimization method to minimize the batch energy, by iteratively reducing the energy $E(\boldsymbol{x}_p)$ as follows:
\begin{equation} \label{eq-8}
    \boldsymbol{x}_p^{(t+1)} \leftarrow \boldsymbol{x}_p^{(t)} - \alpha_p^{(t)} \boldsymbol{H}_p^{(t)} g(\boldsymbol{x}_p^{(t)}), 
\end{equation}
where superscript $t$ indicates the algorithm at the $t$-th iteration, $\boldsymbol{H}_p^{(t)}$ is an approximation Hessian matrix $\left[ \nabla^2 E(\boldsymbol{x}_p^{(t)}) \right]^{-1}$, $\alpha$ is the step length obtained by the line search method, and the function $g(\boldsymbol{x})$ is the gradient of the energy, i.e.,  $\nabla E(\boldsymbol{x})$. 
The update of $\boldsymbol{H}_p^{(t+1)}$ and the calculation of $\alpha_p^{(t)}$ are performed as follows:
\begin{gather}
    \alpha_p^{(t)} = \underset{\alpha \in \mathbb{R}^{+}}{\arg\min} \ E\left(\boldsymbol{x}_p^{(t)} - \alpha \boldsymbol{H}_p^{(t)} g(\boldsymbol{x}_p^{(t)})\right), \label{eq-9} \\
    \boldsymbol{u}_p^{(t)} = \boldsymbol{x}_p^{(t+1)} - \boldsymbol{x}_p^{(t)}, \quad \boldsymbol{v}_p^{(t)} = g(\boldsymbol{x}_p^{(t+1)}) - g(\boldsymbol{x}_p^{(t)}), \quad \beta = (\boldsymbol{v}_p^{(t)})^{\mathrm{T}} \boldsymbol{u}_p^{(t)} \label{eq-10}, \\
    \boldsymbol{H}_p^{(t+1)} = \left( \boldsymbol{I} - \frac{ \boldsymbol{u}_p^{(t)} (\boldsymbol{v}_p^{(t)})^{\mathrm{T}}}{\beta} \right)  \boldsymbol{H}_p^{(t)} \left( \boldsymbol{I} - \frac{ \boldsymbol{v}_p^{(t)} (\boldsymbol{u}_p^{(t)})^{\mathrm{T}}}{\beta} \right) + \frac{ \boldsymbol{u}_p^{(t)} (\boldsymbol{u}_p^{(t)})^{\mathrm{T}} }{\beta}. \label{eq-11}
\end{gather}


\begin{figure}[t]
    \centering
    \begin{minipage}[b]{0.47\linewidth}
        \centering
        \subfloat[][Sector]{\includegraphics[width=1\linewidth]{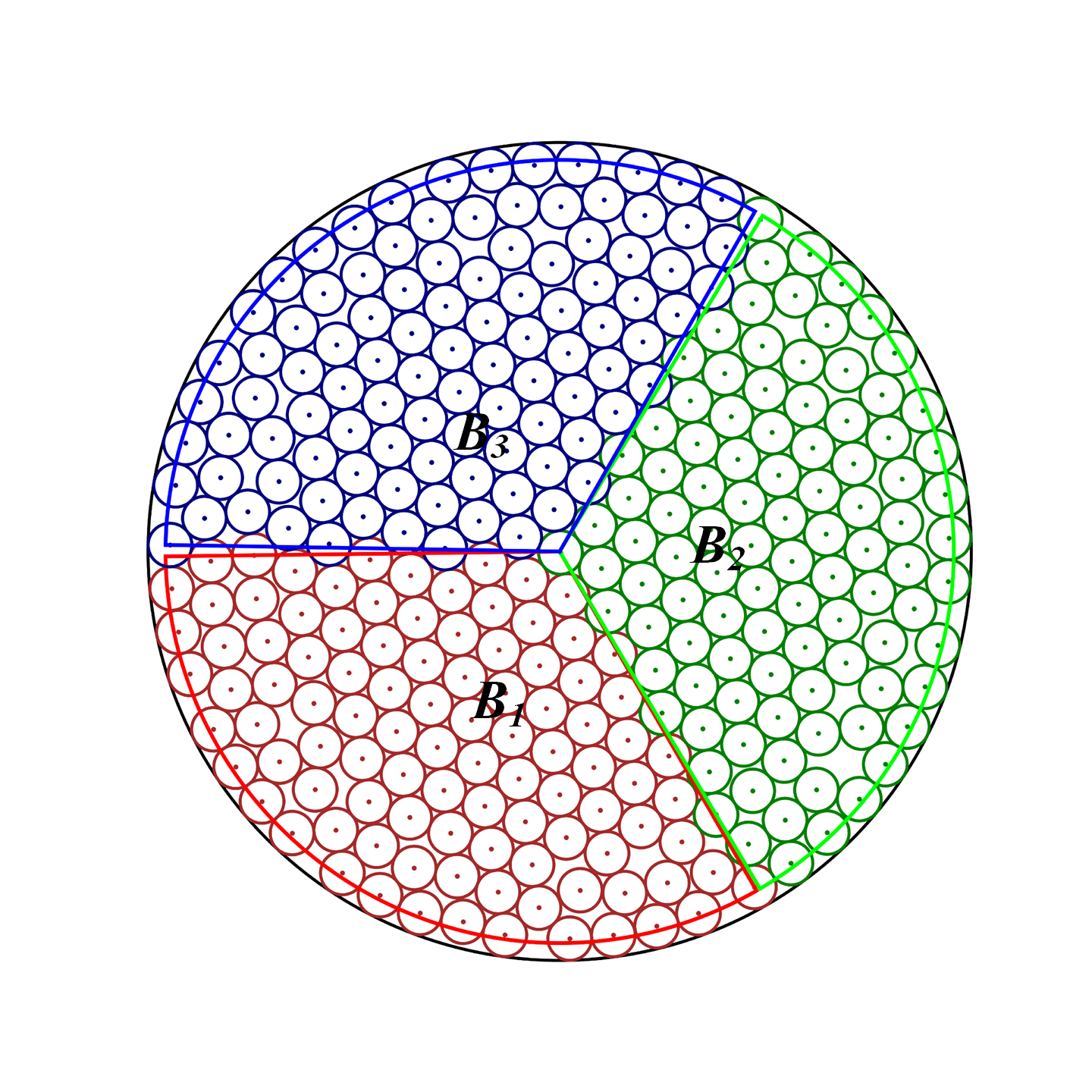}}
    \end{minipage}
    \begin{minipage}[b]{0.47\linewidth}
        \centering
        \subfloat[][Annulus]{\includegraphics[width=1\linewidth]{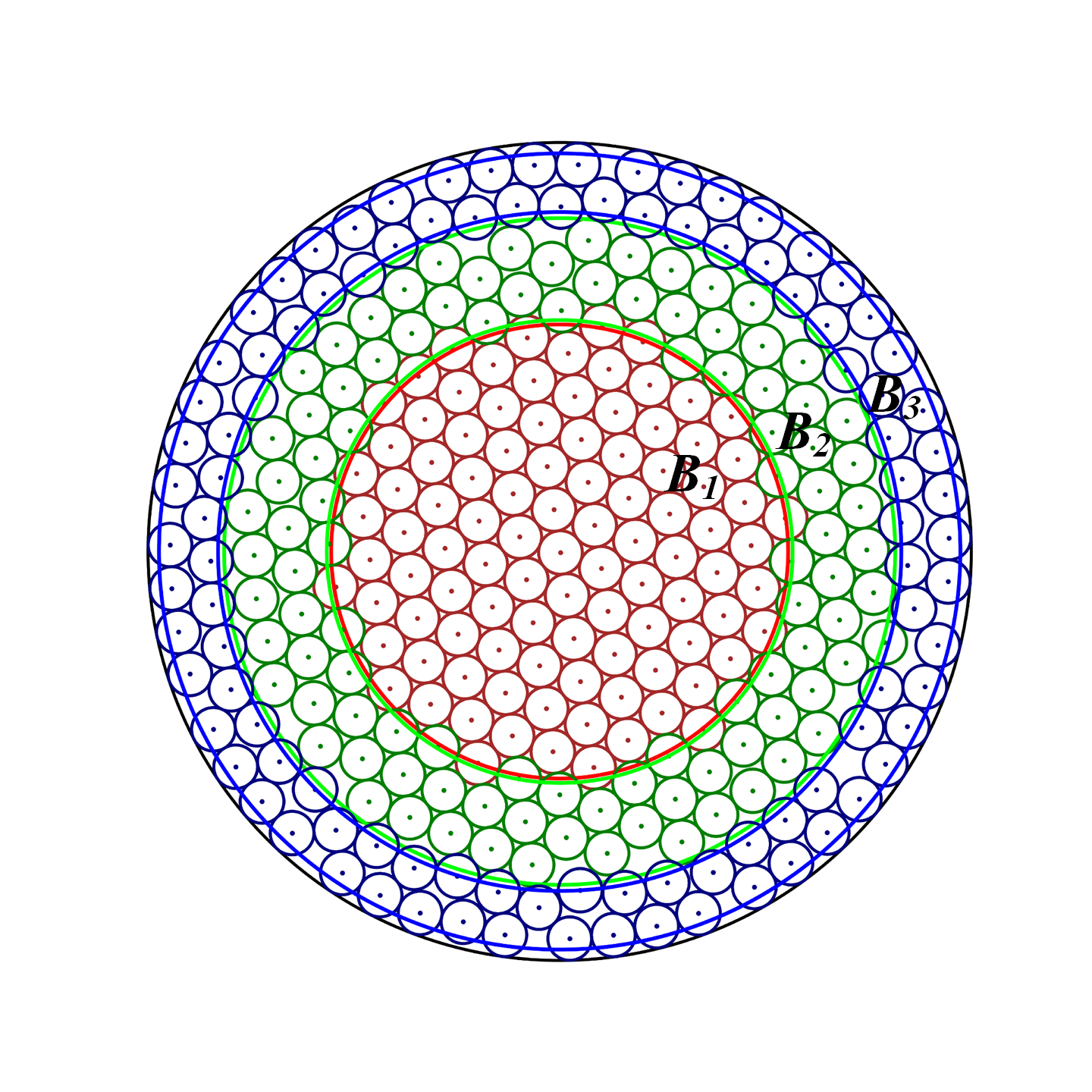}}
    \end{minipage}    
    \begin{minipage}[b]{0.47\linewidth}
        \centering
        \subfloat[][Fence]{\includegraphics[width=1\linewidth]{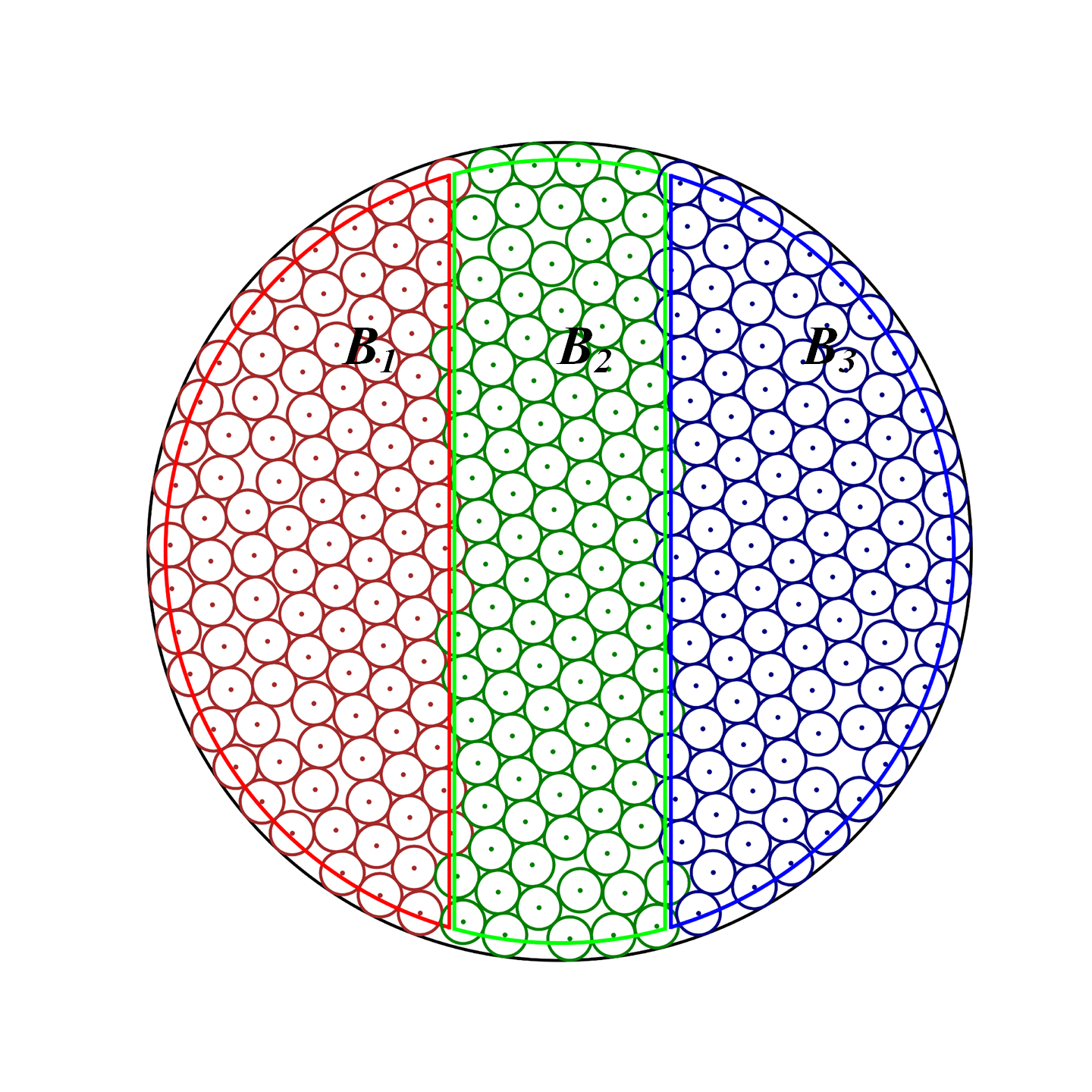}}
    \end{minipage}
    \begin{minipage}[b]{0.47\linewidth}
        \centering
        \subfloat[][Random]{\includegraphics[width=1\linewidth]{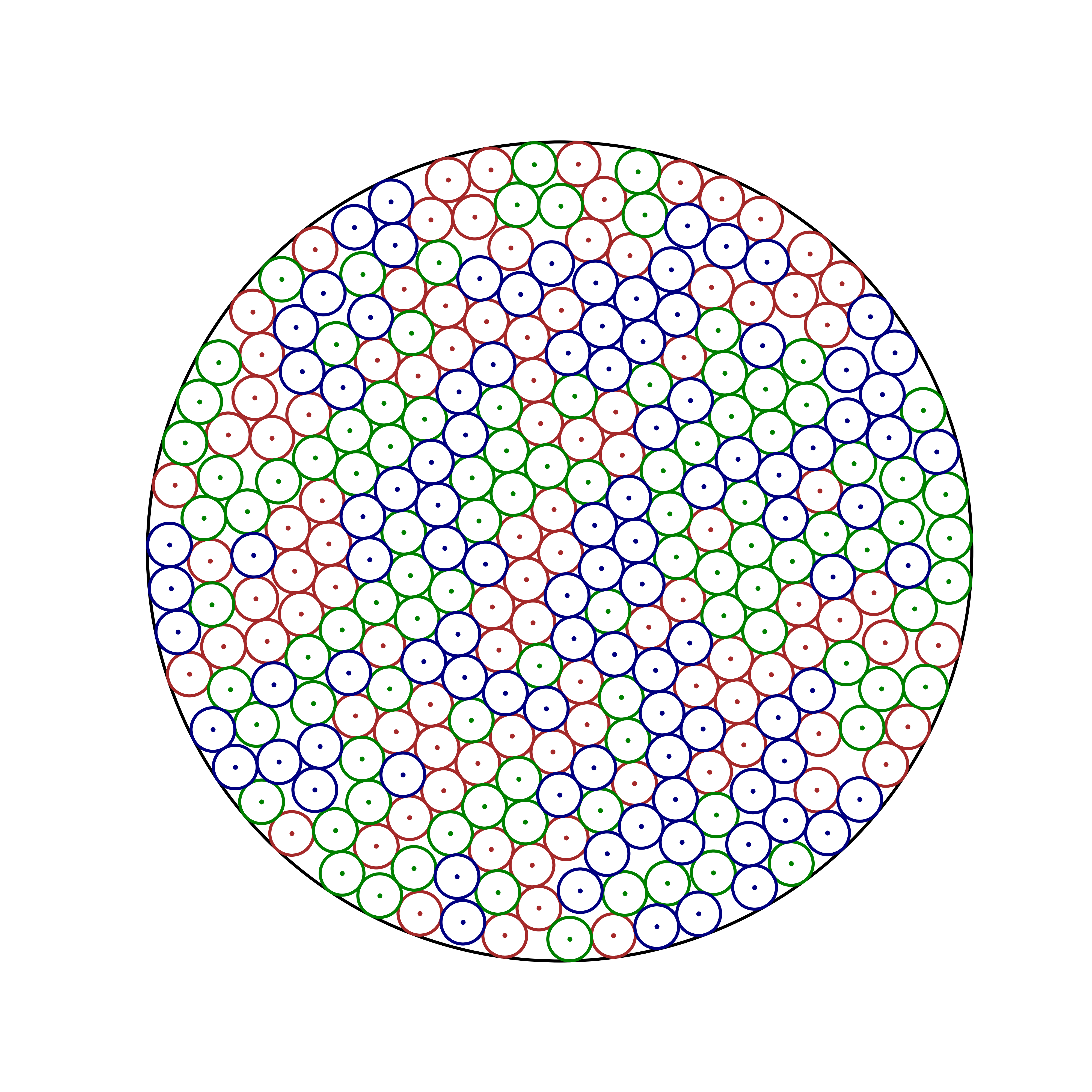}}
    \end{minipage}

    \caption{Illustration of the four partitions of the \name method. The illustrative example has 300 circles ($n = 300$), which is evenly partitioned into three batches ($k = 3$) colored by three different colors where each batch having 100 circles.}
    \label{batch-partition}
\end{figure}

In this way, the PECC problem is converted to an unconstrained non-convex continuous optimization problem with a fixed container radius. 
The pseudocode of our proposed \name algorithm is presented in Algorithm~\ref{alg_batch}.  
Given a candidate solution $\boldsymbol{x}$, a fixed container radius $R$, algorithm ${\rm \name}(\boldsymbol{x}, R)$ returns a solution $\boldsymbol{x}^{*}$ such that energy $E(\boldsymbol{x}^{*})$ reaches a local minimum. For this purpose, \name first partitions the $n$ unit circles into $k$ batches (line 1), then \name applies the BFGS method to minimize the energy on each batch iteratively (lines 6-11), the iteration process terminates when the maximum iteration step is reached (line 5) or the sum of the $L_2$-norm of $k$-batch gradients is tiny enough (lines 13-15). Finally, \name returns a local minimum solution $\boldsymbol{x}^{*}$ obtained by merging the $k$-batch circles (line 17). 
Though it is hard to provide a mathematical analysis on the convergence of the proposed \name algorithm, we observe that it always meets the convergence condition (line 13) when the maximum iteration step is set large enough.

To minimize the total computational complexity of the \name method, each batch groups nearly the same number of circles. Concretely, the size of each batch is either $\lfloor \frac{n}{k} \rfloor$ or $\lceil \frac{n}{k} \rceil$. We 
provide several geometric $k$-batch partition strategies for solving the PECC problem, described as follows: 
\begin{itemize}
    \item \textbf{1) Sector (default)}: Sort all circles in the container in ascending order of the angle ${\theta}_i = \arctan{\frac{y_i}{x_i}}$, then partition them into $k$ subsets successively. 
    \item \textbf{2) Annulus}: Sort all circles in ascending order of the distance ${dis}_i = \sqrt{{x_i}^2 + {y_i}^2}$, then partition them into $k$ subsets successively.
    \item \textbf{3) Fence}: Sort all circles in ascending order of the coordinate $x_i$ , then partition them into $k$ subsets successively.
    \item \textbf{4) Random}: Partition all circles into $k$ sets randomly.
\end{itemize}
Our experiments show that the sector partition outperforms other partitions, so the default partition is set as sector in this work. Figure~\ref{batch-partition} illustrates the above four partitions of the \name method. 

\subsection{Optimization Method for Adjusting the Container} \label{ssec:04-02adj}

Starting from a random layout (all the circles are randomly placed in the container and overlapping is allowed), a local minimum solution is obtained by applying the \name method, possibly containing overlaps and thus being infeasible. One strategy for obtaining a feasible solution is to expand the container radius and adjust the layout until the overlaps are eliminated.  
The most intuitive and common way to achieve this goal is to adopt the binary search~\citep{huang2011global, he2018efficient}. Recently, a smart and significantly faster method is proposed in~\citep{lai2022iterated} to find a feasible solution with a local minimum container radius $R$, which we present as follows. 

Let vector $\boldsymbol{z} = [ x_1, y_1, x_2, y_2, ..., x_n, y_n, R ]^{\mathrm{T}}$, $\boldsymbol{z} \in \mathbb{R}^{2n+1}$, be a candidate solution with the container radius $R$ as a variable. 
A new elastic energy $U$ can be reformulated as follows:
\begin{equation} \label{eq-12} 
    U(\boldsymbol{z}, \lambda) = \sum_{i=1}^{n} \sum_{j=i+1}^{n} d_{ij}^2 + \sum_{i=1}^{n} d_{i0}^2 + \lambda R^2,
\end{equation}
where $\lambda R^2$ is a penalty term, and $\lambda$ is a penalty coefficient. By employing the optimization method to minimize the energy $U$, the container radius prefers to shrink when $\lambda$ increases, and it prefers to expand when $\lambda$ is decreasing. Therefore, the model converts the current goal to an unconstrained continuous optimization problem. By adjusting $\lambda$, it could allow to obtain a feasible solution when minimizing the energy $U$ with a local minimum container radius. 

We adopt the main idea of adjusting container radius of~\citep{lai2022iterated} 
and control 
$\lambda R^2$ to minimize the container size. 
The pseudocode of our container radius adjusting method is depicted in Algorithm~\ref{alg_adjcont}. 
Given a candidate solution $(\boldsymbol{x}, R)$, Algorithm~\ref{alg_adjcont} initializes the coefficient $\lambda$ to an empirically fixed value $10^{-4}$ and combines $(\boldsymbol{x}, R)$ to obtain a new candidate solution $\boldsymbol{z}$; then the algorithm performs several iterations to obtain a feasible solution. At each iteration, the algorithm employs the BFGS method to minimize the energy $U(\boldsymbol{z}, \lambda)$ and updates the candidate solution $\boldsymbol{z}$. Then, the algorithm halves $\lambda$ and continually adjusts the candidate solution $\boldsymbol{z}$ at the next iteration. After several iterations, the energy $U(\boldsymbol{z}, \lambda)$ converges to $0$, so that the overlaps are tiny enough and the candidate solution $\boldsymbol{z}$ can be regarded as a feasible solution with a local minimum container radius. 
Finally, the method returns the solution $(\boldsymbol{x}^*, R^*)$ as the result.


\begin{algorithm}[b]
\caption{adjust\_container($\boldsymbol{x}, R$)} \label{alg_adjcont}
\textbf{Input}: A candidate solution $\boldsymbol{x}$ and the container radius $R$\\
\textbf{Output}: A feasible solution with local minimum container radius $(\boldsymbol{x}^{*}, R^{*})$ 
\begin{algorithmic}[1] 
\STATE $\boldsymbol{z}^{*} \leftarrow \mathrm{combine}(\boldsymbol{x}, R), \ \lambda \leftarrow 10^{-4}$
\FOR {$t$ from $1$ to $35$} 
\STATE employ BFGS to minimize the energy $U(\boldsymbol{z}^*, \lambda)$
\STATE $\lambda \leftarrow 0.5 \times \lambda$
\ENDFOR
\STATE $(\boldsymbol{x}^{*}, R^{*})$ $\leftarrow$ $\mathrm{divide}(\boldsymbol{z}^{*})$
\STATE \textbf{return} $(\boldsymbol{x}^{*}, R^{*})$
\end{algorithmic}
\end{algorithm}

Note that the term $\lambda R^2$ is tiny enough with sufficient iteration steps. The algorithm forces to converge the energy $E(\boldsymbol{x})=0$ ($U$ degenerates to $E$ without penalty term $\lambda R^2$) with non-fixed container radius.

\subsection{The Neighbor Structure of Circles for Optimization}
\label{ssec:04-03nei}

\citet{he2018efficient} first propose an efficient neighbor structure, which can efficiently calculate the energies $E$ and $U$ and its gradient functions of a candidate solution, and the state-of-the-art algorithm IDTS~\citep{lai2022iterated} also adopts this method. In this subsection, we first introduce the neighbor structure. Then, we present our proposed ANM method and discuss the advantage of our method over the existing methods.

Let $l_{ij}$ denote the distance between the centers of two unit circles $c_i$ and $c_j$:
\begin{equation} 
    l_{ij} = \sqrt{(x_i - x_j)^2 + (y_i - y_j)^2}, \nonumber
\end{equation}

Recall that $d_{ij}$ denotes the energy between two unit circles $c_i$ and $c_j$ and is defined by Eq.~\eqref{eq-3}. It is clear that $d_{ij} > 0$ when $l_{ij} < 2$, and $d_{ij} = 0$ otherwise. We define the neighbor $\Gamma(i)$ of circle $c_i$ to be a subset of $n$ unit circles as follows, using a distance controlling hyperparameter $l_{cut}$:
\begin{equation}
    \Gamma(i) = \left\{ c_j \mid \forall j: 1 \leq j \leq n, \ i \neq j, \ l_{ij} < l_{cut} \right\}, \nonumber
\end{equation}
 When $l_{cut} = 2$, all the circles 
$\{c_j\}$ overlapping with circle $c_i$ are contained in $\Gamma(i)$. Therefore the energies concerning circle $c_i$ can be calculated by enumerating the circles $c_j$ in $\Gamma(i)$ instead of enumerating all the packing circles. So, the batch energy $E$ (Eq.~\eqref{eq-7}) and the energy $U$ (Eq.~\eqref{eq-12}) can be reformulated as follows:
\begin{gather}
    E(\boldsymbol{x}_p) = \sum_{i \in B_p} \sum_{j \in \Gamma(i)} d_{ij}^2 [j \notin B_p \vee j < i] + \sum_{i=1} d_{i0}^2, \nonumber \\
    U(\boldsymbol{z}, \lambda) = \sum_{i=1}^{n} \sum_{j \in \Gamma(i)} d_{ij}^2 [j < i] + \sum_{i=1}^{n} d_{i0}^2 + \lambda R^2, \nonumber
\end{gather}

Note that the statements ``$j \notin B_p \vee j < i$'' and ``$j < i$'' guarantee the circle-circle overlaps be calculated only once. 

\begin{figure}[]
    \centering
    \includegraphics[width=0.7\columnwidth]{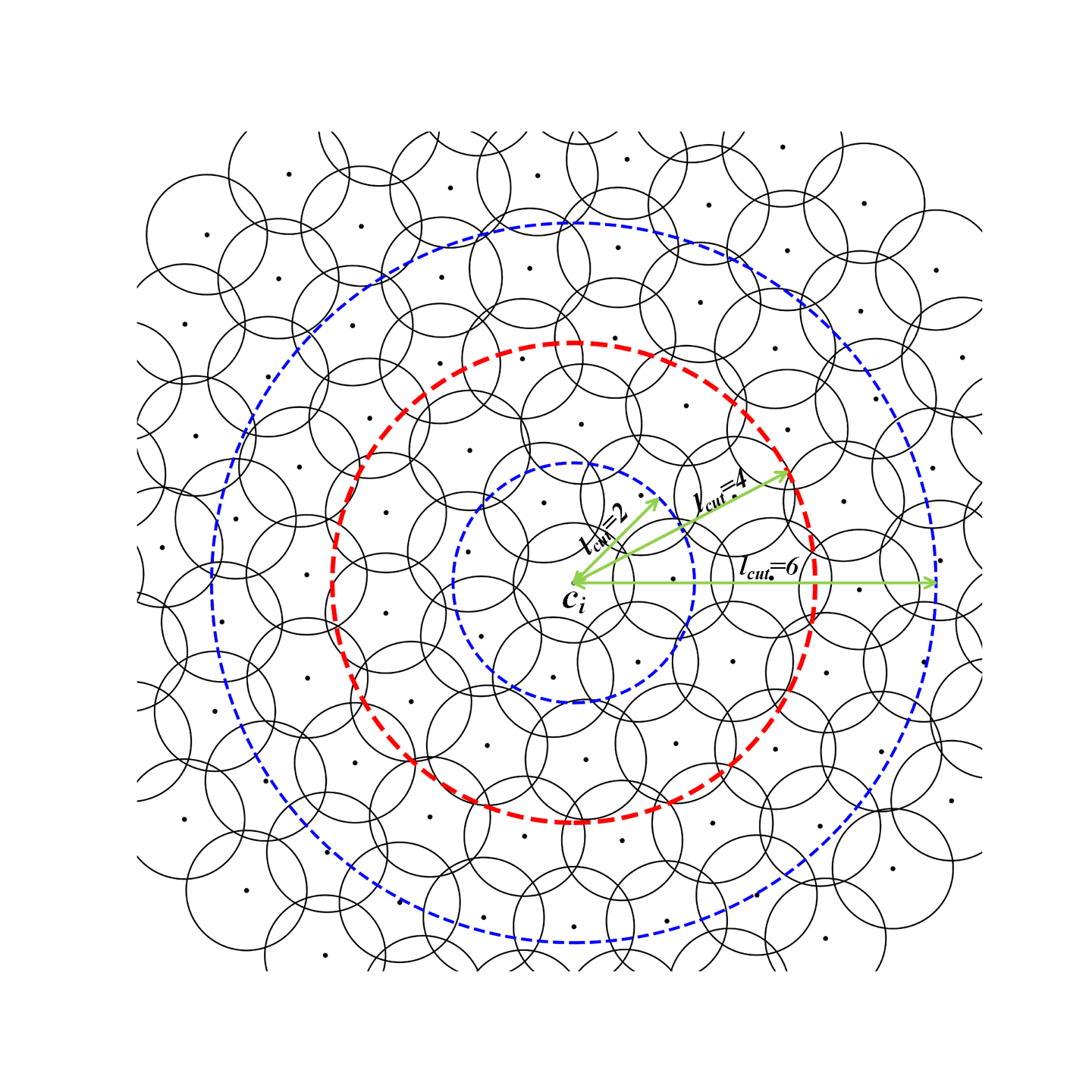}
    \caption{Illustration of the neighbors of a circle with different $l_{cut}$ settings. This illustrative example gives three settings for $l_{cut} = $ 2, 4 or 6 on a conflicting layout. We empirically set $l_{cut} = 4$ as a trade-off setting in this work.}
    \label{fig-neighbor}
\end{figure}

However, when $l_{cut} = 2$, the neighbor structure needs to be maintained even if the circles have minor shifts in the layout, otherwise, the correctness of the energy and gradient computation can not be guaranteed. Therefore, if we properly increase the value of $l_{cut}$, more adjacent circles are contained in the neighbor, so that the correctness is guaranteed without maintaining the neighbor structure when the circles have minor shifts in the layout. But if $l_{cut}$ is set too large, the neighbor contains many unnecessary circles, increasing the time cost of the energy and gradient computation. Therefore, we empirically set $l_{cut} = 4$ as a trade-off setting. Figure~\ref{fig-neighbor} gives an example to show the neighbors of a circle and the different settings of $l_{cut}$. Note that calculating the energies $E$ and $U$ and their gradients by enumerating the pairwise circles is of $O(n^2)$ complexity. By adopting the neighbor structure, the complexity can be reduced to $O(n)$~\citep{he2018efficient}.

\begin{algorithm}[t]
\caption{ANM ($cnt, len, \Gamma$)} \label{alg_mainei}
\textbf{Input}: A deferring counter $cnt$; A deferring length $len$; A neighbor $\Gamma$\\
\textbf{Output}: An updated counter $cnt$; A updated length $len$; A updated neighbor $\Gamma$ 
\begin{algorithmic}[1] 
\STATE $cnt \leftarrow cnt + 1$
\IF {$cnt \geq len$}
\STATE construct the current neighbor $\Gamma'$ 
\IF {$\Gamma \neq \Gamma'$}
\STATE $cnt \leftarrow 0, \ len \leftarrow 1, \ \Gamma \leftarrow \Gamma'$
\ELSE 
\STATE $cnt \leftarrow 0, \ len \leftarrow 2 \times len$
\ENDIF
\ENDIF
\STATE \textbf{return} $(cnt, len, \Gamma)$
\end{algorithmic}
\end{algorithm}

Our proposed ANM method is depicted in Algorithm~\ref{alg_mainei}. ANM maintains three parameters, i.e., the counter $cnt$, deferring length $len$, and historical neighbor $\Gamma$. Initially, the counter is set to $0$, the length is set to $1$, and the algorithm calculates an initial neighbor $\Gamma$ (refer to Algorithm~\ref{alg_batch} lines 3-4). In the continuous optimization process, the ANM is called after each iteration (see in Algorithm~\ref{alg_batch} line 12), and the counter will increase by $1$ when ANM is called. If the counter $cnt$ equals  the deferring length $len$, ANM reconstructs a new neighbor $\Gamma'$ and compares it with the historical neighbor $\Gamma$. If two neighbors $\Gamma'$ and $\Gamma$ are the same, the counter is set to $0$ and the deferring length $len$ is multiplied by $2$, otherwise, the counter is set to $0$, the deferring length is set to $1$ and the neighbor $\Gamma$ is updated by $\Gamma'$. 

The basic idea of ANM to defer maintaining the neighbor structure is based on the operation of the counter $cnt$ and the deferring length $len$. If the layout is unstable and the neighbor is changed, ANM reconstructs the neighbor in every iteration. If the layout is stable, the deferring length $len$ grows exponentially, which makes ANM defers maintaining the neighbor. It is worth noting that the Voronoi diagram approach can also be used to construct the neighbor, which can gain an excellent time complexity $O(n \log n)$. In this work, we adopt the scan line approach ($O(n \sqrt{n})$) to construct the neighbor, which is efficient enough and easy to implement. 

\citet{he2018efficient} use a simple strategy to maintain the neighbor, consisting in reconstructing the neighbor every 10 iterations. However, reconstructing the neighbor is unnecessary when the layout is stable, and it will waste computational resources in this case. \citet{lai2022iterated} propose a two-phase strategy. In the first phase, they calculate the energy and gradient by enumerating all the pairwise circles without using the neighbor structure. When the condition $\| g(\boldsymbol{x})\|_{\infty} < 10^{-2}$ is met, they change to the second phase. In the second phase, they construct the neighbor structure only at the beginning of the phase; then they use the neighbor structure to obtain the energy and gradient until the algorithm finds a local minimum solution. This two-phase strategy has three disadvantages: 1) The enumeration method in the first phase is computationally expensive, especially on large scale instances. 2) The threshold of the condition needs to be fine-tuned on the different scales of instances. 3) The neighbor structure is not updated in the second phase, so that if the solution falls into a saddle point, the correctness of this strategy can not be guaranteed. Our ANM method 
can well handle these issues.  

It is worth 
mentioning that our ANM is an adaptive method, ANM can handle dynamic problems where the number and the radius of the packing items can 
 be changed, such as the online packing problems~\citep{hokama2016bounded, fekete2017online, fekete2019online, lintzmayer2019online}.  


 


\subsection{The Complexity Analysis of \name}
\label{ssec:04-04comlex}

We now provide the time and space complexity analysis of our \name based on the BFGS optimization method.

\textbf{Time complexity.} We analyze the time complexity of each iteration. From Algorithm~\ref{alg_batch} lines 6 to 15, it is obvious that each iteration has 6 components, including the batch gradient $g(\boldsymbol{x}_p)$ calculation, the batch step length $\alpha_p$ calculation, the batch vector $\boldsymbol{x}_p$ update, the batch Hessian matrix $\boldsymbol{H}_p$ update, the maintenance module ANM, and the sum of the $k$-batch gradient norm. Since we adopt the efficient neighbor structure~\citep{he2018efficient} (discussed in Section~\ref{ssec:04-03nei}), the energy and gradient functions can be executed in time complexity $O(n)$. So, the time complexity of the batch gradient calculation, the step length calculation and the sum of $k$-batch gradient norm is $O(\frac{n}{k})$, $O(\beta\frac{n}{k})$ and $O(n)$, respectively, where the constant $\beta$ approximates the recursion depth of line search approach. Each batch has $\frac{n}{k}$ packing items, so the size of the batch Hessian matrix is $O((\frac{n}{k})^2)$. The batch vector update (Eq.~\eqref{eq-8}) and the batch Hessian matrix update (Eqs.~\eqref{eq-10} and \eqref{eq-11}) involve matrix and vector multiplication, so both of their time complexity are $O((\frac{n}{k})^2)$. We employ the scan line approach to construct the neighbor structure in the ANM module (discussed in Section~\ref{ssec:04-03nei}), its time complexity is $O(n\sqrt{n})$ (using Voronoi diagram approach can gain a better time complexity of $O(n\log{n})$). Finally, the time complexity of each iteration is $O(k(\frac{n}{k} + \beta\frac{n}{k} + (\frac{n}{k})^2) + n + n\sqrt{n})$, which can be simplified as $O(\beta{n} + n\sqrt{n} + \frac{n^2}{k})$, and it becomes $O(\beta{n} + \frac{n^2}{k})$ if the ANM module defers the maintaining process.

\textbf{Space complexity.} The memory requirement of the \name method is mainly used to store the $k$ batch Hessian matrices. Therefore, the space complexity of the \name method is $O(k(\frac{n}{k})^2)$, which can be simplified as $O(\frac{n^2}{k})$.

As discussed above, our \name method has lower time and space complexity ($O(\beta{n} + \frac{n^2}{k})$ and $O(\frac{n^2}{k})$) than the classic BFGS method ($O(n^2)$ and $O(n^2)$). It degenerates to the BFGS method when $k=1$.

\section{Search Heuristic} \label{sec:05heur}
\begin{algorithm}[tb]
\caption{The framework for solving PECC} \label{alg_frame}
\textbf{Input}: A number of unit circles $n$; A best-known container radius $R_b$; The cut-off time $T_{cut}$ \\
\textbf{Output}: A feasible solution with the minimal container radius $(\boldsymbol{x}^{*}, R^{*})$
\begin{algorithmic}[1] 
\STATE $\boldsymbol{x}^{*} \leftarrow \mathrm{random\_solution}(n, R_b)$
\STATE $\boldsymbol{x}^{*} \leftarrow \mathrm{\name}(\boldsymbol{x}^{*}, R_b)$
\STATE $(\boldsymbol{x}^{*}, R^{*}) \leftarrow \mathrm{adjust\_container}(\boldsymbol{x}^{*}, R_b)$
\WHILE {$\mathrm{time}() \leq T_{cut}$}
\STATE $R \leftarrow \min(R_b, R^{*})$
\STATE $\boldsymbol{x} \leftarrow \mathrm{SED}(n, R)$
\STATE $(\boldsymbol{x}, R) \leftarrow \mathrm{adjust\_container}(\boldsymbol{x}, R)$
\IF {$R < R^{*}$}
\STATE $\boldsymbol{x}^{*} \leftarrow \boldsymbol{x}, \ R^{*} \leftarrow R$
\ENDIF 
\ENDWHILE
\STATE \textbf{return} $(\boldsymbol{x}^{*}, R^{*})$
\end{algorithmic}
\end{algorithm}

The algorithms for solving the PECC problem based on the elastic model can be divided into two phases. In the first phase, the container radius is fixed, and the goal is to find 
a feasible solution or an infeasible solution with as few overlaps as possible (i.e., the energy $E$ being as minimal as possible). In the second phase, the algorithms expand or shrink the container radius to obtain a feasible solution with the container radius being as minimum as possible. Starting from a random candidate solution, although we can obtain a feasible solution by employing the \name method (Section~\ref{ssec:04-01GBO}) to accomplish the first phase and employing the container adjustment method (Section~\ref{ssec:04-02adj}) to accomplish the second phase. However, the quality of the solution obtained in this way is still unsatisfactory. 

Through sufficient experiments, we observe that the quality of the final solution is directly impacted by the solution obtained in the first phase. If the energy $E(\boldsymbol{x})$, corresponding to the overlapping area, of the obtained solution in the first phase is large, the expanded difference of the radius adjustment in the second phase is also large. On the other hand, if a feasible solution is found in the first phase, the radius can be shrunk in the second phase. Therefore, a minimum energy solution discovered in the first phase is important to obtain a final high-quality feasible solution. However, the elastic model converts the PECC problem to a non-convex optimization problem as discussed in Section~\ref{ssec:03-02QPQH}, and it is extremely difficult to find a global minimum solution. Therefore, we propose an efficient Solution-space Exploring and Descent (SED) heuristic for the first phase to discover a solution with as minimum energy as possible.


\subsection{The Framework for Solving PECC} \label{ssec:05-01frame}

We first introduce our algorithm framework, of which the pseudocode is depicted in Algorithm~\ref{alg_frame}.

Initially, the algorithm adopts the best-known radius~\citep{Spechtweb} $R_b$ as the initial fixed container radius and generates a random layout as the initial solution (line 1). The center $(x_i, y_i)$ of each circle satisfies $x_i, y_i \in \mathcal{U}(-R_b, R_b)$ ($\mathcal{U}$ is denoted as the continuous uniform distribution). Then, the algorithm employs our \name method (Section~\ref{ssec:04-01GBO}) to minimize the energy $E(\boldsymbol{x}^*)$ (line 2) to obtain an initial feasible solution  $\boldsymbol{x}^*$ (line 3) by employing the container adjusting method (Section~\ref{ssec:04-02adj}). Next, the algorithm performs an iterative process to improve the best recorded solution until the cutoff time $T_{cut}$ is reached (lines 4-11). 

At each iteration, the algorithm sets the target radius $R$ as min($R_b$, $R^*$) (line 5), where $R_b$ is the best-known radius $R_b$ and $R^*$ is the best feasible radius $R^*$ found so far. Then, it adopts our proposed SED heuristic (Section~\ref{ssec:05-02SED}) to discover a candidate solution $\boldsymbol{x}$ with as low energy as possible (line 6).
Subsequently, the algorithm computes a new feasible solution by applying the container radius adjusting method to the candidate solution (line 7). If the new feasible radius is smaller than the best recorded radius, indicating a better feasible solution is found, then the best record solution is updated (lines 8-10).
Finally, the algorithm returns the best recorded solution as the result.

Note that our goal is to solve large scale instances, which is an incredibly big computational challenge. Using the best-known radius instead of an approximate radius or a radius obtained by an initial method as the baseline can reduce the computational difficulty, quickly discover a high-quality solution and improve the algorithm performance. Therefore, it is an efficient quick-start method. 

\subsection{The Solution-Space Exploring and Descent Heuristic} \label{ssec:05-02SED}

Our proposed SED heuristic aims to solve a problem described as follows. 
Given a fixed container radius, the problem determines whether there is a feasible solution, and it is essentially a decision PECC problem. If SED discovers a feasible solution, it returns the solution immediately; otherwise, 
it returns an infeasible solution with the smallest energy during the search process. 
First, we define a new metric function $J$, formulated as follows:
\begin{gather}
    J(\boldsymbol{x}) = \lceil -\log_{10} E(\boldsymbol{x}) \rceil, \nonumber
\end{gather}
where function $J$ is the ceiling of the negative logarithm of the function $E$. It maps the energy $E$ to an integer. And the value of function $J$ is applied to control the exploring number in the heuristic process. 

\begin{algorithm}[tb]
\caption{SED($n$, $R$)} \label{alg_SED}
\textbf{Input:} A number of unit circles $n$; A container radius $R$ \\
\textbf{Output:} A smallest energy solution found so far $\boldsymbol{x}^{*}$
\begin{algorithmic}[1] 
\STATE $\boldsymbol{x} \leftarrow \mathrm{random\_solution}(n, R)$
\STATE $\boldsymbol{x} \leftarrow \mathrm{\name}(\boldsymbol{x}, R)$
\STATE $\boldsymbol{x}^* \leftarrow \boldsymbol{x}$
\FOR {$i$ from $1$ to $S_{iter}$}
\IF {$E(\boldsymbol{x}^{*}) \leq 10^{-25}$}
\STATE \textbf{break}
\ENDIF
\STATE $m \leftarrow \max(1, J(\boldsymbol{x})), \  U \leftarrow \emptyset$
\FOR {$j$ from $1$ to $m$} 
\STATE $\boldsymbol{x}' \leftarrow \mathrm{perturbing}(\boldsymbol{x})$
\STATE $\boldsymbol{x}' \leftarrow \mathrm{\name}(\boldsymbol{x}')$
\STATE $U \leftarrow U \cup \{\boldsymbol{x}'\}$
\ENDFOR
\STATE $\boldsymbol{x} \leftarrow \mathrm{select}(U)$
\IF {$E(\boldsymbol{x}) < E(\boldsymbol{x}^{*})$}
\STATE $\boldsymbol{x}^{*} \leftarrow \boldsymbol{x}$
\ENDIF
\ENDFOR
\STATE \textbf{return} $\boldsymbol{x}^{*}$
\end{algorithmic}
\end{algorithm}

Now, we introduce our proposed SED heuristic depicted in Algorithm~\ref{alg_SED}.
Starting from a fixed container radius $R$, SED first generates a random configuration as the initial solution and minimizes the energy of the initial solution $\boldsymbol{x}$ by adopting our \name method (lines 1-2). Then, SED performs an iterative process to discover a feasible solution (lines 4-18). 
At each iteration, SED obtains a value $m$ by function $J$ as the perturbing number (line 8), a candidate solution set $U$, $|U| = m$, is created where the candidate solution $\boldsymbol{x}'$ in $U$ is obtained by perturbing the current operated solution $\boldsymbol{x}$ and minimizing energy $E(\boldsymbol{x}')$ by adopting \name (lines 9-13). Then SED adopts a $select$ operator to choose a candidate solution from set $U$ to replace the solution $\boldsymbol{x}$. If the energy of new solution $\boldsymbol{x}$ is smaller than the energy of the best record solution $\boldsymbol{x}^*$, then $\boldsymbol{x}^*$ is updated (lines 15-17). If the energy of $\boldsymbol{x}^*$ is tiny enough, indicating a feasible solution is discovered, SED returns the 
feasible solution $\boldsymbol{x}^*$ immediately (lines 5-7), otherwise, SED returns an infeasible solution with the smallest energy during the iteration process when reaching the maximum number of iteration steps $S_{iter}$. 

To obtain a perturbed solution $\boldsymbol{x}'$, we randomly shift the coordinate of the circles in the operated solution $\boldsymbol{x}$ which can be described as follows, $x'_i \leftarrow x_i + r$ and $y'_i \leftarrow y_i + r \ \ (1 \leq i \leq n)$, where $r$ is a random number, $r \in \mathcal{U}(-0.8, 0.8)$. 

The strategy of the $select$ operator is described as follows:
\begin{equation} \label{eq-18}
    \mathrm{select}(U) =
    \begin{cases}
    \underset{\boldsymbol{y} \in U}{\arg\min} \ E(\boldsymbol{y}), &
    \mathrm{if} \ \ \underset{\boldsymbol{y} \in U}{\min} \ E(\boldsymbol{y}) < E(\boldsymbol{x}) \\
    P \left( X = \boldsymbol{y} \mid p_{\boldsymbol{y}} = softmax(J(\boldsymbol{y}))  \right), &
    \mathrm{otherwise} 
    \end{cases} \nonumber
\end{equation}
\begin{gather}
    softmax(J(\boldsymbol{y})) = \frac{\exp(J(\boldsymbol{y}))}{\sum_{\boldsymbol{y}' \in U} \exp(J(\boldsymbol{y}'))}.  \nonumber
\end{gather}
This operator first compares the candidate solution with the smallest energy in set $U$ to the current operated solution $\boldsymbol{x}$. If the energy of the candidate solution is smaller than the operated solution, the operator replaces the operated solution as the candidate solution; otherwise, the operator selects a candidate solution from set $U$ according to the probability of the softmax function.



\section{Experiments} \label{sec:06exp}
For experiments, we first evaluate the performance of our proposed \name method on the different number of batches and different geometric partition methods, then we present the comparisons between the results and also give the parameter studies.

\subsection{Experimental Setup} \label{ssec:06-01set}
Our algorithm was implemented in C++ and compiled using g++ 5.4.0.
Experiments were performed on a server with Intel® Xeon® E5-2650 v3 CPU and 256 GBytes RAM, running the Linux OS. 
Due to the randomness, we run our algorithm  
multiple times independently with different random seeds (CPU timestamps). 
To evaluate the performance of our algorithm thoroughly, we select three instance scales as our benchmarks, described as follows:

\begin{itemize}
    \item \textbf{Moderate scale}: $300 \leq n \leq 320$, for comparing with the state-of-the-art algorithm IDTS~\citep{lai2022iterated}. We set $k = 3$ for the batch number of \name, 12 hours for cut-off time $T_{cut}$ of our overall search. The algorithm performs 20 times independently, where the settings of cut-off time and performing time are the same as in the IDTS work.
    \item \textbf{Large scale I}: $500 \leq n < 800$, for comparing with the best-known results~\citep{Spechtweb}. We set $k = 5$ for the batch number, 24 hours for cut-off time $T_{cut}$. The algorithm performs 10 times independently. 
    \item \textbf{Large scale II}: $800 \leq n \leq 1000$, for comparing with the best-known results~\citep{Spechtweb}. We set $k = 5$ for the batch number, 48 hours for cut-off time $T_{cut}$. The algorithm performs 10 times independently. 
\end{itemize}

The rest of the parameters are consistently set as follows. The geometrical partition method is sector (Section~\ref{ssec:04-01GBO}), the maximum iteration steps of \name $MaxIter = 5000$ (Section~\ref{ssec:04-01GBO}), and the maximum iteration steps of SED $S_{iter} = 500$ (Section~\ref{ssec:05-02SED}). The parameters tuning and analysis are presented in Section~\ref{ssec:06-05para}.

\subsection{Comparison on the Well-Studied Moderate Scale Instances} \label{ssec:06-02moderate}

\begin{sidewaystable} 
\centering
\caption{Comparison between the best-known results, IDTS and our proposed algorithms SED (1-batch \name) and (5-batch \name) on the 21 well-studied moderate scale instances. The improved best results $R_{best}$ found by our proposed algorithms appear in bold.}
\label{tb-cmp-moderate}
\resizebox{1.0\textwidth}{!}{
\begin{tabular}{llllllllllllllllllllllllllll}
\toprule
\multirow{2}{*}{$n$} & \multirow{2}{*}{$R^*$} &           & \multirow{2}{*}{IDST} &  & \multicolumn{11}{l}{SED (1-batch \name)}                                                                       &  & \multicolumn{11}{l}{SED (3-batch \name)}                                                                                \\ \cline{6-16} \cline{18-28} 
                   &                     &           &                       &  & $R_{best}$               &           & $R_{avg}$       &           & $R_{best}-R^*$ &  & $RR$    &  & $HR$    &  & $time~(s)$ &  & $R_{best}$               &           & $R_{avg}$                &           & $R_{best}-R^*$ &  & $RR$    &  & $HR$    &  & $time~(s)$ \\ \midrule
300                & 18.813153706        & \textbf{} & 18.813153706          &  & 18.813157576          & \textbf{} & 18.813191071 & \textbf{} & 3.87E-06   &  & 0/20  &  & 1/20  &  & 26216   &  & 18.813189941          & \textbf{} & 18.813198345          & \textbf{} & 3.62E-05   &  & 0/20  &  & 12/20 &  & 33237   \\
301                & 18.843463507        & \textbf{} & 18.843463507          &  & 18.843463507          & \textbf{} & 18.844498079 & \textbf{} & 0.00E+00   &  & 2/20  &  & 2/20  &  & 26222   &  & 18.843463507          & \textbf{} & 18.843551084          & \textbf{} & 0.00E+00   &  & 1/20  &  & 1/20  &  & 38618   \\
302                & 18.891782255        & \textbf{} & 18.891782255          &  & \textbf{18.891781604} & \textbf{} & 18.892064228 & \textbf{} & -6.51E-07  &  & 2/20  &  & 1/20  &  & 20792   &  & 18.891782255          & \textbf{} & 18.892033313          & \textbf{} & 0.00E+00   &  & 2/20  &  & 2/20  &  & 23987   \\
303                & 18.929749153        & \textbf{} & 18.929749153          &  & 18.929749153          & \textbf{} & 18.930326363 & \textbf{} & 0.00E+00   &  & 1/20  &  & 1/20  &  & 23549   &  & 18.929750618          & \textbf{} & 18.930328723          & \textbf{} & 1.47E-06   &  & 0/20  &  & 1/20  &  & 23652   \\
304                & 18.964441751        & \textbf{} & 18.964441751          &  & \textbf{18.964297557} & \textbf{} & 18.964819754 & \textbf{} & -1.44E-04  &  & 3/20  &  & 1/20  &  & 24562   &  & \textbf{18.963620323} & \textbf{} & 18.964664008          & \textbf{} & -8.21E-04  &  & 5/20  &  & 1/20  &  & 18852   \\
305                & 19.001754565        & \textbf{} & 19.001754565          &  & \textbf{19.001744832} & \textbf{} & 19.002856526 & \textbf{} & -9.73E-06  &  & 1/20  &  & 1/20  &  & 24542   &  & \textbf{19.001726813} & \textbf{} & 19.002734687          & \textbf{} & -2.78E-05  &  & 2/20  &  & 1/20  &  & 21922   \\
306                & 19.030389407        & \textbf{} & 19.030389407          &  & 19.031079983          & \textbf{} & 19.031763629 & \textbf{} & 6.91E-04   &  & 0/20  &  & 1/20  &  & 21131   &  & 19.030651242          & \textbf{} & 19.031391719          & \textbf{} & 2.62E-04   &  & 0/20  &  & 2/20  &  & 25226   \\
307                & 19.060160922        & \textbf{} & 19.060160922          &  & 19.061100857          & \textbf{} & 19.062150163 & \textbf{} & 9.40E-04   &  & 0/20  &  & 1/20  &  & 23579   &  & 19.060841920          & \textbf{} & 19.061955433          & \textbf{} & 6.81E-04   &  & 0/20  &  & 1/20  &  & 26981   \\
308                & 19.104991437        & \textbf{} & 19.104991437          &  & 19.109083748          & \textbf{} & 19.110722905 & \textbf{} & 4.09E-03   &  & 0/20  &  & 1/20  &  & 26804   &  & 19.107239953          & \textbf{} & 19.109820607          & \textbf{} & 2.25E-03   &  & 0/20  &  & 1/20  &  & 23291   \\
309                & 19.142573165        & \textbf{} & 19.142573165          &  & \textbf{19.141335827} & \textbf{} & 19.143371324 & \textbf{} & -1.24E-03  &  & 5/20  &  & 1/20  &  & 25662   &  & 19.142625039          & \textbf{} & 19.143620434          & \textbf{} & 5.19E-05   &  & 0/20  &  & 1/20  &  & 24278   \\
310                & 19.178928265        & \textbf{} & 19.178928265          &  & \textbf{19.178560517} & \textbf{} & 19.179857308 & \textbf{} & -3.68E-04  &  & 2/20  &  & 1/20  &  & 18906   &  & \textbf{19.178419320} & \textbf{} & 19.179442843          & \textbf{} & -5.09E-04  &  & 6/20  &  & 1/20  &  & 24718   \\
311                & 19.212365036        & \textbf{} & 19.212365036          &  & \textbf{19.210669807} & \textbf{} & 19.213148503 & \textbf{} & -7.17E-04  &  & 5/20  &  & 1/20  &  & 22109   &  & \textbf{19.210564074} & \textbf{} & 19.212561780          & \textbf{} & -8.22E-04  &  & 4/20  &  & 2/20  &  & 21153   \\
312                & 19.233585653        & \textbf{} & 19.233585653          &  & 19.233585653          & \textbf{} & 19.234863780 & \textbf{} & 0.00E+00   &  & 1/20  &  & 1/20  &  & 23934   &  & 19.233585653          & \textbf{} & 19.234529499          & \textbf{} & 0.00E+00   &  & 4/20  &  & 4/20  &  & 30386   \\
313                & 19.257103014        & \textbf{} & 19.257103014          &  & \textbf{19.256994660} & \textbf{} & 19.258015876 & \textbf{} & -1.08E-04  &  & 1/20  &  & 1/20  &  & 21251   &  & 19.257103014          & \textbf{} & 19.258284406          & \textbf{} & 0.00E+00   &  & 1/20  &  & 1/20  &  & 25457   \\
314                & 19.286195141        & \textbf{} & 19.286195141          &  & \textbf{19.286190236} & \textbf{} & 19.286514591 & \textbf{} & -4.91E-06  &  & 4/20  &  & 4/20  &  & 25125   &  & \textbf{19.286190236} & \textbf{} & 19.286476954          & \textbf{} & -4.91E-06  &  & 1/20  &  & 1/20  &  & 26386   \\
315                & 19.302288067        & \textbf{} & 19.302288067          &  & \textbf{19.302273991} & \textbf{} & 19.302586043 & \textbf{} & -1.41E-05  &  & 17/20 &  & 11/20 &  & 25517   &  & \textbf{19.302273991} & \textbf{} & \textbf{19.302286175} & \textbf{} & -1.41E-05  &  & 18/20 &  & 3/20  &  & 24488   \\
316                & 19.334041754        & \textbf{} & 19.334041754          &  & 19.334041754          & \textbf{} & 19.334805176 & \textbf{} & 0.00E+00   &  & 9/20  &  & 9/20  &  & 25370   &  & 19.334041754          & \textbf{} & 19.334584281          & \textbf{} & 0.00E+00   &  & 7/20  &  & 7/20  &  & 22840   \\
317                & 19.367595672        & \textbf{} & 19.367595672          &  & 19.367595672          & \textbf{} & 19.367930543 & \textbf{} & 0.00E+00   &  & 10/20 &  & 10/20 &  & 23713   &  & 19.367595672          & \textbf{} & 19.367871121          & \textbf{} & 0.00E+00   &  & 9/20  &  & 9/20  &  & 23129   \\
318                & 19.391566091        & \textbf{} & 19.391566091          &  & 19.391566091          & \textbf{} & 19.391909306 & \textbf{} & 0.00E+00   &  & 15/20 &  & 15/20 &  & 22836   &  & 19.391566091          & \textbf{} & 19.391631572          & \textbf{} & 0.00E+00   &  & 17/20 &  & 17/20 &  & 27084   \\
319                & 19.424277830        & \textbf{} & 19.424277830          &  & 19.424277830          & \textbf{} & 19.424920970 & \textbf{} & 0.00E+00   &  & 15/20 &  & 15/20 &  & 18382   &  & 19.424277830          & \textbf{} & 19.425310867          & \textbf{} & 0.00E+00   &  & 12/20 &  & 12/20 &  & 24782   \\
320                & 19.456230764        & \textbf{} & 19.451649630          &  & \textbf{19.451583741} & \textbf{} & 19.453206681 & \textbf{} & -4.65E-03  &  & 20/20 &  & 1/20  &  & 27713   &  & 19.451734176          & \textbf{} & 19.454455527          & \textbf{} & -4.50E-03  &  & 19/20 &  & 3/20  &  & 30039   \\ \midrule
\#Improve          &                     &           &                       &  & 10                    &           &              &           &            &  &       &  &       &  &         &  & 6                     &           &                       &           &            &  &       &  &       &  &         \\
\#Equal            &                     &           &                       &  & 7                     &           &              &           &            &  &       &  &       &  &         &  & 8                     &           &                       &           &            &  &       &  &       &  &         \\
\#Worse            &                     &           &                       &  & 4                     &           &              &           &            &  &       &  &       &  &         &  & 7                     &           &                       &           &            &  &       &  &       &  &         \\ \bottomrule
\end{tabular}
}
\end{sidewaystable}

We perform our proposed SED (3-batch \name) algorithm and its variant SED (1-batch \name) on the moderate scale instances ($300 \leq n \leq 320$). The comparisonal results of two algorithms with best-known results~\citep{Spechtweb} and IDTS~\citep{lai2022iterated} are shown in Table~\ref{tb-cmp-moderate}. Note that SED (1-batch \name) only changes the number of batches from $k=3$ to $k=1$ for SED (3-batch \name), and the 1-batch \name degenerates to the classic BFGS method.

In Table~\ref{tb-cmp-moderate}, $n$ corresponds to the number of items in the instance, 
$R^{*}$ is for the best-known results from the Packomania website~\citep{Spechtweb} (download data 2022/10/1), followed by the best results of IDTS~\citep{lai2022iterated}. 
SED (1-batch \name) and SED (3-batch \name) correspond to the results of our methods:  $R_{best}$ shows the best result of 20 independent runs, $R_{avg}$ shows the average result of 20 independent runs, $R_{best}-R^*$ is the difference between $R_{best}$ and $R^*$ (a negative value indicates an improved best result), $RR$ is the ratio of better than or equal to the best-known result $R^*$, $HR$ is the ratio of hitting the best value $R_{best}$, and $time~(s)$ shows the average time of obtaining a best solution in seconds. 
At the bottom of the table, ``\#Improve'', ``\#Equal'' and ``\#Worse'' indicates that for our two algorithms SED (1-batch \name) and SED (3-batch \name), the number of instances that our algorithm obtained better, equal, or worse result than the best of $R^*$ and IDTS.

From Table~\ref{tb-cmp-moderate}, we can draw several conclusions as follows:
\begin{itemize}
    \item [(1)] Our proposed heuristic SED with 1-batch \name (i.e., classic BFGS) has 10 improved best results, 7 equal best results and 4 worse best results to IDTS on the 21 moderate scale instances. It demonstrates that SED outperforms IDTS. Note that the best-known results are same as IDTS, excluding $n = 320$, and IDTS adopts L-BFGS as the optimization method, which is also a Quasi-Newton method.
    \item [(2)] SED (3-batch \name) has 6 improved best results, 8 equal best results and 7 worse best results on the 21 moderate scale instances. The results show that the 3-batch \name does not outperform the 1-batch \name (classic BFGS method). It implies the multi-batch \name does not work well on moderate scale instances. 
    \item [(3)] From the ratio of hitting the best result $R_{best}$ ($HR$), there are few ratios of $HR$ high than 10/20, the most of the ratios of HR are equal to 1/20. From the IDTS work, we also observe that all the ratios of $HR$ in the IDTS work for $300 \leq n \leq 320$ are less than 6/20, and there are 14 ratios of $HR$ equal to 1/20. These results demonstrate the moderate scale instances are well-studied and obtaining an improved best result is very difficult.  
\end{itemize}

\begin{figure}
    \centering
    \begin{minipage}[b]{0.33\linewidth}
        \centering
        \subfloat[][$n = 302$]{\includegraphics[width=1\linewidth]{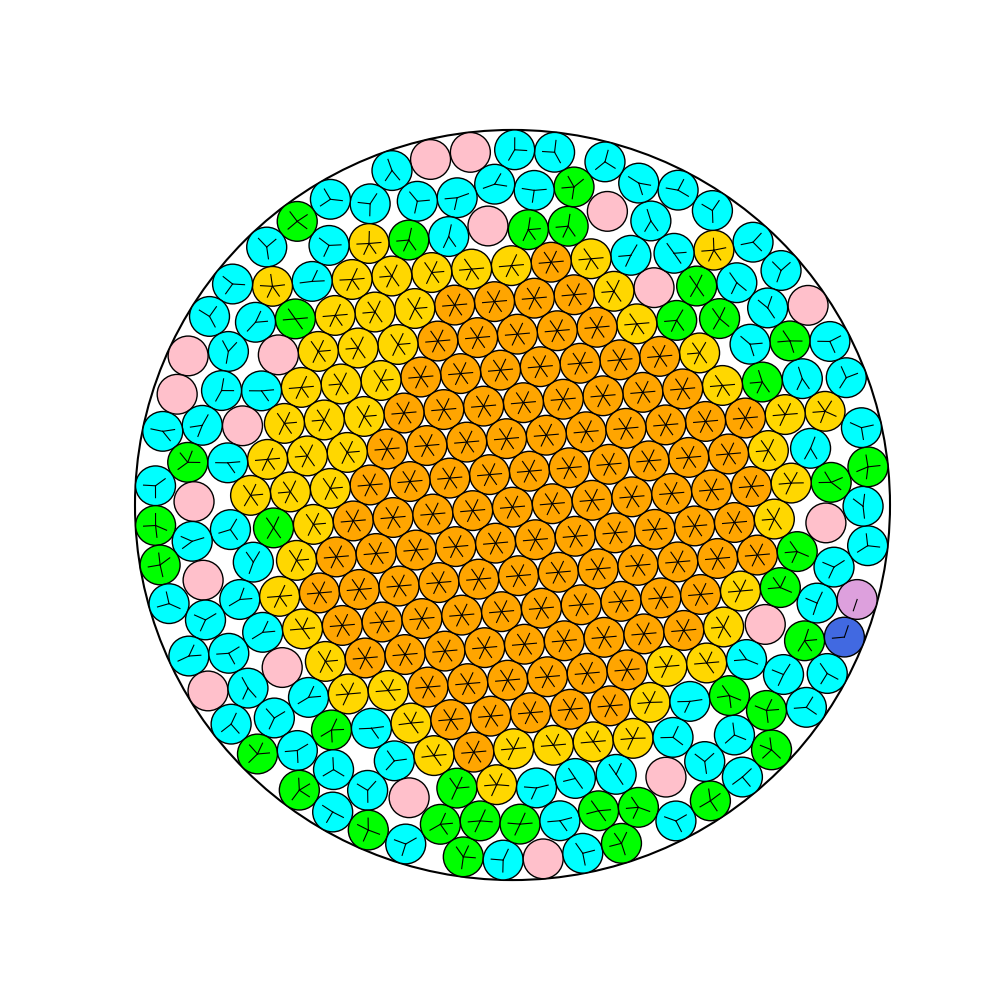}}
    \end{minipage}
    \begin{minipage}[b]{0.33\linewidth}
        \centering
        \subfloat[][$n = 304$]{\includegraphics[width=1\linewidth]{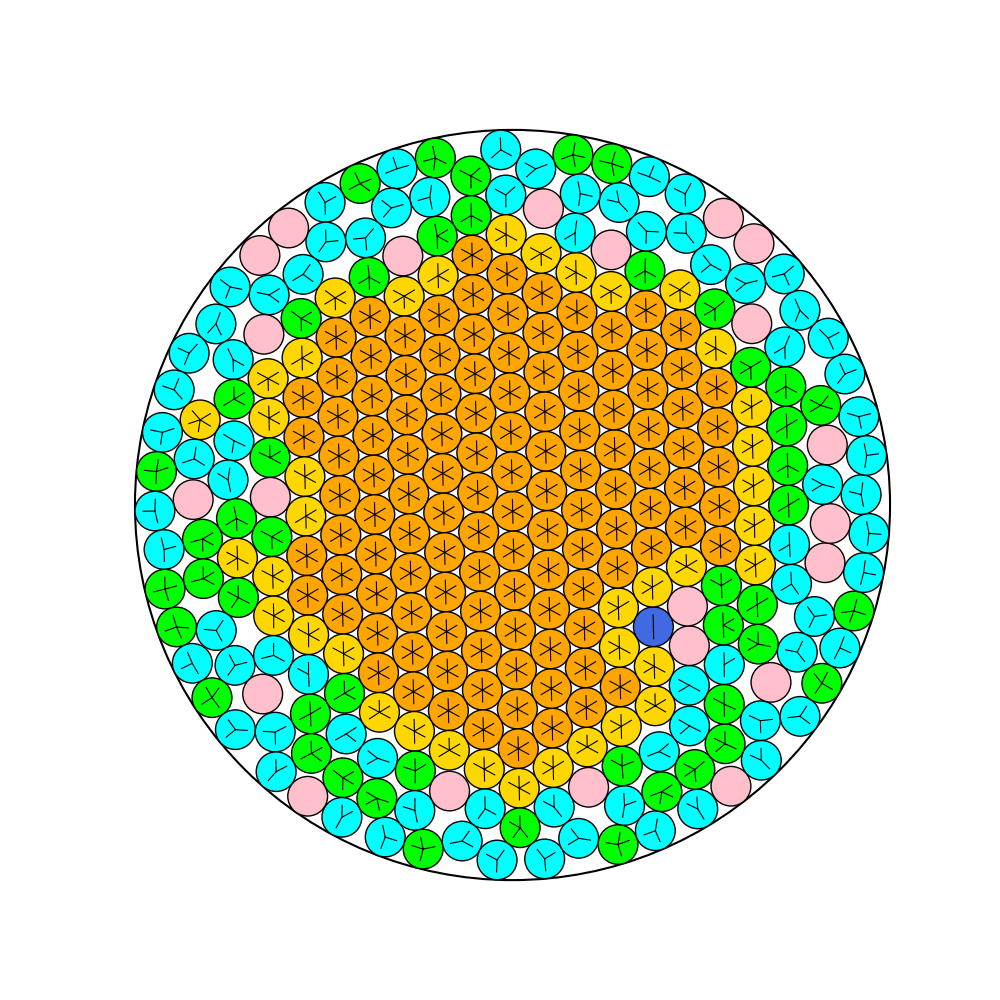}}
    \end{minipage}    
    \begin{minipage}[b]{0.33\linewidth}
        \centering
        \subfloat[][$n = 305$]{\includegraphics[width=1\linewidth]{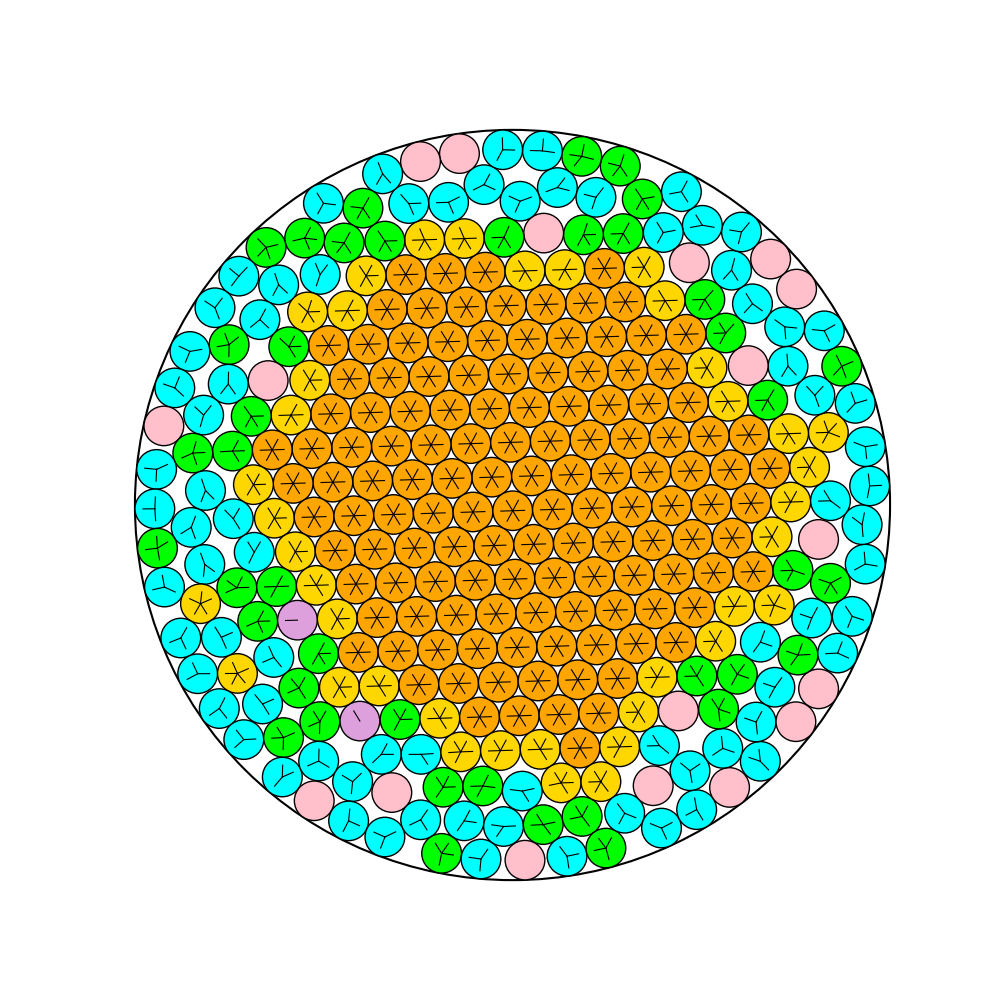}}
    \end{minipage}
    \begin{minipage}[b]{0.33\linewidth}
        \centering
        \subfloat[][$n = 309$]{\includegraphics[width=1\linewidth]{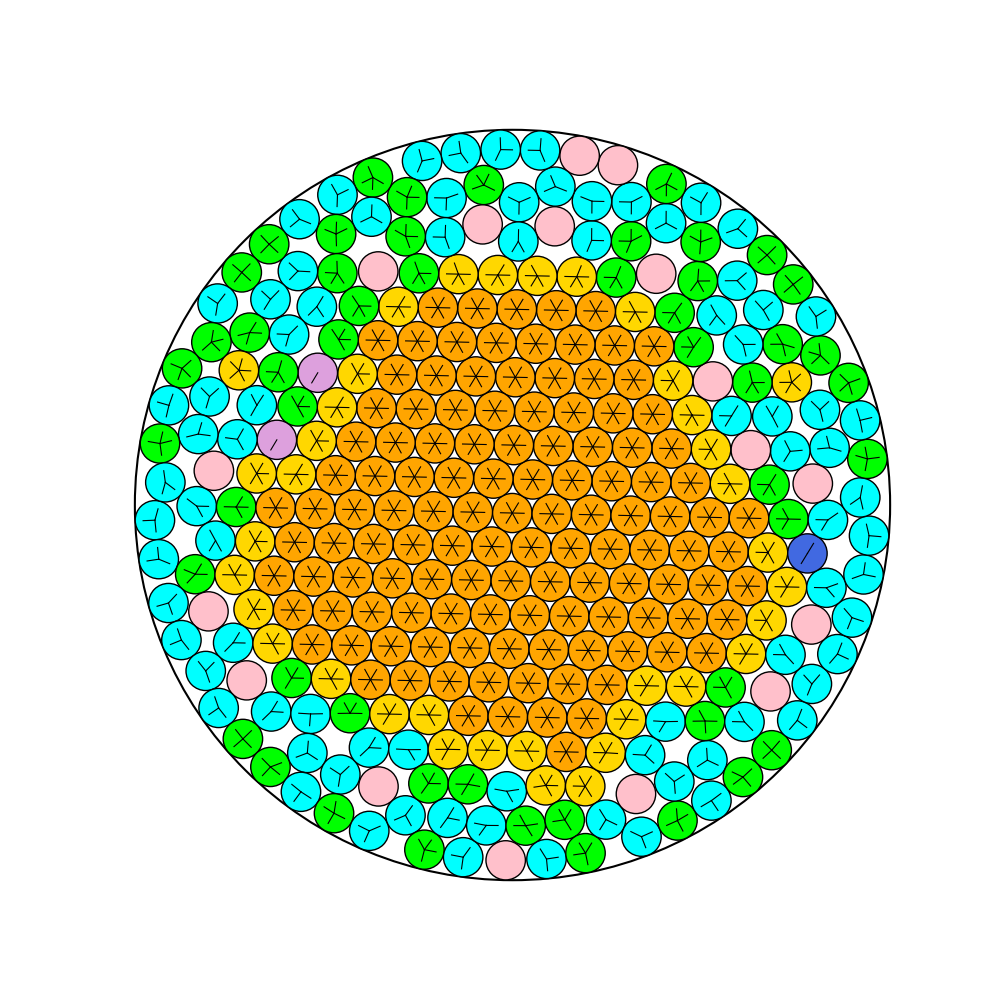}}
    \end{minipage}
    \begin{minipage}[b]{0.33\linewidth}
        \centering
        \subfloat[][$n = 310$]{\includegraphics[width=1\linewidth]{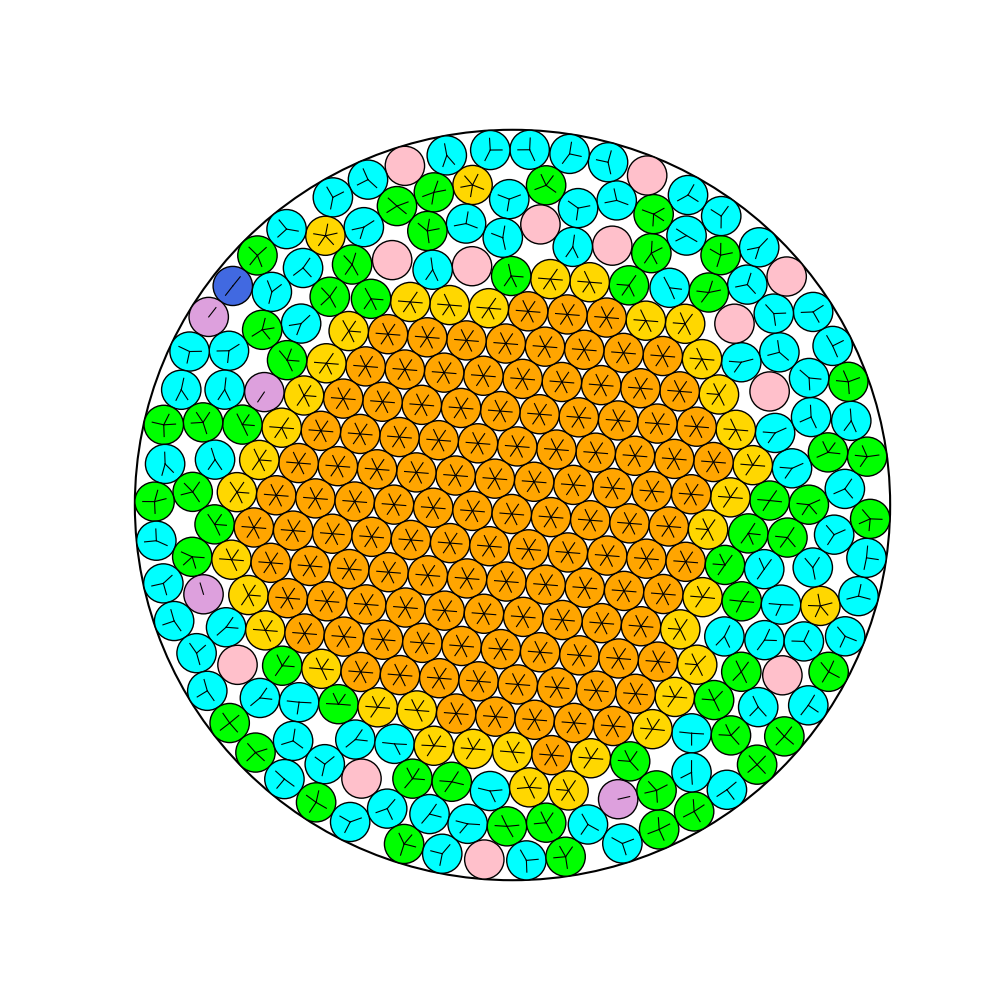}}
    \end{minipage}    
    \begin{minipage}[b]{0.33\linewidth}
        \centering
        \subfloat[][$n = 311$]{\includegraphics[width=1\linewidth]{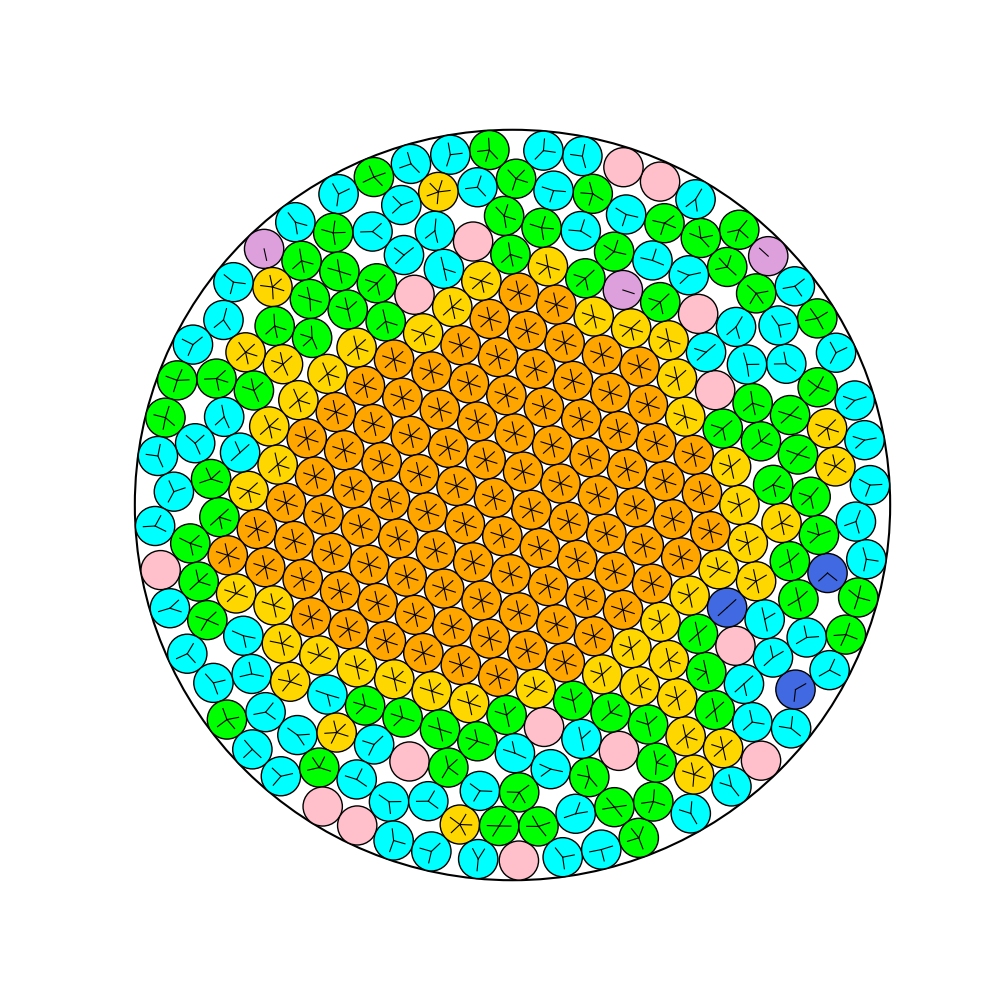}}
    \end{minipage}
    \begin{minipage}[b]{0.33\linewidth}
        \centering
        \subfloat[][$n = 313$]{\includegraphics[width=1\linewidth]{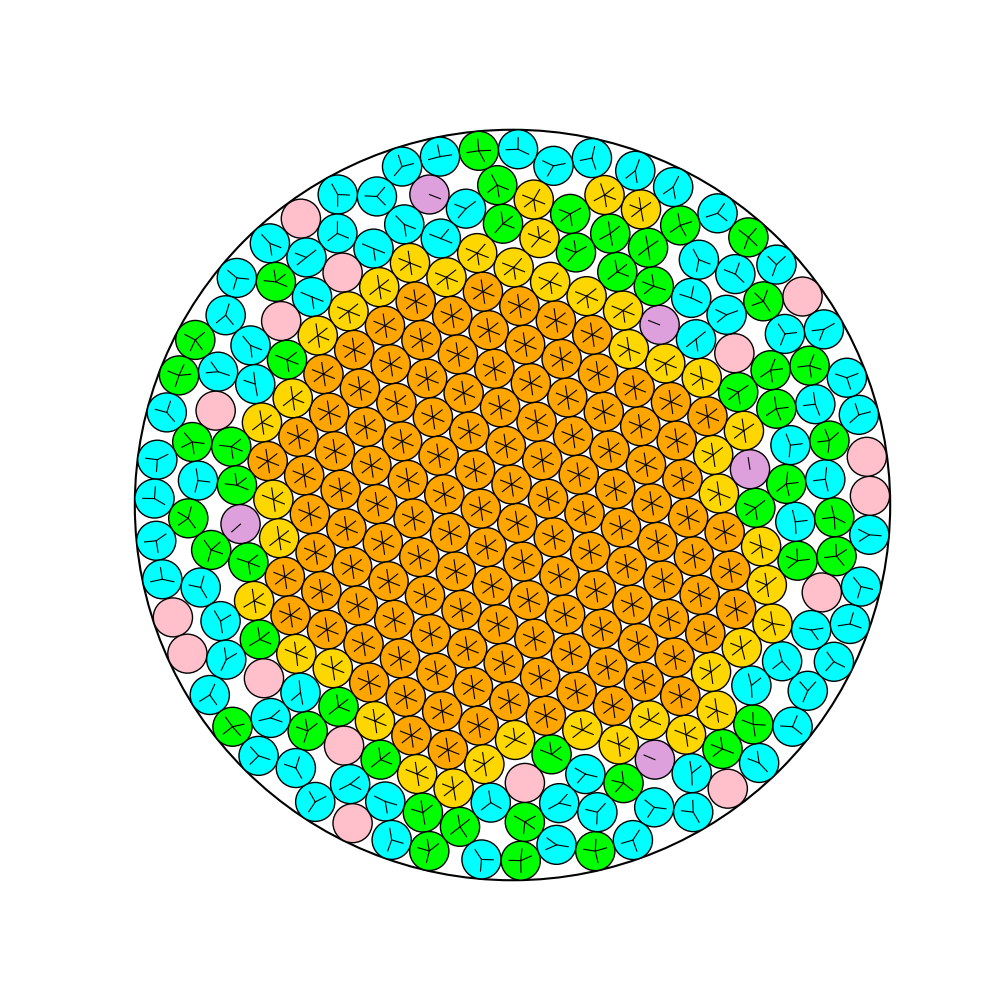}}
    \end{minipage}
    \begin{minipage}[b]{0.33\linewidth}
        \centering
        \subfloat[][$n = 314$]{\includegraphics[width=1\linewidth]{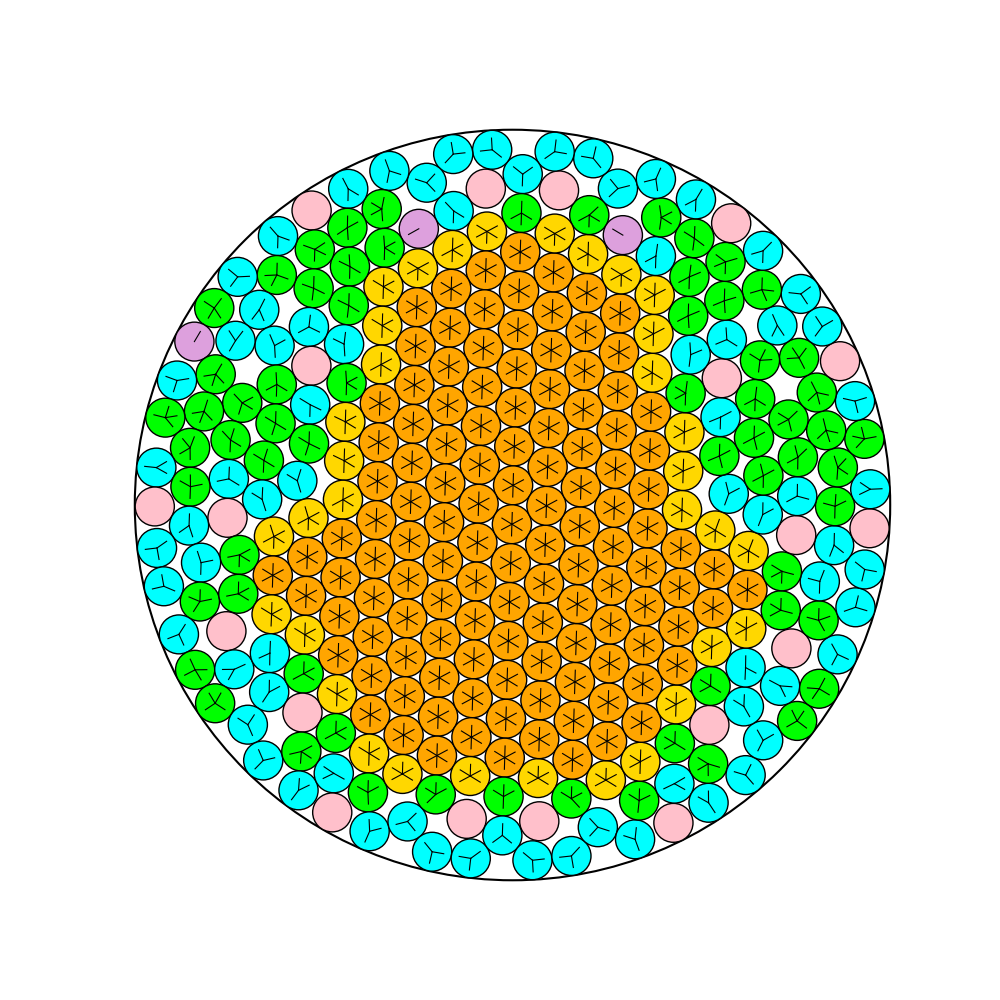}}
    \end{minipage}    
    \begin{minipage}[b]{0.33\linewidth}
        \centering
        \subfloat[][$n = 315$]{\includegraphics[width=1\linewidth]{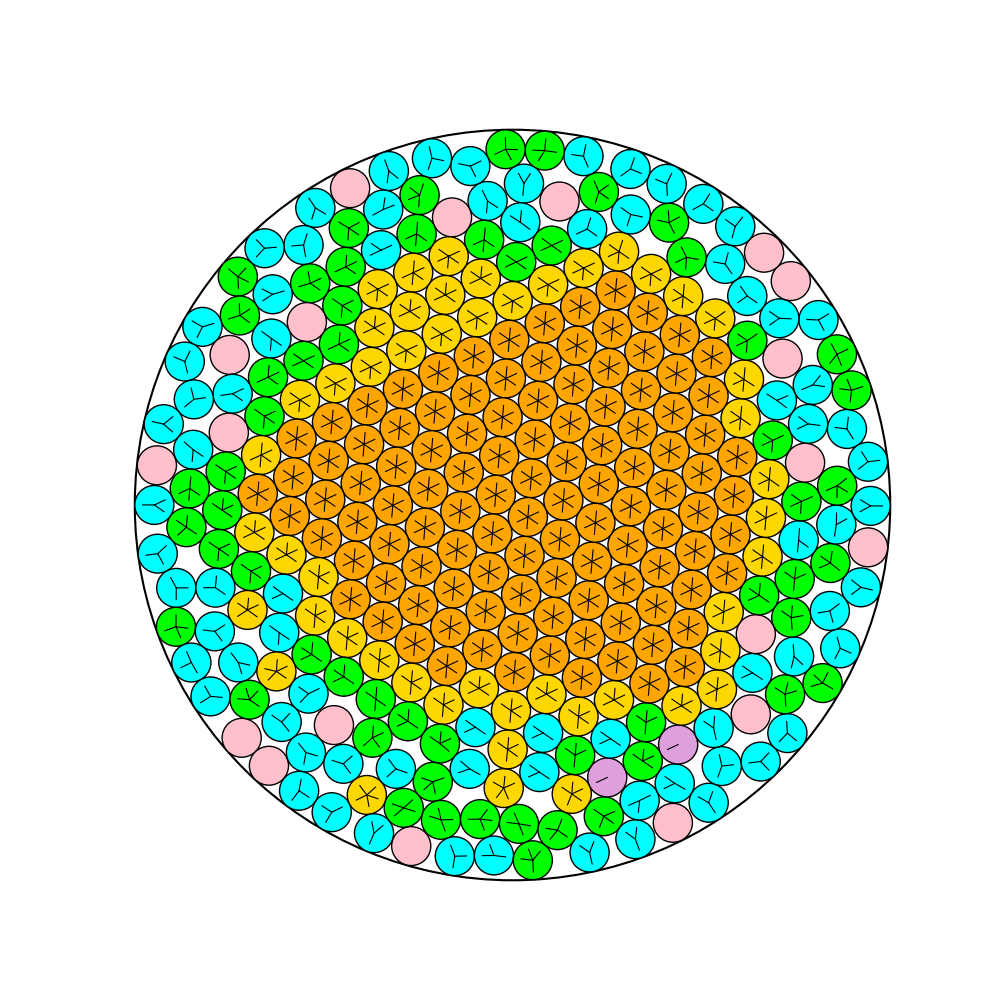}}
    \end{minipage}
    
    \caption{New improved solutions found by our proposed algorithm on the moderate scale instances $(300 \leq n \leq 320)$. The circles are colored by various colors according to the number of contact circles, where two circles $c_i$ and $c_j$ are considered to contact each other if the distance $d(c_i, c_j)$ between two circle centers satisfies $d(c_i, c_j) \leq 2 + 10^{-10}$.}
    \label{moderate-results}
\end{figure}

\subsection{Comparison on Large Scale Instances} \label{ssec:06-03large}

\begin{sidewaystable} 
\centering
\caption{Comparison between the best-known results and SED (5-batch \name) on the 51 large scale instances of the regular number. The improved best results of $R_{best}$ and $R_{avg}$ appear in bold.}
\label{tb-cmp-large-re}
\resizebox{1.0\textwidth}{!}{
\begin{tabular}{lllllllllllllllllllllllllllllll}
\toprule
\multirow{2}{*}{$n$} & \multirow{2}{*}{$R^*$} &  & \multicolumn{11}{l}{SED (5-batch \name)}                                                                                 &  &  &  & \multirow{2}{*}{$n$} & \multirow{2}{*}{$R^*$} &  & \multicolumn{11}{l}{SED (5-batch \name)}                                                                                 \\ \cline{4-14} \cline{21-31} 
                   &                     &  & $R_{best}$                &           & $R_{avg}$                 &           & $R_{best}-R^*$ &  & $RR$    &  & $HR$   &  & $time~(s)$ &  &  &  &                    &                     &  & $R_{best}$                &           & $R_{avg}$                 &           & $R_{best}-R^*$ &  & $RR$    &  & $HR$   &  & $time~(s)$ \\ \midrule
500                & 24.1329376240       &  & \textbf{24.1312529426} & \textbf{} & \textbf{24.1313788210} & \textbf{} & -1.68E-03  &  & 10/10 &  & 1/10 &  & 39561   &  &  &  & 800                & 30.4212133790       &  & \textbf{30.4198645288} & \textbf{} & 30.4250741893          & \textbf{} & -1.35E-03  &  & 2/10  &  & 1/10 &  & 132389  \\
510                & 24.4365629292       &  & \textbf{24.4210537570} & \textbf{} & \textbf{24.4249135890} & \textbf{} & -1.55E-02  &  & 10/10 &  & 1/10 &  & 60259   &  &  &  & 810                & 30.6017659657       &  & \textbf{30.5956736421} & \textbf{} & \textbf{30.6002028281} & \textbf{} & -6.09E-03  &  & 6/10  &  & 1/10 &  & 92738   \\
520                & 24.6609522831       &  & \textbf{24.6258073657} & \textbf{} & \textbf{24.6314124118} & \textbf{} & -3.51E-02  &  & 10/10 &  & 1/10 &  & 69426   &  &  &  & 820                & 30.7666908826       &  & \textbf{30.7489836784} & \textbf{} & \textbf{30.7533477503} & \textbf{} & -1.77E-02  &  & 10/10 &  & 1/10 &  & 125921  \\
530                & 24.8482376878       &  & \textbf{24.8455280225} & \textbf{} & \textbf{24.8472603022} & \textbf{} & -2.71E-03  &  & 7/10  &  & 1/10 &  & 44648   &  &  &  & 830                & 30.9418117602       &  & \textbf{30.9297617239} & \textbf{} & \textbf{30.9337956547} & \textbf{} & -1.21E-02  &  & 10/10 &  & 1/10 &  & 116582  \\
540                & 25.0884399378       &  & \textbf{25.0855911217} & \textbf{} & \textbf{25.0877127273} & \textbf{} & -2.85E-03  &  & 8/10  &  & 1/10 &  & 45255   &  &  &  & 840                & 31.1208576445       &  & \textbf{31.1194578886} & \textbf{} & 31.1224527712          & \textbf{} & -1.40E-03  &  & 2/10  &  & 1/10 &  & 88791   \\
550                & 25.3384484709       &  & \textbf{25.3342535707} & \textbf{} & \textbf{25.3357628704} & \textbf{} & -4.19E-03  &  & 10/10 &  & 1/10 &  & 43783   &  &  &  & 850                & 31.3552353388       &  & \textbf{31.3432578820} & \textbf{} & \textbf{31.3503256930} & \textbf{} & -1.20E-02  &  & 9/10  &  & 1/10 &  & 117055  \\
560                & 25.5167889934       &  & \textbf{25.5122424885} & \textbf{} & \textbf{25.5146796829} & \textbf{} & -4.55E-03  &  & 9/10  &  & 1/10 &  & 59827   &  &  &  & 860                & 31.5114350783       &  & \textbf{31.5070879233} & \textbf{} & 31.5135835953          & \textbf{} & -4.35E-03  &  & 5/10  &  & 1/10 &  & 114252  \\
570                & 25.7224085766       &  & \textbf{25.7134847249} & \textbf{} & \textbf{25.7138542411} & \textbf{} & -8.92E-03  &  & 10/10 &  & 1/10 &  & 63852   &  &  &  & 870                & 31.6807723726       &  & \textbf{31.6747965238} & \textbf{} & \textbf{31.6790476977} & \textbf{} & -5.98E-03  &  & 8/10  &  & 1/10 &  & 125813  \\
580                & 25.9623218516       &  & \textbf{25.9511074848} & \textbf{} & \textbf{25.9527379554} & \textbf{} & -1.12E-02  &  & 10/10 &  & 1/10 &  & 69140   &  &  &  & 880                & 31.8536138755       &  & \textbf{31.8463147084} & \textbf{} & \textbf{31.8512855368} & \textbf{} & -7.30E-03  &  & 8/10  &  & 1/10 &  & 107682  \\
590                & 26.2105443770       &  & \textbf{26.2025041590} & \textbf{} & \textbf{26.2069156613} & \textbf{} & -8.04E-03  &  & 9/10  &  & 1/10 &  & 62246   &  &  &  & 890                & 32.0482429714       &  & \textbf{32.0438440391} & \textbf{} & 32.0485774225          & \textbf{} & -4.40E-03  &  & 6/10  &  & 1/10 &  & 88801   \\
600                & 26.4274162694       &  & \textbf{26.4176880113} & \textbf{} & \textbf{26.4216830234} & \textbf{} & -9.73E-03  &  & 10/10 &  & 1/10 &  & 58273   &  &  &  & 900                & 32.2330843545       &  & \textbf{32.2199533426} & \textbf{} & \textbf{32.2301042219} & \textbf{} & -1.31E-02  &  & 8/10  &  & 1/10 &  & 92746   \\
610                & 26.6310600018       &  & \textbf{26.6227736610} & \textbf{} & \textbf{26.6256150993} & \textbf{} & -8.29E-03  &  & 10/10 &  & 1/10 &  & 55976   &  &  &  & 910                & 32.3661258161       &  & \textbf{32.3658580937} & \textbf{} & 32.3778653220          & \textbf{} & -2.68E-04  &  & 1/10  &  & 1/10 &  & 79106   \\
620                & 26.8618811252       &  & \textbf{26.8431235737} & \textbf{} & \textbf{26.8457652476} & \textbf{} & -1.88E-02  &  & 10/10 &  & 1/10 &  & 65063   &  &  &  & 920                & 32.5489524357       &  & \textbf{32.5347534619} & \textbf{} & \textbf{32.5391088907} & \textbf{} & -1.42E-02  &  & 10/10 &  & 1/10 &  & 121667  \\
630                & 27.0340036487       &  & 27.0525778864          & \textbf{} & 27.0618588183          & \textbf{} & 1.86E-02   &  & 0/10  &  & 1/10 &  & 44900   &  &  &  & 930                & 32.7013004786       &  & \textbf{32.6957714244} & \textbf{} & 32.7024754370          & \textbf{} & -5.53E-03  &  & 7/10  &  & 1/10 &  & 119341  \\
640                & 27.2419589706       &  & \textbf{27.2387282736} & \textbf{} & 27.2421894917          & \textbf{} & -3.23E-03  &  & 6/10  &  & 1/10 &  & 59358   &  &  &  & 940                & 32.9143189848       &  & \textbf{32.9009771761} & \textbf{} & \textbf{32.9052353405} & \textbf{} & -1.33E-02  &  & 10/10 &  & 1/10 &  & 109627  \\
650                & 27.4458279070       &  & \textbf{27.4382286041} & \textbf{} & \textbf{27.4422058882} & \textbf{} & -7.60E-03  &  & 10/10 &  & 1/10 &  & 71570   &  &  &  & 950                & 33.1232153862       &  & \textbf{33.1032839823} & \textbf{} & \textbf{33.1093226461} & \textbf{} & -1.99E-02  &  & 10/10 &  & 1/10 &  & 136179  \\
660                & 27.6680891671       &  & \textbf{27.6589093888} & \textbf{} & \textbf{27.6622225342} & \textbf{} & -9.18E-03  &  & 10/10 &  & 1/10 &  & 69611   &  &  &  & 960                & 33.2729042511       &  & \textbf{33.2554924317} & \textbf{} & \textbf{33.2639325269} & \textbf{} & -1.74E-02  &  & 10/10 &  & 1/10 &  & 111880  \\
670                & 27.9068676439       &  & \textbf{27.9008215002} & \textbf{} & \textbf{27.9030904780} & \textbf{} & -6.05E-03  &  & 10/10 &  & 1/10 &  & 65422   &  &  &  & 970                & 33.4357278371       &  & \textbf{33.4225668260} & \textbf{} & \textbf{33.4271703375} & \textbf{} & -1.32E-02  &  & 10/10 &  & 1/10 &  & 114131  \\
680                & 28.0951980575       &  & \textbf{28.0877536042} & \textbf{} & \textbf{28.0903964995} & \textbf{} & -7.44E-03  &  & 10/10 &  & 1/10 &  & 63392   &  &  &  & 980                & 33.6049866471       &  & \textbf{33.5827959909} & \textbf{} & \textbf{33.5885379042} & \textbf{} & -2.22E-02  &  & 10/10 &  & 1/10 &  & 154687  \\
690                & 28.2458655422       &  & \textbf{28.2447186303} & \textbf{} & 28.2638787279          & \textbf{} & -1.15E-03  &  & 5/10  &  & 1/10 &  & 55853   &  &  &  & 990                & 33.7831428326       &  & \textbf{33.7627857460} & \textbf{} & \textbf{33.7698434841} & \textbf{} & -2.04E-02  &  & 10/10 &  & 1/10 &  & 132277  \\
700                & 28.4958443164       &  & \textbf{28.4839888638} & \textbf{} & \textbf{28.4877680865} & \textbf{} & -1.19E-02  &  & 10/10 &  & 1/10 &  & 59161   &  &  &  & 1000               & 33.9571409147       &  & \textbf{33.9457725483} & \textbf{} & \textbf{33.9500559701} & \textbf{} & -1.14E-02  &  & 10/10 &  & 1/10 &  & 115406  \\
710                & 28.7110433153       &  & \textbf{28.6956216568} & \textbf{} & \textbf{28.7030160501} & \textbf{} & -1.54E-02  &  & 10/10 &  & 1/10 &  & 51258   &  &  &  &                    &                     &  &                        &           &                        &           &            &  &       &  &      &  &         \\
720                & 28.8599089374       &  & \textbf{28.8547583326} & \textbf{} & \textbf{28.8594487318} & \textbf{} & -5.15E-03  &  & 6/10  &  & 1/10 &  & 49792   &  &  &  &                    &                     &  &                        &           &                        &           &            &  &       &  &      &  &         \\
730                & 29.0380889370       &  & \textbf{29.0368792578} & \textbf{} & 29.0438783072          & \textbf{} & -1.21E-03  &  & 1/10  &  & 1/10 &  & 39872   &  &  &  &                    &                     &  &                        &           &                        &           &            &  &       &  &      &  &         \\
740                & 29.2501613747       &  & \textbf{29.2418887324} & \textbf{} & \textbf{29.2440743866} & \textbf{} & -8.27E-03  &  & 10/10 &  & 1/10 &  & 61698   &  &  &  &                    &                     &  &                        &           &                        &           &            &  &       &  &      &  &         \\
750                & 29.4806882503       &  & \textbf{29.4704602348} & \textbf{} & \textbf{29.4725052490} & \textbf{} & -1.02E-02  &  & 10/10 &  & 1/10 &  & 51594   &  &  &  &                    &                     &  &                        &           &                        &           &            &  &       &  &      &  &         \\
760                & 29.6611069657       &  & \textbf{29.6529205123} & \textbf{} & 29.6626249312          & \textbf{} & -8.19E-03  &  & 5/10  &  & 1/10 &  & 45894   &  &  &  &                    &                     &  &                        &           &                        &           &            &  &       &  &      &  &         \\
770                & 29.8480812041       &  & \textbf{29.8415580215} & \textbf{} & \textbf{29.8451997650} & \textbf{} & -6.52E-03  &  & 9/10  &  & 1/10 &  & 47679   &  &  &  &                    &                     &  &                        &           &                        &           &            &  &       &  &      &  &         \\
780                & 30.0188056551       &  & \textbf{30.0138757407} & \textbf{} & \textbf{30.0164282850} & \textbf{} & -4.93E-03  &  & 9/10  &  & 1/10 &  & 59540   &  &  &  &                    &                     &  &                        &           &                        &           &            &  &       &  &      &  &         \\
790                & 30.2195970061       &  & \textbf{30.2169949810} & \textbf{} & 30.2208074501          & \textbf{} & -2.60E-03  &  & 4/10  &  & 1/10 &  & 44975   &  &  &  &                    &                     &  &                        &           &                        &           &            &  &       &  &      &  &         \\ \midrule
\#Improve          &                     &  & 29                     &           & 24                     &           &            &  &       &  &      &  &         &  &  &  &                    &                     &  & 21                     &           & 15                     &           &            &  &       &  &      &  &         \\
\#Equal            &                     &  & 0                      &           & 0                      &           &            &  &       &  &      &  &         &  &  &  &                    &                     &  & 0                      &           & 0                      &           &            &  &       &  &      &  &         \\
\#Worse            &                     &  & 1                      &           & 6                      &           &            &  &       &  &      &  &         &  &  &  &                    &                     &  & 0                      &           & 6                      &           &            &  &       &  &      &  &         \\ \bottomrule
\end{tabular}
}
\end{sidewaystable}

\begin{sidewaystable} 
\centering
\caption{Comparison between the best-known results and SED (5-batch \name) on the 50 large scale instances of the irregular number. The improved best results of $R_{best}$ and $R_{avg}$ appear in bold.}
\label{tb-cmp-large-ir}
\resizebox{1.0\textwidth}{!}{
\begin{tabular}{llllllllllllllllllllllllllllllll}
\toprule
\multirow{2}{*}{$n$} & \multirow{2}{*}{$R^*$} &  & \multicolumn{11}{l}{SED (5-batch \name)}                                                                                 &  &  &  & \multirow{2}{*}{$n$} &  & \multirow{2}{*}{$R^*$} &  & \multicolumn{11}{l}{SED (5-batch \name)}                                                                                 \\ \cline{4-14} \cline{22-32} 
                   &                     &  & $R_{best}$                &           & $R_{avg}$                 &           & $R_{best}-R^*$ &  & $RR$    &  & $HR$   &  & $time~(s)$ &  &  &  &                    &  &                     &  & $R_{best}$                &           & $R_{avg}$                 &           & $R_{best}-R^*$ &  & $RR$    &  & $HR$   &  & $time~(s)$ \\ \midrule
505                & 24.2933415273       &  & \textbf{24.2888919603} & \textbf{} & \textbf{24.2901907584} & \textbf{} & -4.45E-03  &  & 10/10 &  & 1/10 &  & 41754   &  &  &  & 818                &  & 30.7297559207       &  & \textbf{30.7181308589} & \textbf{} & \textbf{30.7214168182} & \textbf{} & -1.16E-02  &  & 10/10 &  & 1/10 &  & 96312   \\
507                & 24.3570322874       &  & \textbf{24.3483441023} & \textbf{} & \textbf{24.3499568679} & \textbf{} & -8.69E-03  &  & 10/10 &  & 1/10 &  & 56666   &  &  &  & 823                &  & 30.7951925938       &  & \textbf{30.7907479760} & \textbf{} & 30.7953689952          & \textbf{} & -4.44E-03  &  & 7/10  &  & 1/10 &  & 102853  \\
511                & 24.4539029279       &  & \textbf{24.4418315424} & \textbf{} & \textbf{24.4443899780} & \textbf{} & -1.21E-02  &  & 10/10 &  & 1/10 &  & 65166   &  &  &  & 828                &  & 30.9026164659       &  & \textbf{30.8916045494} & \textbf{} & \textbf{30.8945942840} & \textbf{} & -1.10E-02  &  & 10/10 &  & 1/10 &  & 129179  \\
513                & 24.4944643498       &  & \textbf{24.4803487672} & \textbf{} & \textbf{24.4836476996} & \textbf{} & -1.41E-02  &  & 10/10 &  & 1/10 &  & 53628   &  &  &  & 846                &  & 31.2210952935       &  & \textbf{31.2169756294} & \textbf{} & 31.2320033828          & \textbf{} & -4.12E-03  &  & 7/10  &  & 1/10 &  & 77582   \\
515                & 24.5352933431       &  & \textbf{24.5160216754} & \textbf{} & \textbf{24.5234661213} & \textbf{} & -1.93E-02  &  & 10/10 &  & 1/10 &  & 64179   &  &  &  & 856                &  & 31.4404613224       &  & 31.4415597563          & \textbf{} & 31.4451742962          & \textbf{} & 1.10E-03   &  & 0/10  &  & 1/10 &  & 93393   \\
517                & 24.5806822590       &  & \textbf{24.5688597754} & \textbf{} & \textbf{24.5708937360} & \textbf{} & -1.18E-02  &  & 10/10 &  & 1/10 &  & 64080   &  &  &  & 861                &  & 31.5340220772       &  & \textbf{31.5286770943} & \textbf{} & \textbf{31.5304575449} & \textbf{} & -5.34E-03  &  & 10/10 &  & 1/10 &  & 121885  \\
539                & 25.0666591245       &  & \textbf{25.0644157476} & \textbf{} & \textbf{25.0649834706} & \textbf{} & -2.24E-03  &  & 10/10 &  & 1/10 &  & 61702   &  &  &  & 872                &  & 31.7129332496       &  & \textbf{31.7028455621} & \textbf{} & \textbf{31.7105946951} & \textbf{} & -1.01E-02  &  & 9/10  &  & 1/10 &  & 93714   \\
547                & 25.2486649453       &  & 25.2630537774          & \textbf{} & 25.2647995297          & \textbf{} & 1.44E-02   &  & 0/10  &  & 1/10 &  & 39954   &  &  &  & 873                &  & 31.7297370844       &  & \textbf{31.7220224328} & \textbf{} & \textbf{31.7251146977} & \textbf{} & -7.71E-03  &  & 9/10  &  & 1/10 &  & 95472   \\
568                & 25.6768452576       &  & \textbf{25.6742746832} & \textbf{} & \textbf{25.6747142118} & \textbf{} & -2.57E-03  &  & 9/10  &  & 2/10 &  & 63844   &  &  &  & 877                &  & 31.8055541388       &  & \textbf{31.7939196678} & \textbf{} & \textbf{31.7991773790} & \textbf{} & -1.16E-02  &  & 10/10 &  & 1/10 &  & 123868  \\
578                & 25.9195920293       &  & \textbf{25.9105502154} & \textbf{} & \textbf{25.9141526168} & \textbf{} & -9.04E-03  &  & 10/10 &  & 1/10 &  & 57305   &  &  &  & 879                &  & 31.8366948955       &  & \textbf{31.8265208472} & \textbf{} & \textbf{31.8312690574} & \textbf{} & -1.02E-02  &  & 10/10 &  & 1/10 &  & 110296  \\
591                & 26.2324777614       &  & \textbf{26.2247762629} & \textbf{} & \textbf{26.2300329942} & \textbf{} & -7.70E-03  &  & 8/10  &  & 1/10 &  & 63852   &  &  &  & 888                &  & 32.0091101899       &  & \textbf{31.9999805643} & \textbf{} & \textbf{32.0056588693} & \textbf{} & -9.13E-03  &  & 10/10 &  & 1/10 &  & 93530   \\
597                & 26.3542748538       &  & \textbf{26.3529416759} & \textbf{} & 26.3549863877          & \textbf{} & -1.33E-03  &  & 5/10  &  & 1/10 &  & 48364   &  &  &  & 892                &  & 32.0884743996       &  & \textbf{32.0850195925} & \textbf{} & 32.0888263967          & \textbf{} & -3.45E-03  &  & 5/10  &  & 1/10 &  & 118031  \\
605                & 26.5365274699       &  & \textbf{26.5233023305} & \textbf{} & \textbf{26.5261325786} & \textbf{} & -1.32E-02  &  & 10/10 &  & 1/10 &  & 72521   &  &  &  & 899                &  & 32.2281486010       &  & \textbf{32.2157001450} & \textbf{} & \textbf{32.2200915147} & \textbf{} & -1.24E-02  &  & 10/10 &  & 1/10 &  & 130749  \\
608                & 26.6012529572       &  & \textbf{26.5828030363} & \textbf{} & \textbf{26.5852665616} & \textbf{} & -1.84E-02  &  & 10/10 &  & 1/10 &  & 73642   &  &  &  & 906                &  & 32.3065651824       &  & 32.3093777750          & \textbf{} & 32.3220781255          & \textbf{} & 2.81E-03   &  & 0/10  &  & 1/10 &  & 131780  \\
613                & 26.6970192350       &  & \textbf{26.6806757801} & \textbf{} & \textbf{26.6857073661} & \textbf{} & -1.63E-02  &  & 10/10 &  & 1/10 &  & 77051   &  &  &  & 911                &  & 32.3720611668       &  & 32.3751529633          & \textbf{} & 32.3910623981          & \textbf{} & 3.09E-03   &  & 0/10  &  & 1/10 &  & 90007   \\
666                & 27.8222921562       &  & \textbf{27.8095070987} & \textbf{} & \textbf{27.8149840911} & \textbf{} & -1.28E-02  &  & 10/10 &  & 1/10 &  & 62523   &  &  &  & 923                &  & 32.6092522621       &  & \textbf{32.5891329001} & \textbf{} & \textbf{32.5927036033} & \textbf{} & -2.01E-02  &  & 10/10 &  & 1/10 &  & 117298  \\
669                & 27.8866609962       &  & \textbf{27.8807032094} & \textbf{} & \textbf{27.8832913945} & \textbf{} & -5.96E-03  &  & 10/10 &  & 1/10 &  & 50575   &  &  &  & 924                &  & 32.6279909132       &  & \textbf{32.6035557397} & \textbf{} & \textbf{32.6110620319} & \textbf{} & -2.44E-02  &  & 10/10 &  & 1/10 &  & 131299  \\
677                & 28.0279735048       &  & \textbf{28.0223920966} & \textbf{} & 28.0294202784          & \textbf{} & -5.58E-03  &  & 3/10  &  & 1/10 &  & 50970   &  &  &  & 945                &  & 33.0363462137       &  & \textbf{33.0028997187} & \textbf{} & \textbf{33.0183695018} & \textbf{} & -3.34E-02  &  & 10/10 &  & 1/10 &  & 138869  \\
678                & 28.0545929509       &  & \textbf{28.0452214253} & \textbf{} & \textbf{28.0486293696} & \textbf{} & -9.37E-03  &  & 10/10 &  & 1/10 &  & 70367   &  &  &  & 964                &  & 33.3393817195       &  & \textbf{33.3204312890} & \textbf{} & \textbf{33.3287459497} & \textbf{} & -1.90E-02  &  & 10/10 &  & 1/10 &  & 115451  \\
737                & 29.1885579873       &  & \textbf{29.1805602263} & \textbf{} & \textbf{29.1837249288} & \textbf{} & -8.00E-03  &  & 10/10 &  & 1/10 &  & 59552   &  &  &  & 977                &  & 33.5337256998       &  & \textbf{33.5256191705} & \textbf{} & \textbf{33.5286656634} & \textbf{} & -8.11E-03  &  & 10/10 &  & 1/10 &  & 117418  \\
741                & 29.2673369351       &  & \textbf{29.2630224724} & \textbf{} & \textbf{29.2656945026} & \textbf{} & -4.31E-03  &  & 8/10  &  & 1/10 &  & 52375   &  &  &  &                    &  &                     &  &                        &           &                        &           &            &  &       &  &      &  &         \\
743                & 29.2989651471       &  & \textbf{29.2950962890} & \textbf{} & 29.3006429415          & \textbf{} & -3.87E-03  &  & 7/10  &  & 1/10 &  & 51812   &  &  &  &                    &  &                     &  &                        &           &                        &           &            &  &       &  &      &  &         \\
755                & 29.5864164492       &  & \textbf{29.5781420320} & \textbf{} & \textbf{29.5813137994} & \textbf{} & -8.27E-03  &  & 10/10 &  & 1/10 &  & 66710   &  &  &  &                    &  &                     &  &                        &           &                        &           &            &  &       &  &      &  &         \\
763                & 29.7207712331       &  & \textbf{29.7156118909} & \textbf{} & 29.7208072730          & \textbf{} & -5.16E-03  &  & 6/10  &  & 1/10 &  & 47539   &  &  &  &                    &  &                     &  &                        &           &                        &           &            &  &       &  &      &  &         \\
764                & 29.7349505495       &  & \textbf{29.7321782541} & \textbf{} & 29.7356589694          & \textbf{} & -2.77E-03  &  & 7/10  &  & 1/10 &  & 42529   &  &  &  &                    &  &                     &  &                        &           &                        &           &            &  &       &  &      &  &         \\
774                & 29.9175478793       &  & \textbf{29.9123498351} & \textbf{} & \textbf{29.9157783799} & \textbf{} & -5.20E-03  &  & 9/10  &  & 1/10 &  & 44943   &  &  &  &                    &  &                     &  &                        &           &                        &           &            &  &       &  &      &  &         \\
778                & 29.9893439763       &  & \textbf{29.9830573225} & \textbf{} & \textbf{29.9844978472} & \textbf{} & -6.29E-03  &  & 10/10 &  & 1/10 &  & 55350   &  &  &  &                    &  &                     &  &                        &           &                        &           &            &  &       &  &      &  &         \\
781                & 30.0278742024       &  & 30.0301280216          & \textbf{} & 30.0358959038          & \textbf{} & 2.25E-03   &  & 0/10  &  & 1/10 &  & 65116   &  &  &  &                    &  &                     &  &                        &           &                        &           &            &  &       &  &      &  &         \\
796                & 30.3480735601       &  & \textbf{30.3399225581} & \textbf{} & \textbf{30.3447190305} & \textbf{} & -8.15E-03  &  & 10/10 &  & 1/10 &  & 49815   &  &  &  &                    &  &                     &  &                        &           &                        &           &            &  &       &  &      &  &         \\
797                & 30.3755692236       &  & \textbf{30.3676919043} & \textbf{} & \textbf{30.3715277628} & \textbf{} & -7.88E-03  &  & 9/10  &  & 1/10 &  & 46237   &  &  &  &                    &  &                     &  &                        &           &                        &           &            &  &       &  &      &  &         \\ \midrule
\#Improve          &                     &  & 28                     &           & 23                     &           &            &  &       &  &      &  &         &  &  &  &                    &  &                     &  & 17                     &           & 14                     &           &            &  &       &  &      &  &         \\
\#Equal            &                     &  & 0                      &           & 0                      &           &            &  &       &  &      &  &         &  &  &  &                    &  &                     &  & 0                      &           & 0                      &           &            &  &       &  &      &  &         \\
\#Worse            &                     &  & 2                      &           & 7                      &           &            &  &       &  &      &  &         &  &  &  &                    &  &                     &  & 3                      &           & 6                      &           &            &  &       &  &      &  &         \\ \bottomrule
\end{tabular}
}
\end{sidewaystable}

For large scale instances, we select $n = 500, 510, 520, ..., 990, 1000$ as our 51 large scale instances of the regular number, then we randomly sample 30 irregular numbers from $500 \leq n < 800$ and 20 irregular numbers from $800 \leq n \leq 1000$ as our 50 large scale instances of the irregular number. We perform SED (5-batch \name) on these instances, and the experimental results of the regular and irregular numbers are shown in Table~\ref{tb-cmp-large-re} and Table~\ref{tb-cmp-large-ir}, respectively. 

In Tables~\ref{tb-cmp-large-re} and \ref{tb-cmp-large-ir}, We provide $n$ for the number of items in the instances, $R^*$ for the best-known results from the Packomania website~\citep{Spechtweb} (download data 2022/10/1), $R_{best}$ for the best result of 10 independent runs, $R_{avg}$ for the average result of 10 independent runs. $R_{best}-R^*$ shows the difference between $R_{best}$ and $R^*$ (a negative value indicates an improved best result). $RR$ shows the ratio of equal or better than the best-known result $R^*$, and $HR$ shows the ratio of hitting the best value $R_{best}$. The last column of $time~(s)$ shows the average time of obtaining a best solution. 
At the bottom of the tables,  ``\#Improve'', ``\#Equal'' and ``\#Worse'' show the number of instances for which SED (5-batch \name) obtained an improved, equal and worse result compared to the best-known results.

From the results, we can draw conclusions as follows:
\begin{itemize}
    \item [(1)] SED (5-batch \name) has 50 improved, 0 equal and 1 worse best results of the 51 large scale instances with regular number, and it has 45 improved, 0 equal and 5 worse best results of the 50 large scale instances with irregular number. The results demonstrate that our proposed SED (5-batch \name) algorithm has excellent performance on large scale instances. 
    \item [(2)] Most ratios of $RR$ are greater than 5/10, and many of them are equal to 10/10. SED (5-batch \name) had 39 improved, 0 equal and 12 worse average results of the 51 large scale instances of the regular number, and it has 37 improved, 0 equal and 13 worse average results of the 50 large scale instances of the irregular number. These results imply many runs of SED (5-batch \name) are better than the best-known results. It also demonstrates that the algorithm has excellent performance on large scale instances.
    \item [(3)] All the ratios of HR are equal to 1/10 except $n = 568$. It shows that obtaining the best results is extremely difficult, and the large scale PECC problem is computationally challenging. 
\end{itemize}

\begin{figure}[b]
    \centering
    \begin{minipage}[b]{0.33\linewidth}
        \centering
        \subfloat[][$n = 505$]{\includegraphics[width=1\linewidth]{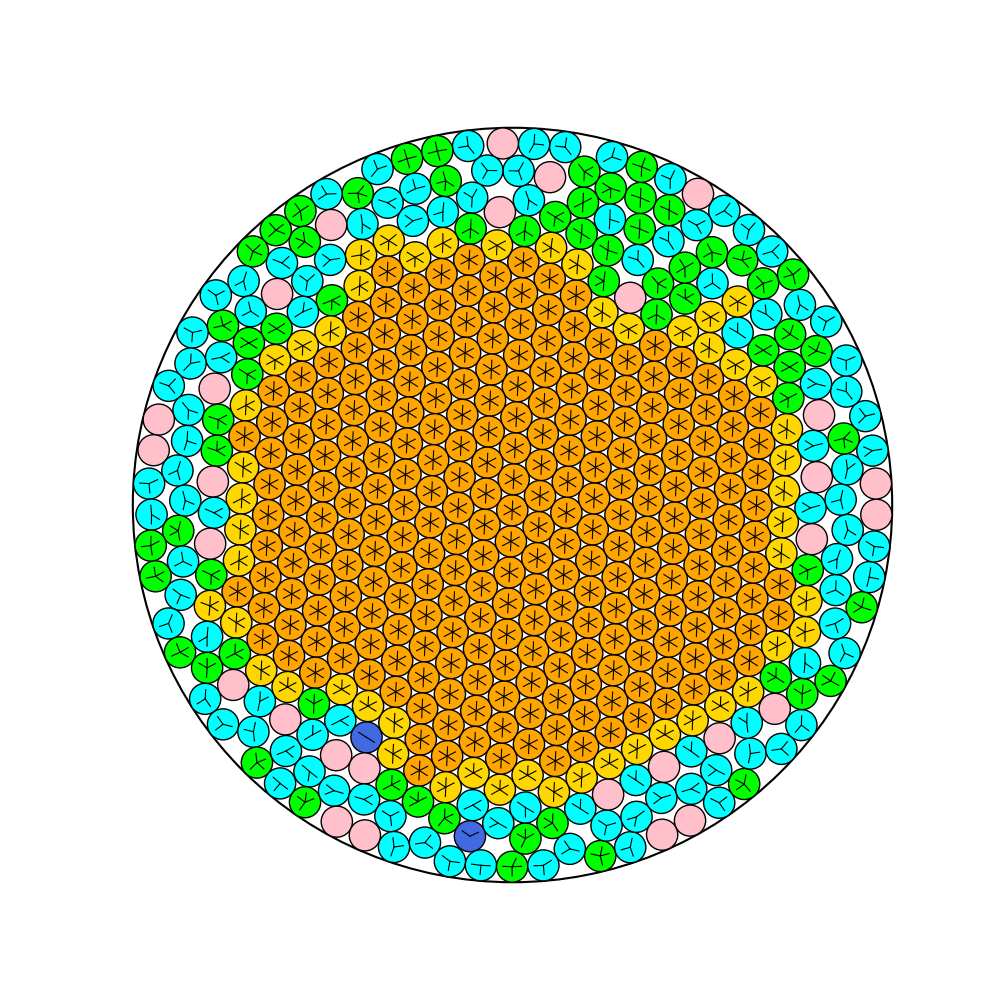}}
    \end{minipage}
    \begin{minipage}[b]{0.33\linewidth}
        \centering
        \subfloat[][$n = 610$]{\includegraphics[width=1\linewidth]{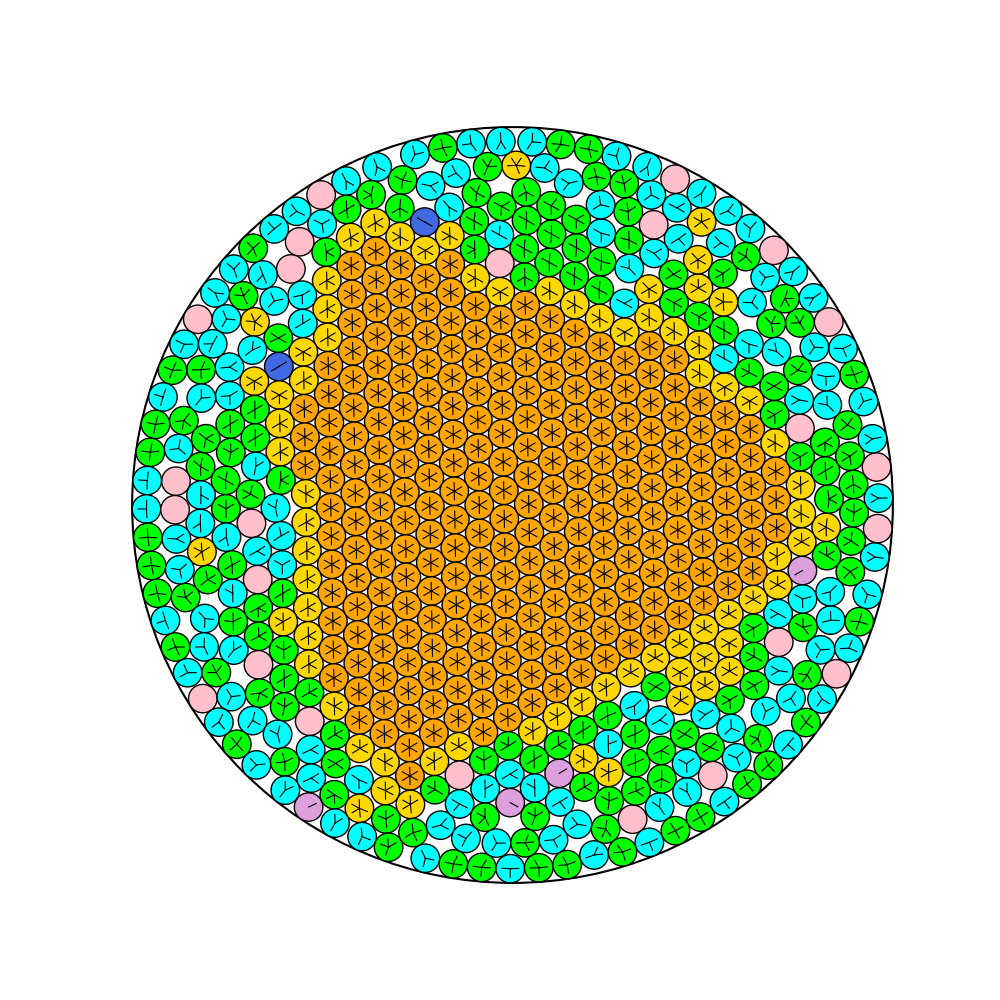}}
    \end{minipage}    
    \begin{minipage}[b]{0.33\linewidth}
        \centering
        \subfloat[][$n = 650$]{\includegraphics[width=1\linewidth]{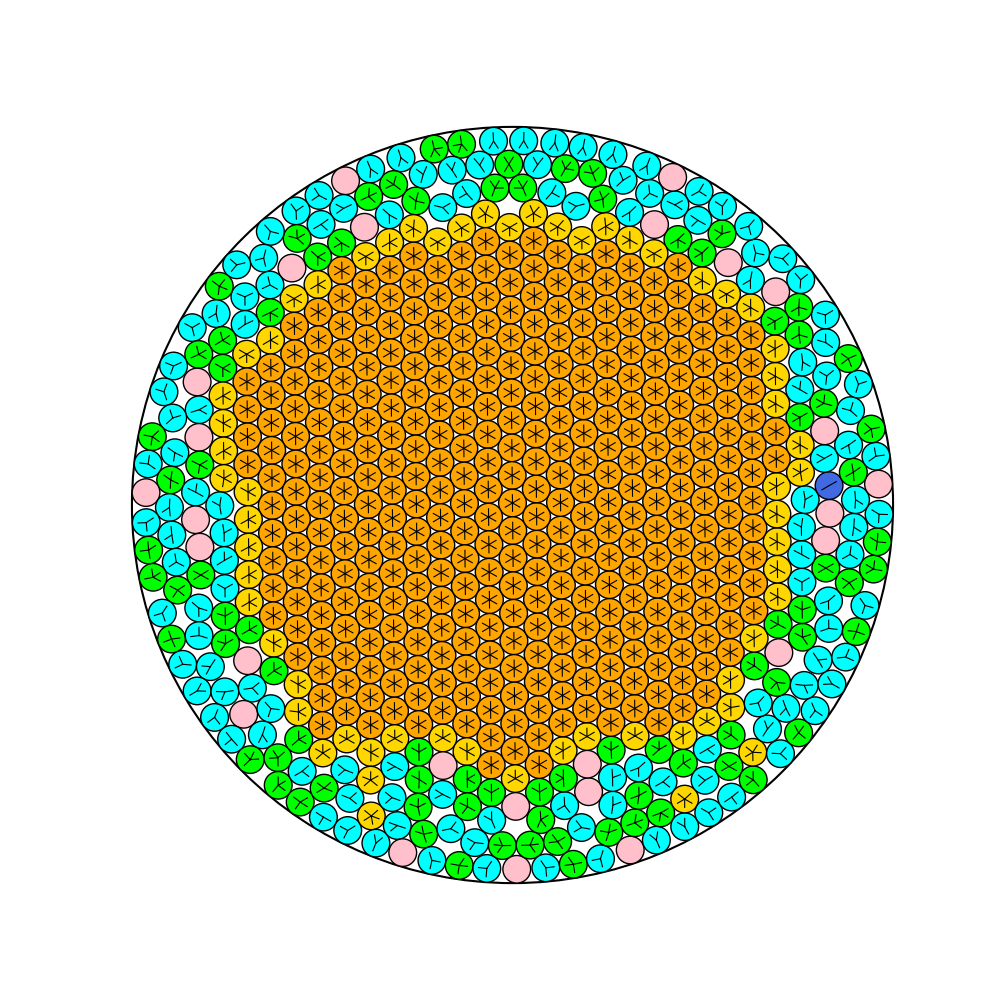}}
    \end{minipage}
    \begin{minipage}[b]{0.33\linewidth}
        \centering
        \subfloat[][$n = 710$]{\includegraphics[width=1\linewidth]{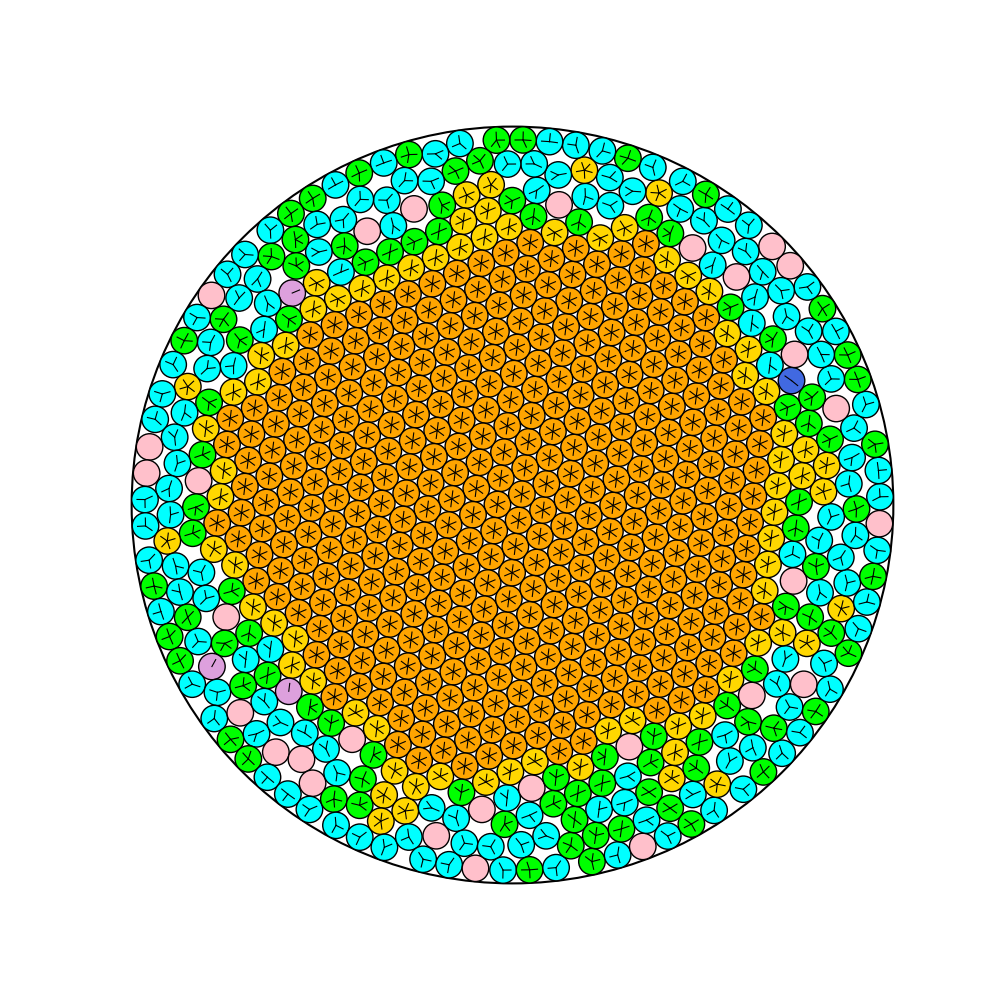}}
    \end{minipage}    
    \begin{minipage}[b]{0.33\linewidth}
        \centering
        \subfloat[][$n = 774$]{\includegraphics[width=1\linewidth]{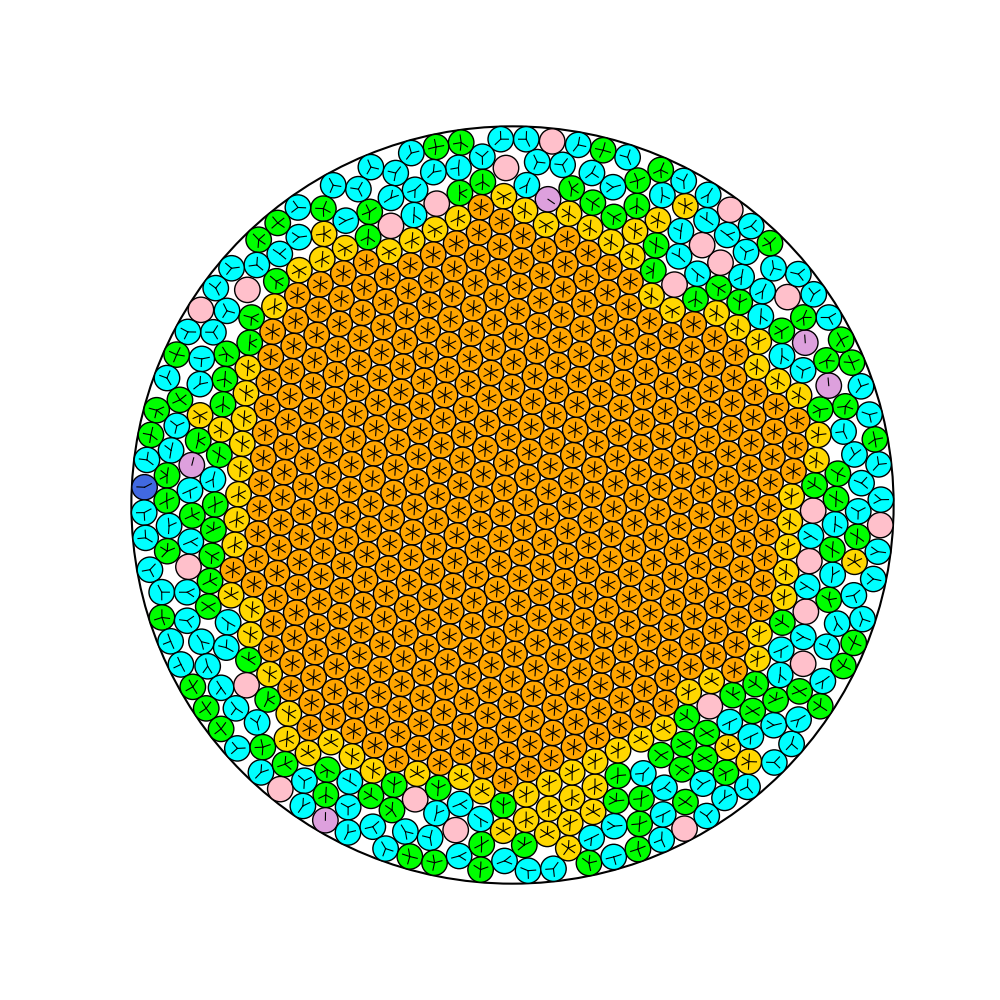}}
    \end{minipage}
    \begin{minipage}[b]{0.33\linewidth}
        \centering
        \subfloat[][$n = 828$]{\includegraphics[width=1\linewidth]{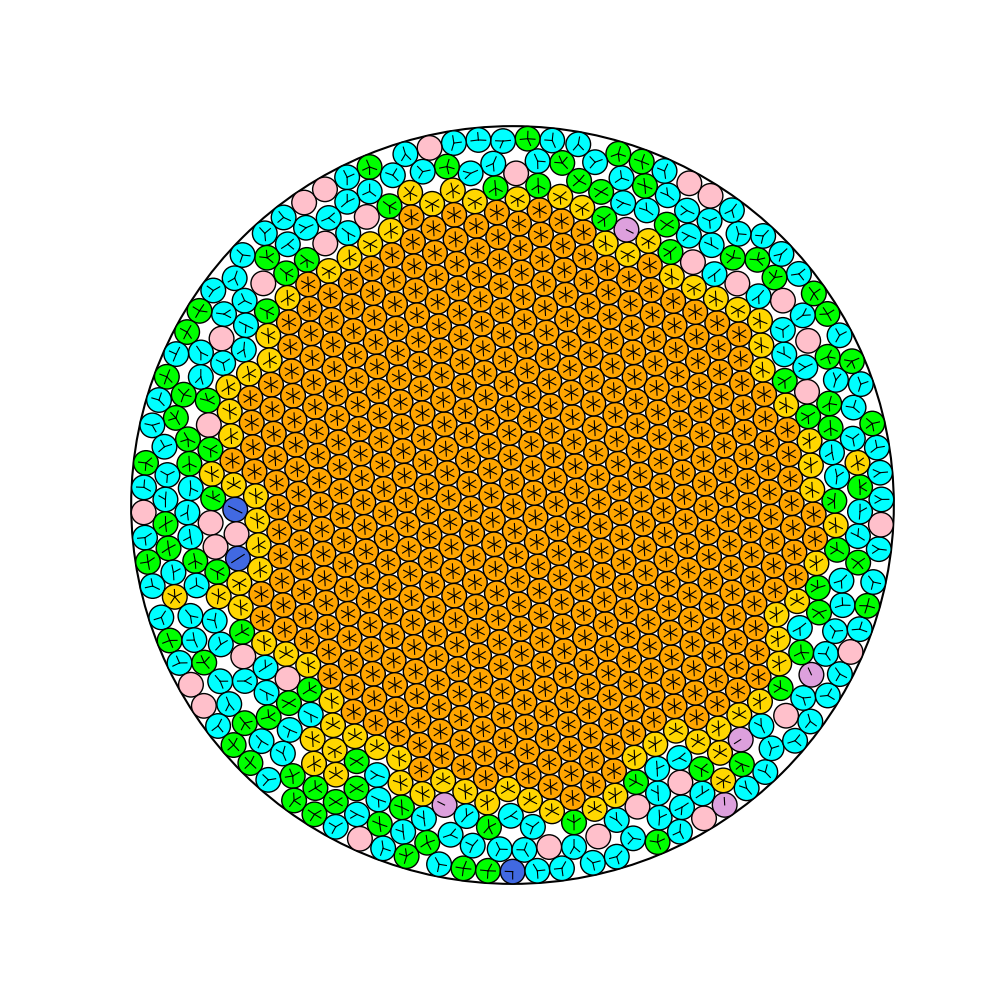}}
    \end{minipage}
    \begin{minipage}[b]{0.33\linewidth}
        \centering
        \subfloat[][$n = 890$]{\includegraphics[width=1\linewidth]{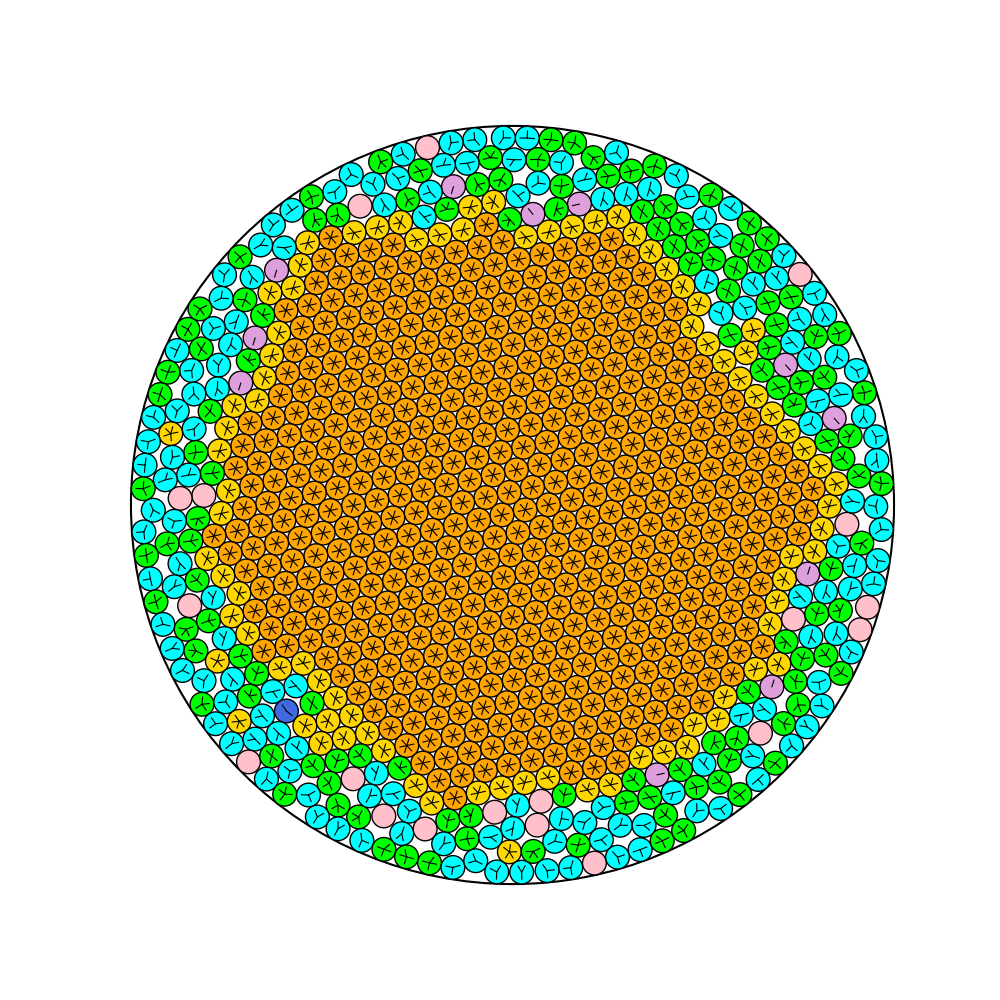}}
    \end{minipage}
    \begin{minipage}[b]{0.33\linewidth}
        \centering
        \subfloat[][$n = 940$]{\includegraphics[width=1\linewidth]{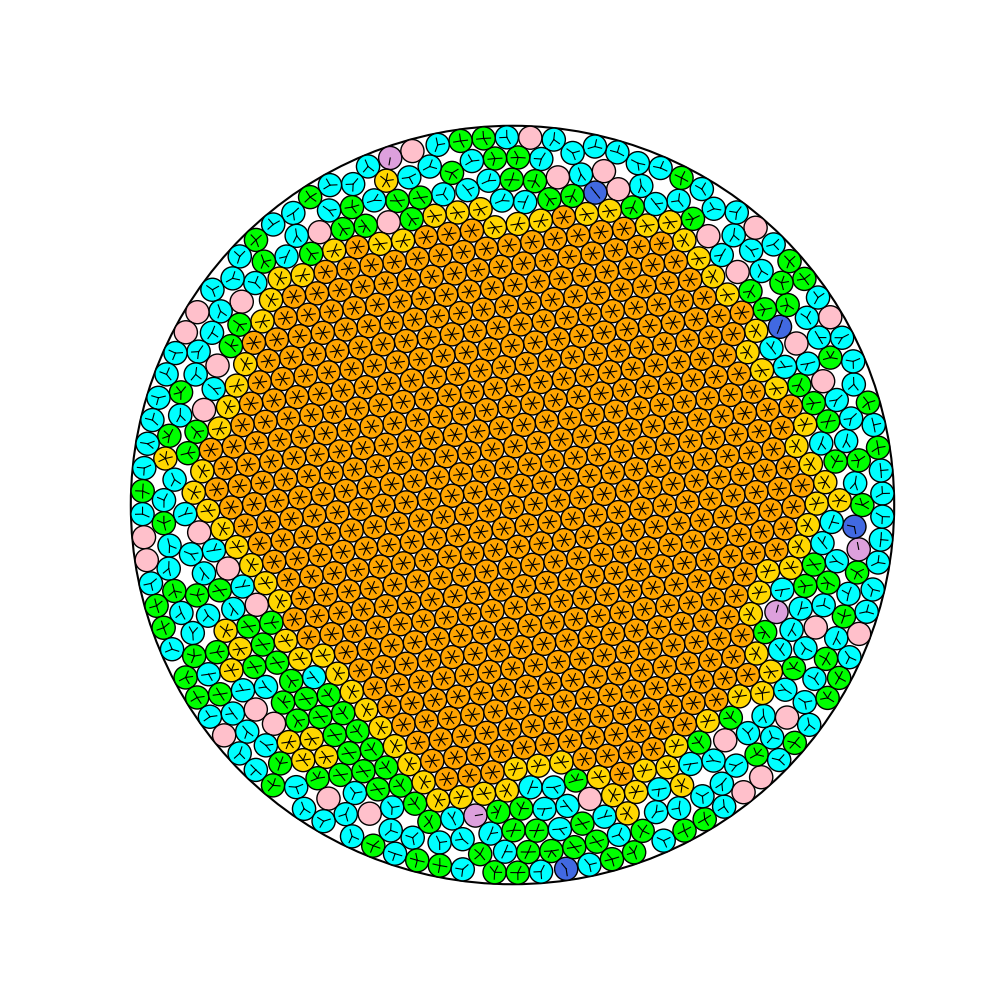}}
    \end{minipage}    
    \begin{minipage}[b]{0.33\linewidth}
        \centering
        \subfloat[][$n = 990$]{\includegraphics[width=1\linewidth]{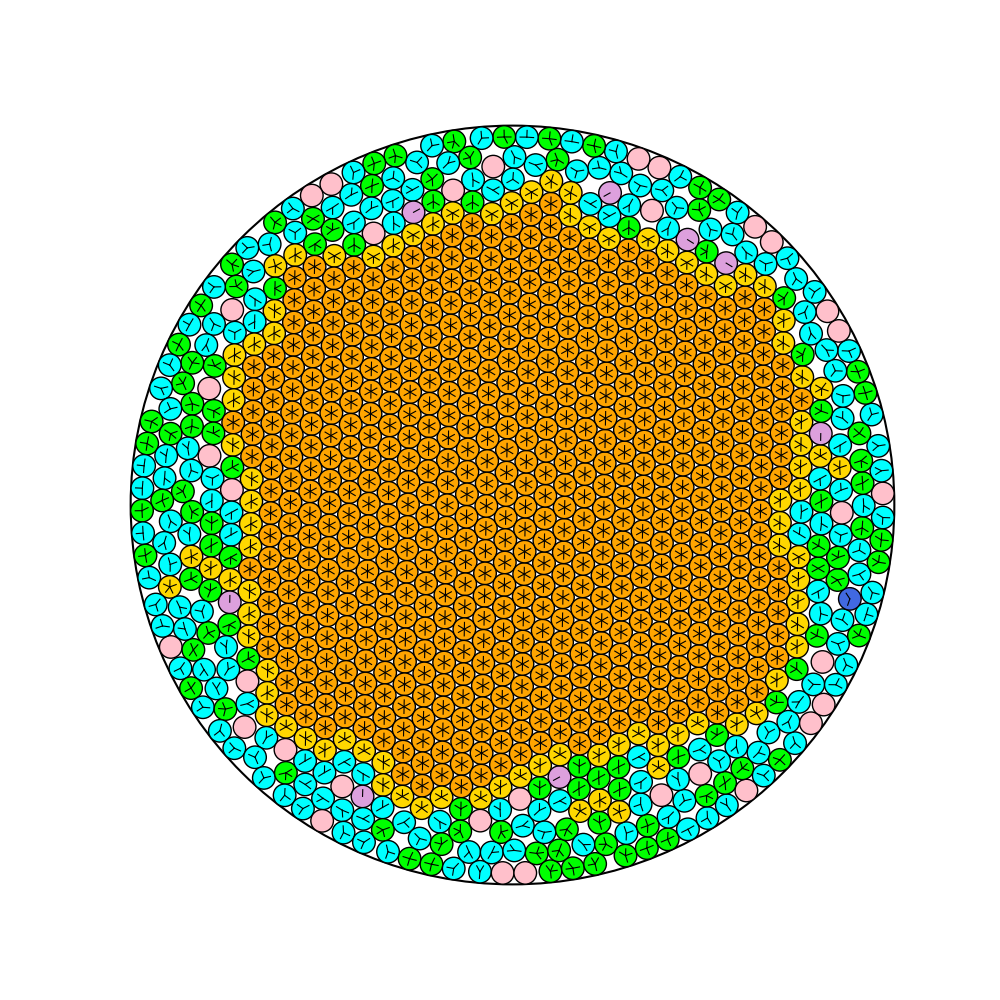}}
    \end{minipage}
    
    \caption{New improved solutions found by our algorithm for some representative instances on the large scale instances, which have the closest packing in the central zone.}
    \label{large-results(CP)}
\end{figure}

\begin{figure}[t]
    \centering
    \begin{minipage}[b]{0.33\linewidth}
        \centering
        \subfloat[][$n = 500$]{\includegraphics[width=1\linewidth]{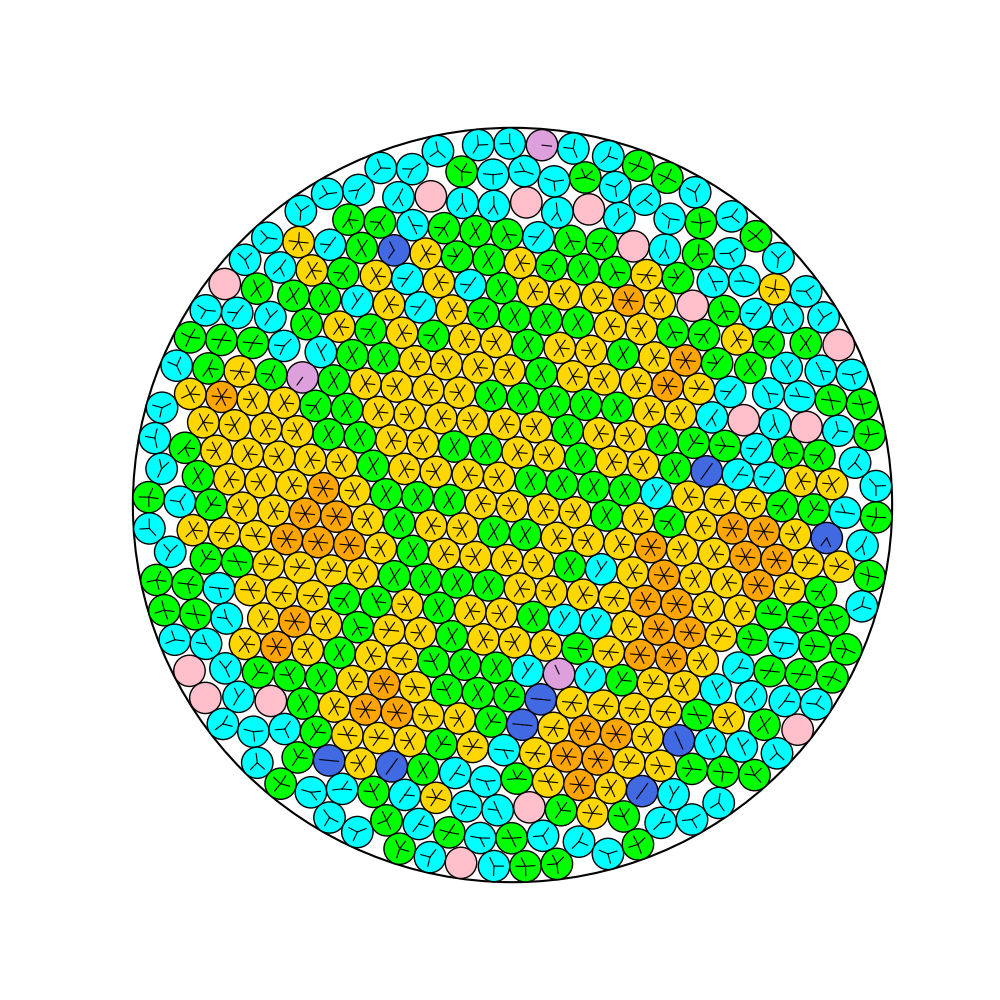}}
    \end{minipage}
    \begin{minipage}[b]{0.33\linewidth}
        \centering
        \subfloat[][$n = 513$]{\includegraphics[width=1\linewidth]{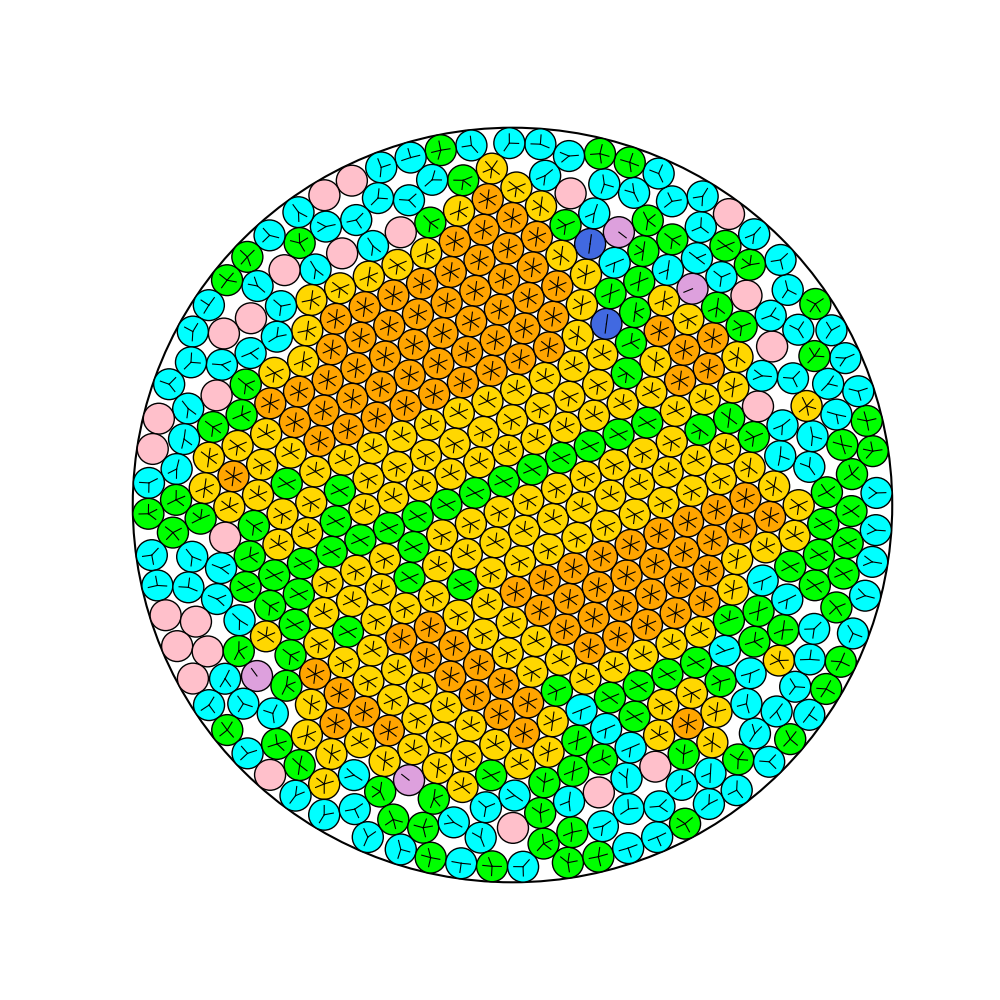}}
    \end{minipage}    
    \begin{minipage}[b]{0.33\linewidth}
        \centering
        \subfloat[][$n = 669$]{\includegraphics[width=1\linewidth]{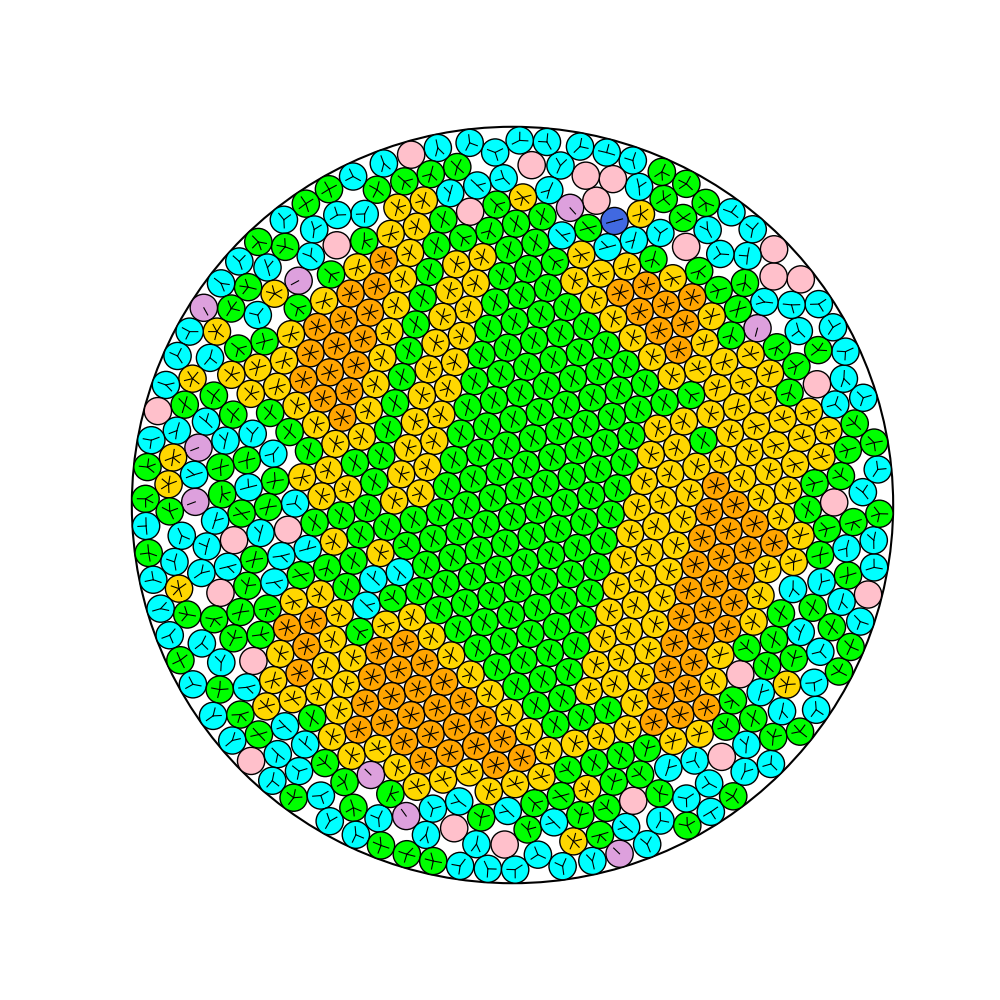}}
    \end{minipage}
    \begin{minipage}[b]{0.33\linewidth}
        \centering
        \subfloat[][$n = 743$]{\includegraphics[width=1\linewidth]{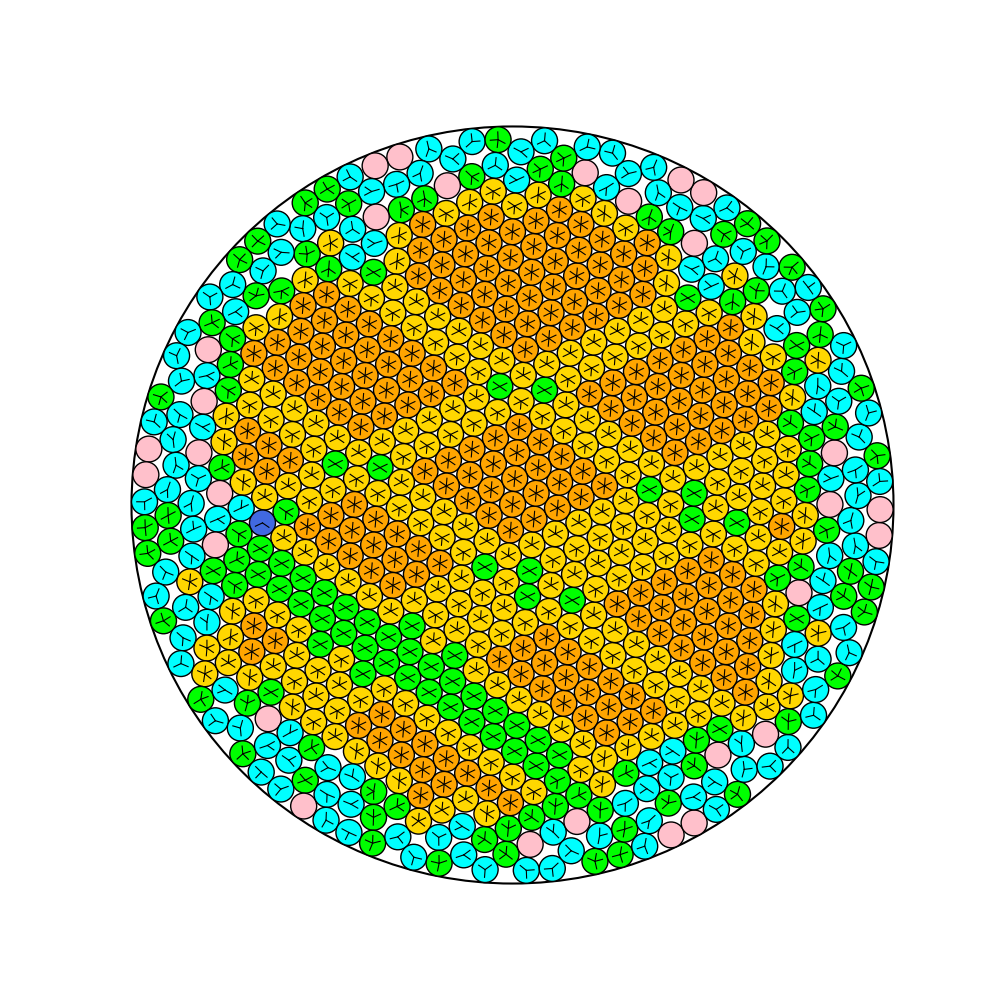}}
    \end{minipage}
    \begin{minipage}[b]{0.33\linewidth}
        \centering
        \subfloat[][$n = 800$]{\includegraphics[width=1\linewidth]{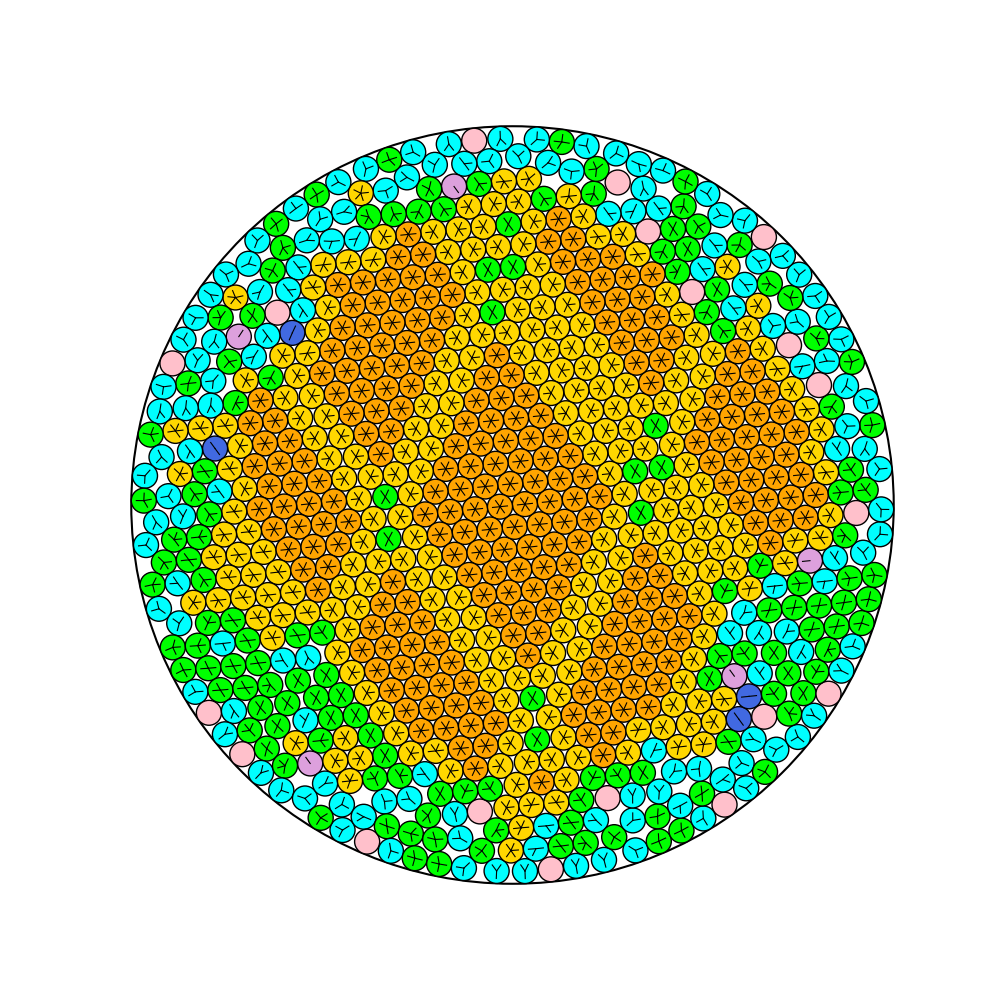}}
    \end{minipage}    
    \begin{minipage}[b]{0.33\linewidth}
        \centering
        \subfloat[][$n = 840$]{\includegraphics[width=1\linewidth]{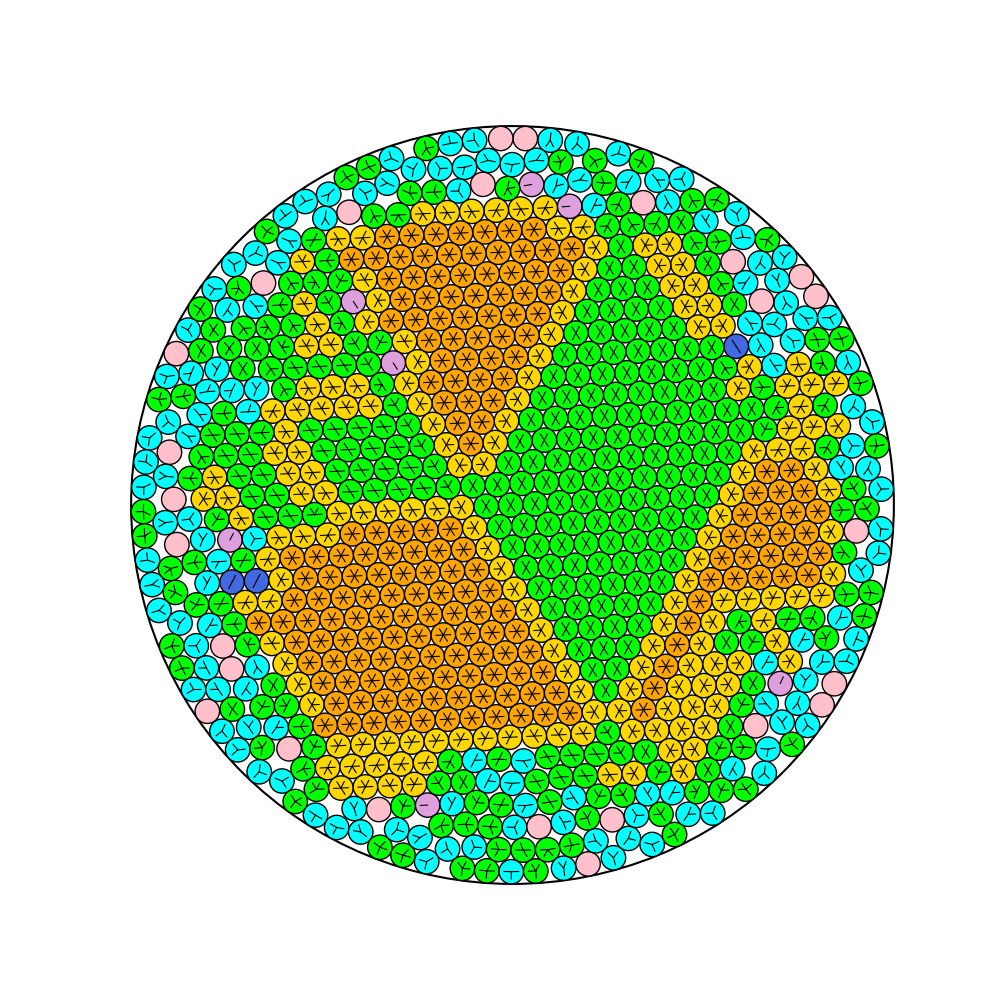}}
    \end{minipage}
    \begin{minipage}[b]{0.33\linewidth}
        \centering
        \subfloat[][$n = 900$]{\includegraphics[width=1\linewidth]{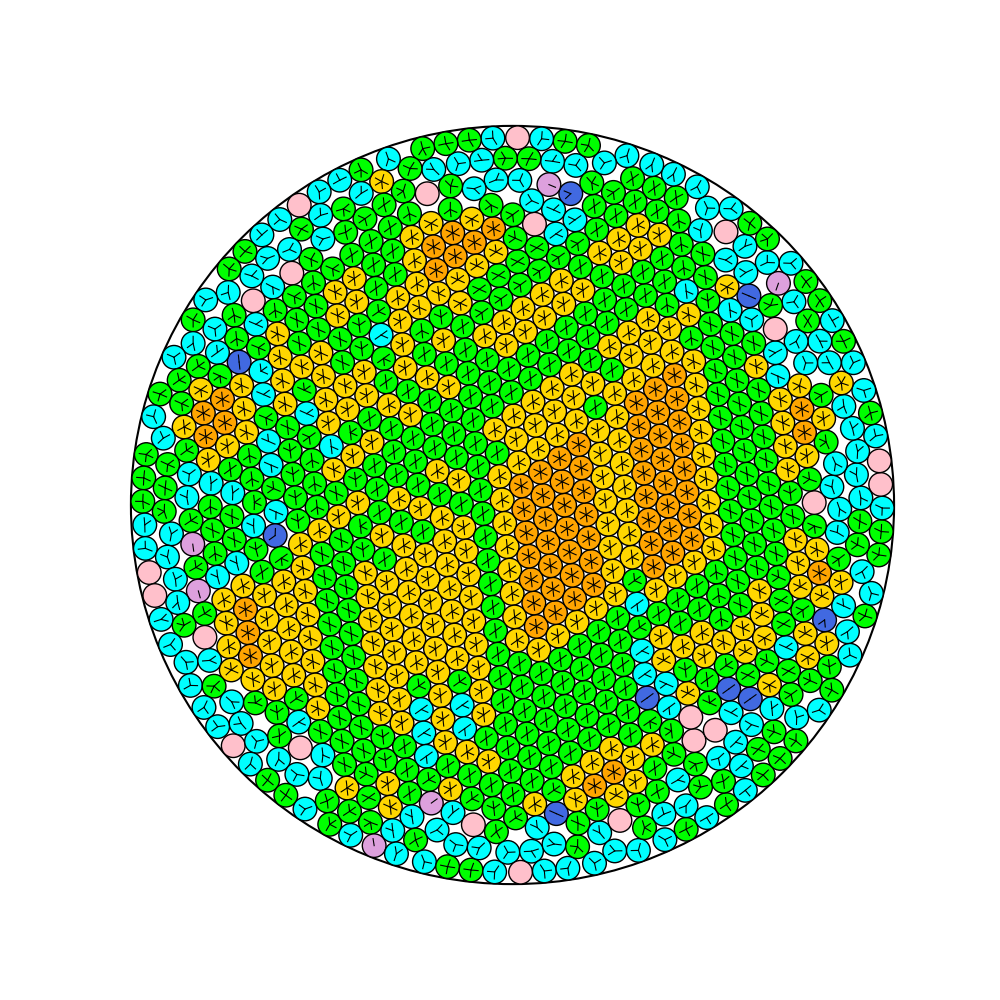}}
    \end{minipage}
    \begin{minipage}[b]{0.33\linewidth}
        \centering
        \subfloat[][$n = 930$]{\includegraphics[width=1\linewidth]{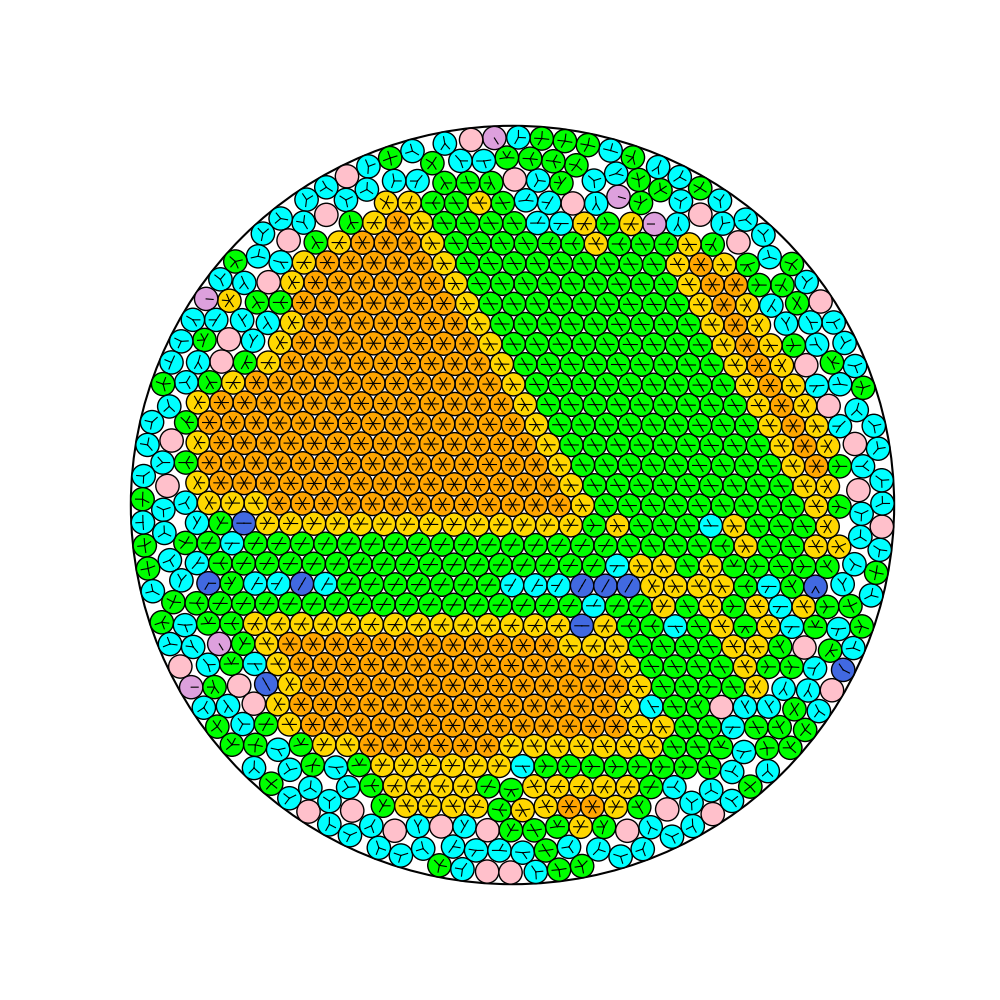}}
    \end{minipage}    
    \begin{minipage}[b]{0.33\linewidth}
        \centering
        \subfloat[][$n = 970$]{\includegraphics[width=1\linewidth]{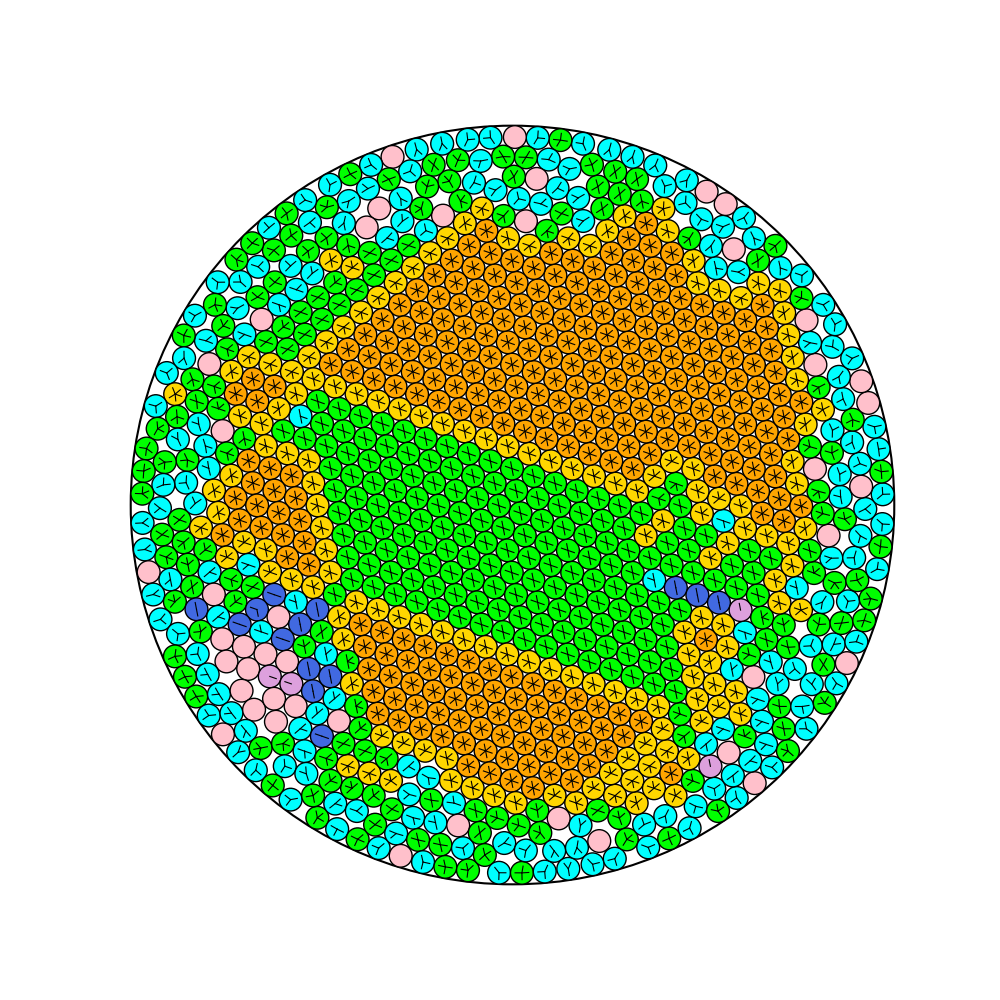}}
    \end{minipage}
    
    \caption{New improved solutions found by our algorithm for some representative instances on the large scale instances, which do not have the closest packing in the central zone.}
    \label{large-results(NCP)}
\end{figure}

\subsection{Comparison of \name (Multi-Batch) and Non-Batch} \label{ssec:06-04abla}

\begin{table}[b]
\centering
\caption{Comparison between 1-batch \name and 5-batch \name on the 20 selected large scale instances. The best results of $R_{best}'$ and $R_{best}$ and the best results of $R_{avg}'$ and $R_{avg}$ appear in bold.}
\label{tb-cmp-ablation1}

\scalebox{0.7}{
\begin{tabular}{lllllllllllllllll}
\toprule
\multirow{2}{*}{$n$} &  & \multicolumn{5}{l}{SED (1-batch \name)}                  &  & \multicolumn{5}{l}{SED (5-batch \name)}                           &  &           &  &           \\ \cline{3-7} \cline{9-13}
                   &  & $R_{best}'$                &  & $R_{avg}'$        &  & $time~(s)$ &  & $R_{best}$                &  & $R_{avg}$                 &  & $time~(s)$ &  & $R_{best} - R_{best}'$         &  & $R_{avg} - R_{avg}'$        \\ \midrule
510                &  & 24.4854461416          &  & 24.5081091304 &  & 48297   &  & \textbf{24.4210537570} &  & \textbf{24.4249135890} &  & 60259   &  & -6.44E-02 &  & -8.32E-02 \\
610                &  & 26.6992073479          &  & 26.7150332110 &  & 58763   &  & \textbf{26.6227736610} &  & \textbf{26.6256150993} &  & 55976   &  & -7.64E-02 &  & -8.94E-02 \\
700                &  & 28.5679045373          &  & 28.5899851016 &  & 52511   &  & \textbf{28.4839888638} &  & \textbf{28.4877680865} &  & 59161   &  & -8.39E-02 &  & -1.02E-01 \\
740                &  & 29.2443717879          &  & 29.2477209407 &  & 56338   &  & \textbf{29.2418887324} &  & \textbf{29.2440743866} &  & 61698   &  & -2.48E-03 &  & -3.65E-03 \\
760                &  & 29.7452134712          &  & 29.7964289470 &  & 37391   &  & \textbf{29.6529205123} &  & \textbf{29.6626249312} &  & 45894   &  & -9.23E-02 &  & -1.34E-01 \\
780                &  & 30.1284909797          &  & 30.1454731784 &  & 36161   &  & \textbf{30.0138757407} &  & \textbf{30.0164282850} &  & 59540   &  & -1.15E-01 &  & -1.29E-01 \\
820                &  & 30.8413365443          &  & 30.8831339363 &  & 87278   &  & \textbf{30.7489836784} &  & \textbf{30.7533477503} &  & 125921  &  & -9.24E-02 &  & -1.30E-01 \\
890                &  & 32.1297407398          &  & 32.1758794016 &  & 80371   &  & \textbf{32.0438440391} &  & \textbf{32.0485774225} &  & 88801   &  & -8.59E-02 &  & -1.27E-01 \\
930                &  & 32.8196370611          &  & 32.8494260033 &  & 104544  &  & \textbf{32.6957714244} &  & \textbf{32.7024754370} &  & 119341  &  & -1.24E-01 &  & -1.47E-01 \\
960                &  & 33.3779501763          &  & 33.4204902518 &  & 100149  &  & \textbf{33.2554924317} &  & \textbf{33.2639325269} &  & 111880  &  & -1.22E-01 &  & -1.57E-01 \\ \midrule
513                &  & 24.4833011422          &  & 24.4880257346 &  & 69720   &  & \textbf{24.4803487672} &  & \textbf{24.4836476996} &  & 53628   &  & -2.95E-03 &  & -4.38E-03 \\
568                &  & \textbf{25.6742746832} &  & 25.6750224839 &  & 58749   &  & \textbf{25.6742746832} &  & \textbf{25.6747142118} &  & 63844   &  & 0.00E+00  &  & -3.08E-04 \\
608                &  & 26.5872589451          &  & 26.5914219805 &  & 63320   &  & \textbf{26.5828030363} &  & \textbf{26.5852665616} &  & 73642   &  & -4.46E-03 &  & -6.16E-03 \\
678                &  & 28.0507101923          &  & 28.0533386356 &  & 49399   &  & \textbf{28.0452214253} &  & \textbf{28.0486293696} &  & 70367   &  & -5.49E-03 &  & -4.71E-03 \\
737                &  & 29.2746711273          &  & 29.3098829156 &  & 46403   &  & \textbf{29.1805602263} &  & \textbf{29.1837249288} &  & 59552   &  & -9.41E-02 &  & -1.26E-01 \\
774                &  & \textbf{29.9122799819} &  & 29.9166270797 &  & 46759   &  & 29.9123498351          &  & \textbf{29.9157783799} &  & 44943   &  & 6.99E-05  &  & -8.49E-04 \\
846                &  & 31.2194588931          &  & 31.2658324893 &  & 96767   &  & \textbf{31.2169756294} &  & \textbf{31.2320033828} &  & 77582   &  & -2.48E-03 &  & -3.38E-02 \\
877                &  & 31.7991261202          &  & 31.8019948228 &  & 105985  &  & \textbf{31.7939196678} &  & \textbf{31.7991773790} &  & 123868  &  & -5.21E-03 &  & -2.82E-03 \\
923                &  & 32.5978399965          &  & 32.6013970912 &  & 128923  &  & \textbf{32.5891329001} &  & \textbf{32.5927036033} &  & 117298  &  & -8.71E-03 &  & -8.69E-03 \\
964                &  & 33.3301336049          &  & 33.3336095372 &  & 110927  &  & \textbf{33.3204312890} &  & \textbf{33.3287459497} &  & 115451  &  & -9.70E-03 &  & -4.86E-03 \\ \bottomrule
\end{tabular}
}
\end{table}

\begin{figure}[tb]
    \centering
    \includegraphics[width=0.7\columnwidth]{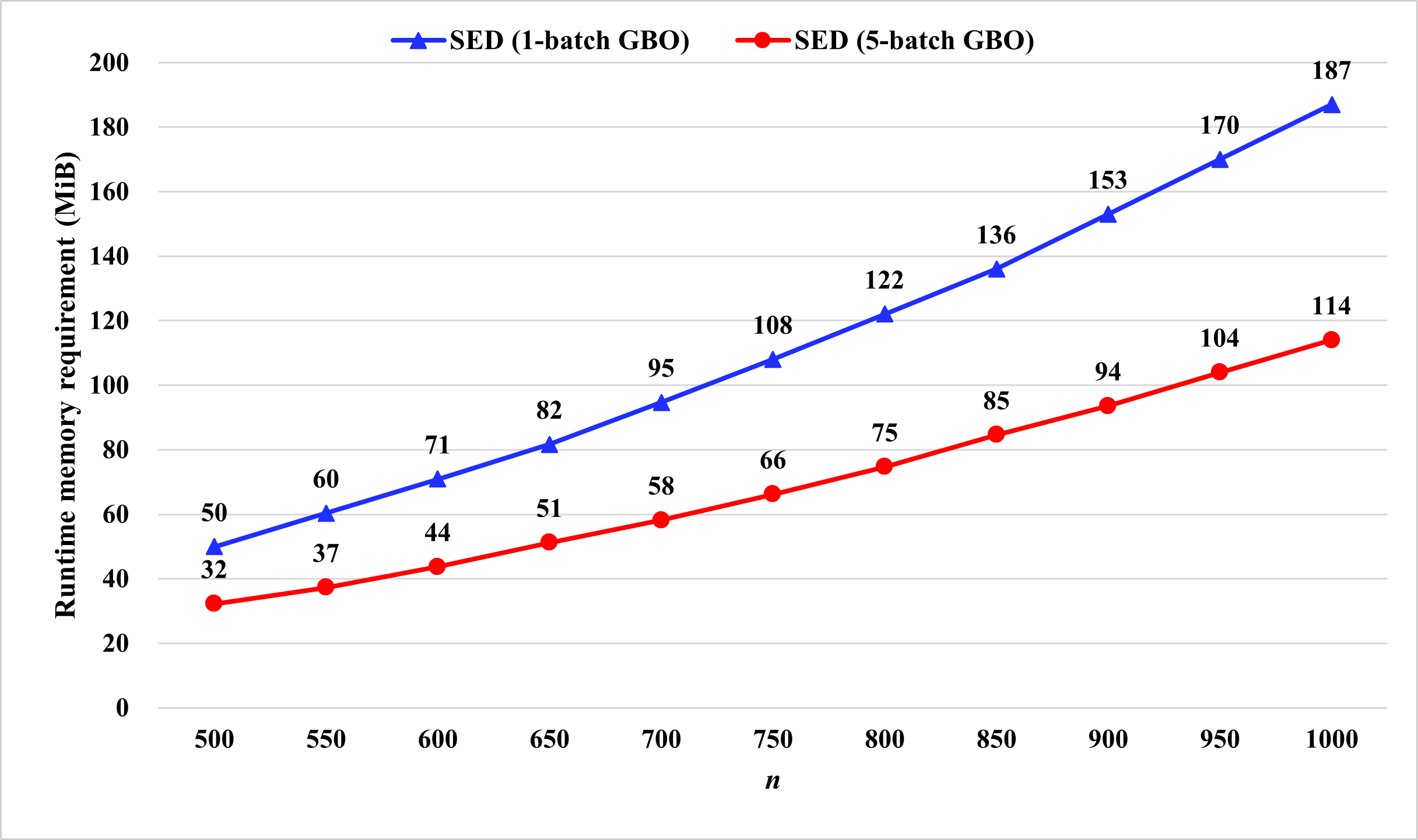} 
    \caption{Comparison of the runtime resident memory requirement of SED (5-batch \name) and SED (1-batch \name) for the instances $n=$ 500, 550, 600, ..., 1000, where the memory requirement is presented in MiB (i.e., MebiByte).}
    \label{batch-fig-momery}
\end{figure}

We further do a comparison to evaluate the performance of our proposed \name module on the large scale instances. We randomly select 10 regular numbers and 10 irregular numbers for the large scale instances, then we perform  SED (5-batch \name) and its variant SED (1-batch \name), which only changes the batch number from $k = 5$ to $k = 1$ for SED (5-batch \name), on the 20 selected instances. Both algorithms run 10 times on each instance independently. The results are shown in Table~\ref{tb-cmp-ablation1}. Note that the 1-batch (non-batch) \name degenerates to the classic BFGS optimization method. 

In the table, we show $n$ for the number of instances, $R_{best}'$ and $R_{avg}'$ for the best results and average results of 10 runs of SED (1-batch \name) respectively, $R_{best}$ and $R_{avg}$ for the best results and average results of 10 runs of SED (5-batch \name) respectively, $time~(s)$ for the average time of obtaining a best solution. $R_{best} - R_{best}'$ and $R_{avg} - R_{avg}'$ show the difference between two types of results represented 
where a negative value indicates SED (5-batch \name) yields better results than SED (1-batch \name). 

From the results, we can observe that SED (5-batch \name) has 18 best results better than SED (1-batch \name), 1 best result equal to the latter and 1 best result worse. All the average results of SED (5-batch \name) are better than SED (1-batch \name). It 
clearly demonstrates that the multi-batch method outperforms the non-batch method on large scale instances, and our proposed \name method has excellent performance on large scale instances.

We also do a comparison of the runtime memory requirement of 5-batch \name compared and non-batch (i.e., 1-batch). We perform SED (5-batch \name) and SED (1-batch \name) for the instances $n=$ 500, 550, 600, ..., 1000, and we record the resident memory requirement during the two programs' runtime. The comparisonal results are presented in Figure~\ref{batch-fig-momery}. 

From the Figure~\ref{batch-fig-momery}, we observe that SED (5-batch \name) requires lower resident runtime memory than SED (1-batch \name). In particular, 5-batch \name only needs 60.96\% to 64.00\% runtime memory of 1-batch \name (i.e., the classic BFGS) for the instances $n=$ 500, 550, 600, ..., 1000. The memory ratio of 5-batch \name to 1-batch \name decreases as $n$ increases. It demonstrates multi-batch \name has advantage of the runtime memory requirement on large scale instances. 


\subsection{Parameter Study} \label{ssec:06-05para}

Three parameters need to be tuned in our proposed algorithm, i.e., the batch strategy of the \name partition (Section~\ref{ssec:04-01GBO}), the batch number $k$ of \name (Section~\ref{ssec:04-01GBO}) and the iteration step $S_{iter}$ of SED heuristic (Section~\ref{ssec:05-02SED}). 
In this subsection, we give the experimental design and comparisonal results to determine a suitable parameter setting.


\begin{figure}[t]
    \centering
    \subfloat[]{\includegraphics[width=0.47\columnwidth]{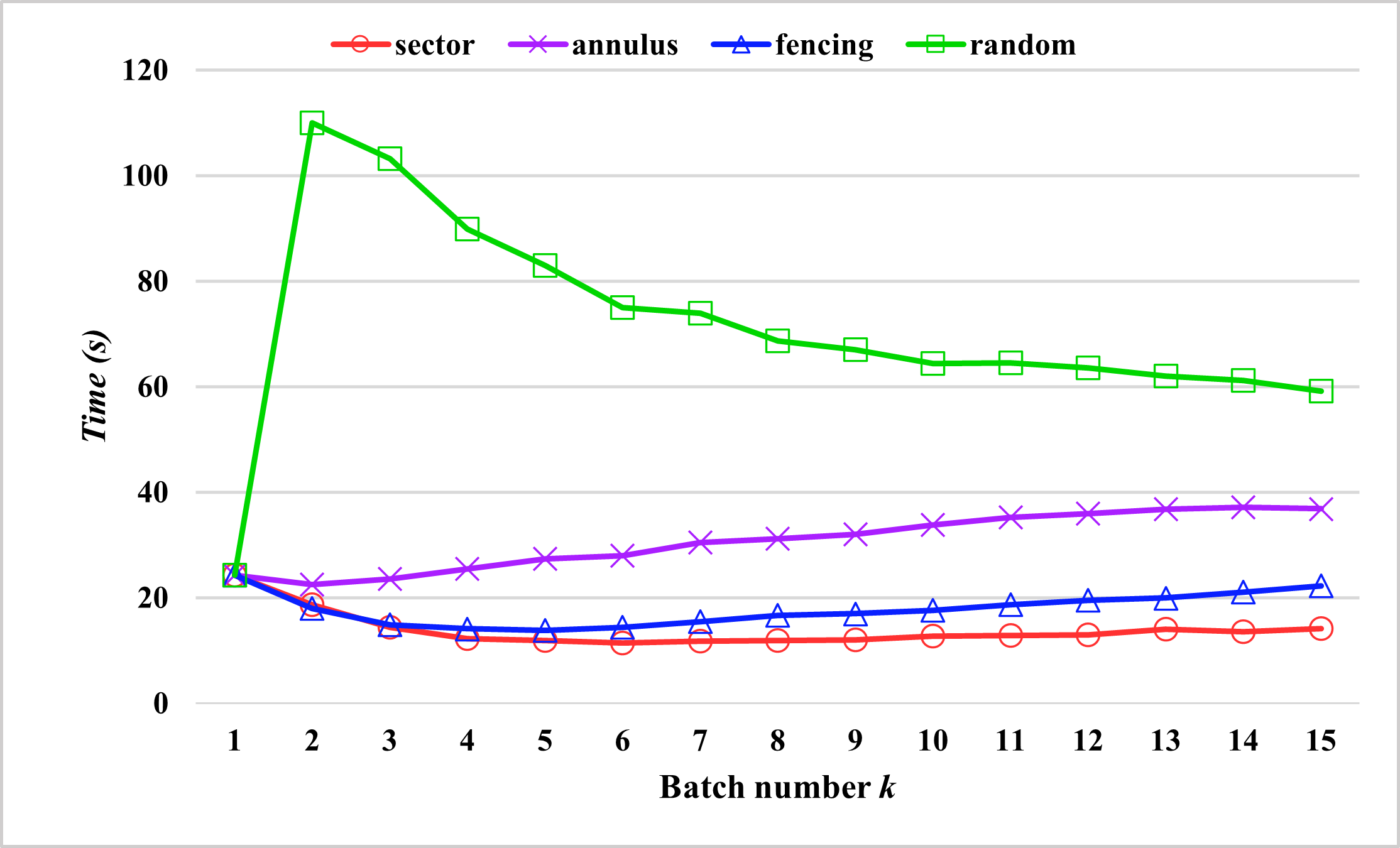} \label{batch-fig-time}}
    \subfloat[]{\includegraphics[width=0.47\columnwidth]{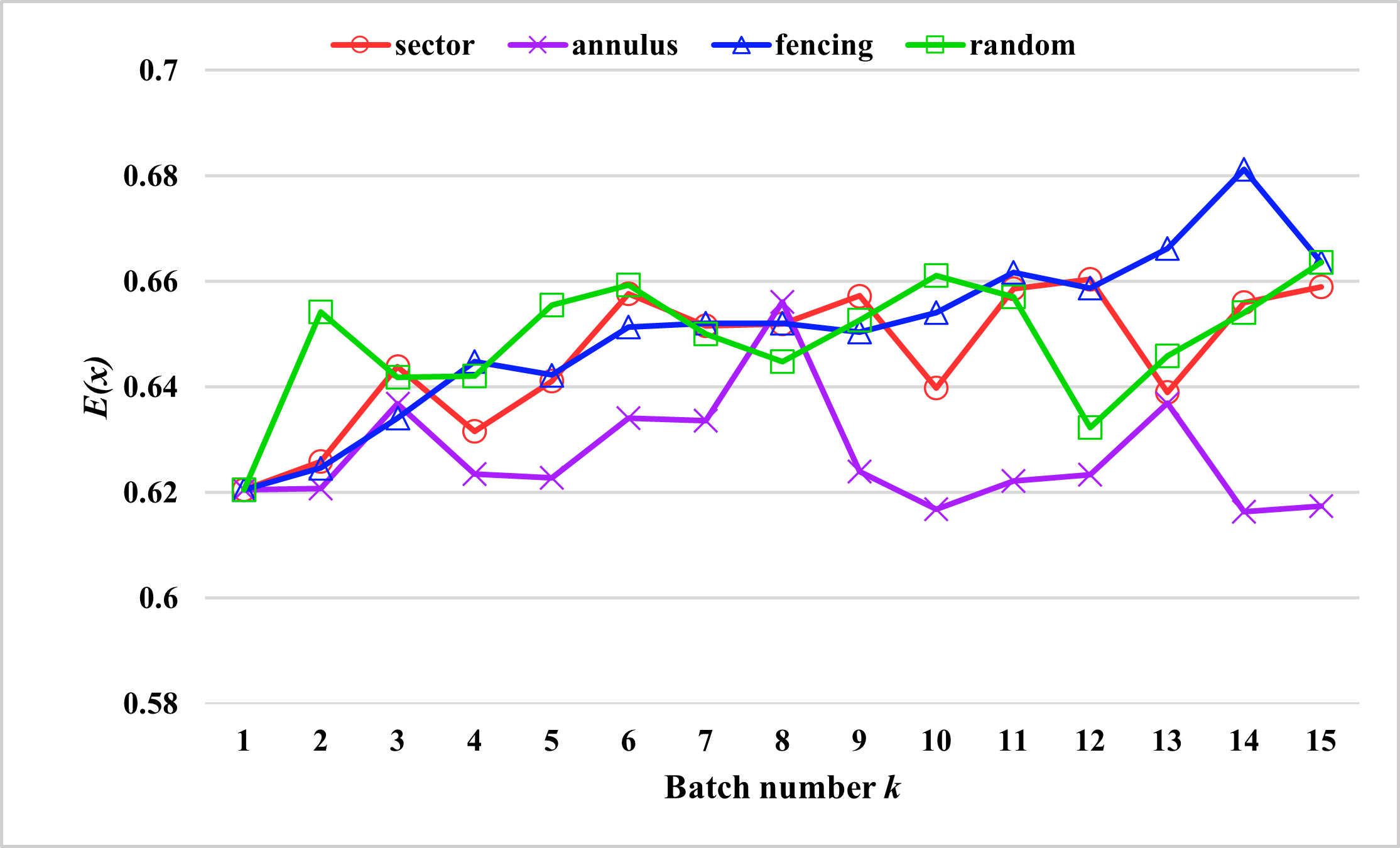} \label{batch-fig-energy}}
    \caption{Comparison between the four partitions of the \name method on large scale instances.}
    \label{batch-fig}
\end{figure}

\textbf{On batch partition strategy.} 
Since \name is a continuous optimization method, it is not sensitive to a specific instance, but it is sensitive to the instance scale. Therefore, we performed the four batch partition strategies (sector, annulus, fence and random) on the $n = 1000$ scale for the batch number from $k = 1$ to $15$, for investigating the performance of the partition strategies on large scale instances. We run each of the settings independently for 1,000 times where each of the runs starts from a random initial layout and terminates at the energy be converged or the maximal iteration step be reached (see in Algorithm~\ref{alg_batch}), and the experimental results of the average time cost and the average converged energy are shown in Figure~\ref{batch-fig}. Note that the \name method degenerates to the classic BFGS method when the batch number $k = 1$ (i.e., the non-batch method). 

Figure~\ref{batch-fig-time} illustrates the comparison of the average time cost of the four partitions where the X-axis indicates the batch number $k$ and the Y-axis indicates the average time cost of 1,000 runs. From the figure we can observe that:
\begin{itemize}
    \item [(1)] The time cost of random partition is significantly higher than non-batch (i.e., $k = 1$), which implies applying the random partition on \name makes the performance worse. Still, the average time cost of the random partition can be decreased as the batch number increases.
    \item [(2)] The three geometric batch partition strategies 
    show a similar trend 
    that the average time cost of the three strategies first decreases and then increases with the increasing batch number. The time cost of the annulus partition is slightly lower than non-batch when $k = 2$ and $3$, and it is higher than non-batch when $k \geq 4$, the time cost of the sector and fence partitions are all lower than non-batch when $k \geq 2$, and the sector partition has the best performance on the large scale. 
    \item [(3)] By comparing the random partition with the sector, annulus and fence partitions, we see that a reasonable geometric partition strategy is necessary for solving the PECC problem instead of using the random partition, and the partition strategy directly impacts the performance of the \name method. 
\end{itemize}

Figure~\ref{batch-fig-energy} gives the comparison of the average converged energy of the four partitions where the X-axis indicates the batch number $k$ and the Y-axis indicates the average converged energy $E(\boldsymbol{x})$ of 1,000 runs. From the figure we can observe that the four average converged energies of most of the batch partition settings are slightly higher than non-batch. However, most of these average converged energies locate between $0.62$ and $0.66$, and we consider the difference as the experimental error because the difference between these average converged energies compared with non-batch does not exceed 7\%.

According to the above discussion, the sector partition of the \name method can obviously boost the convergence speed, and 
there is no essential difference between sector partition and non-batch in the convergence result. Therefore, we select the sector partition as the optimal setting of the batch strategy.  


\begin{figure}[t]
    \centering
    \subfloat[]{\includegraphics[width=0.47\columnwidth]{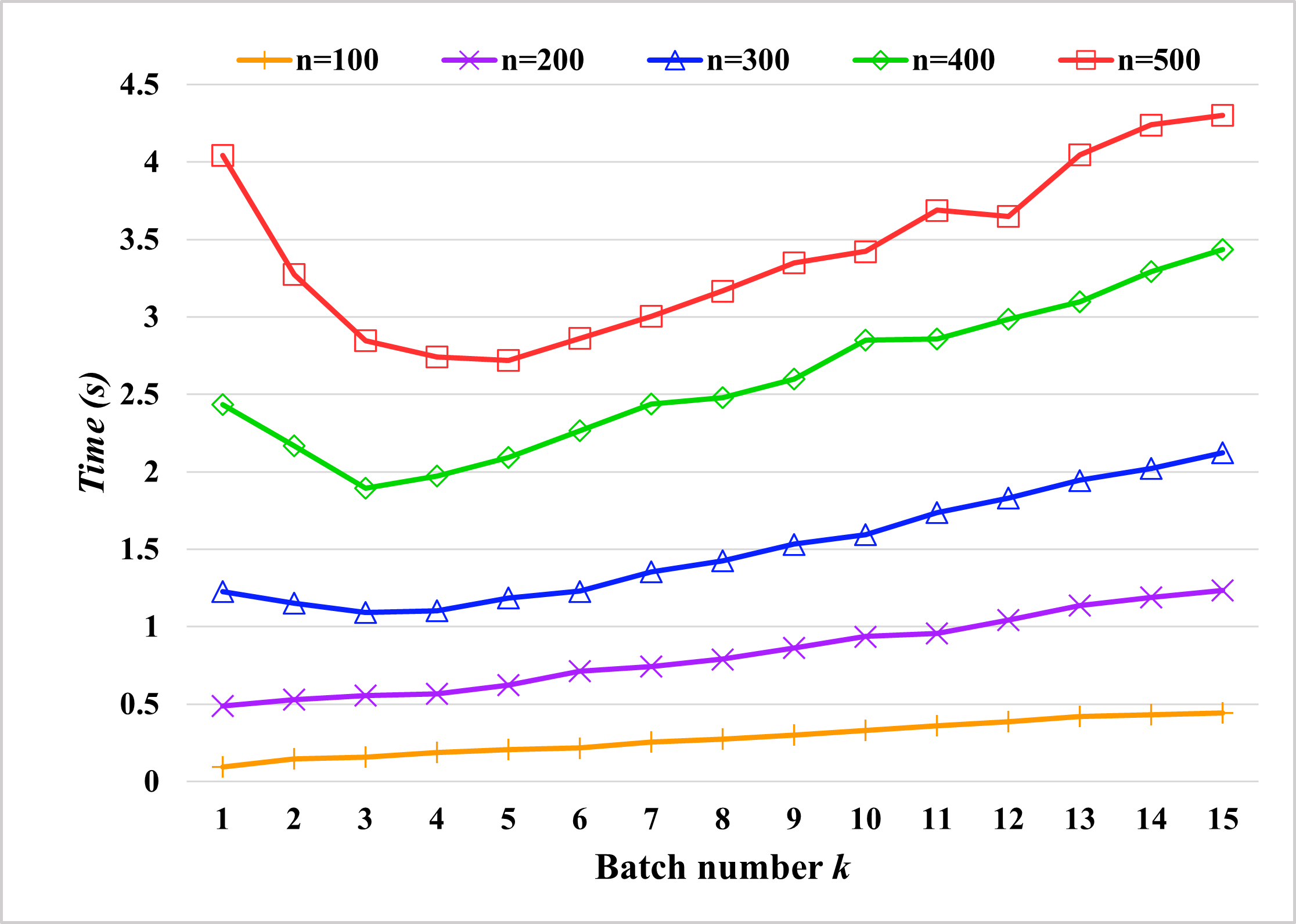} \label{batch-fig-scale-s}}
    \subfloat[]{\includegraphics[width=0.47\columnwidth]{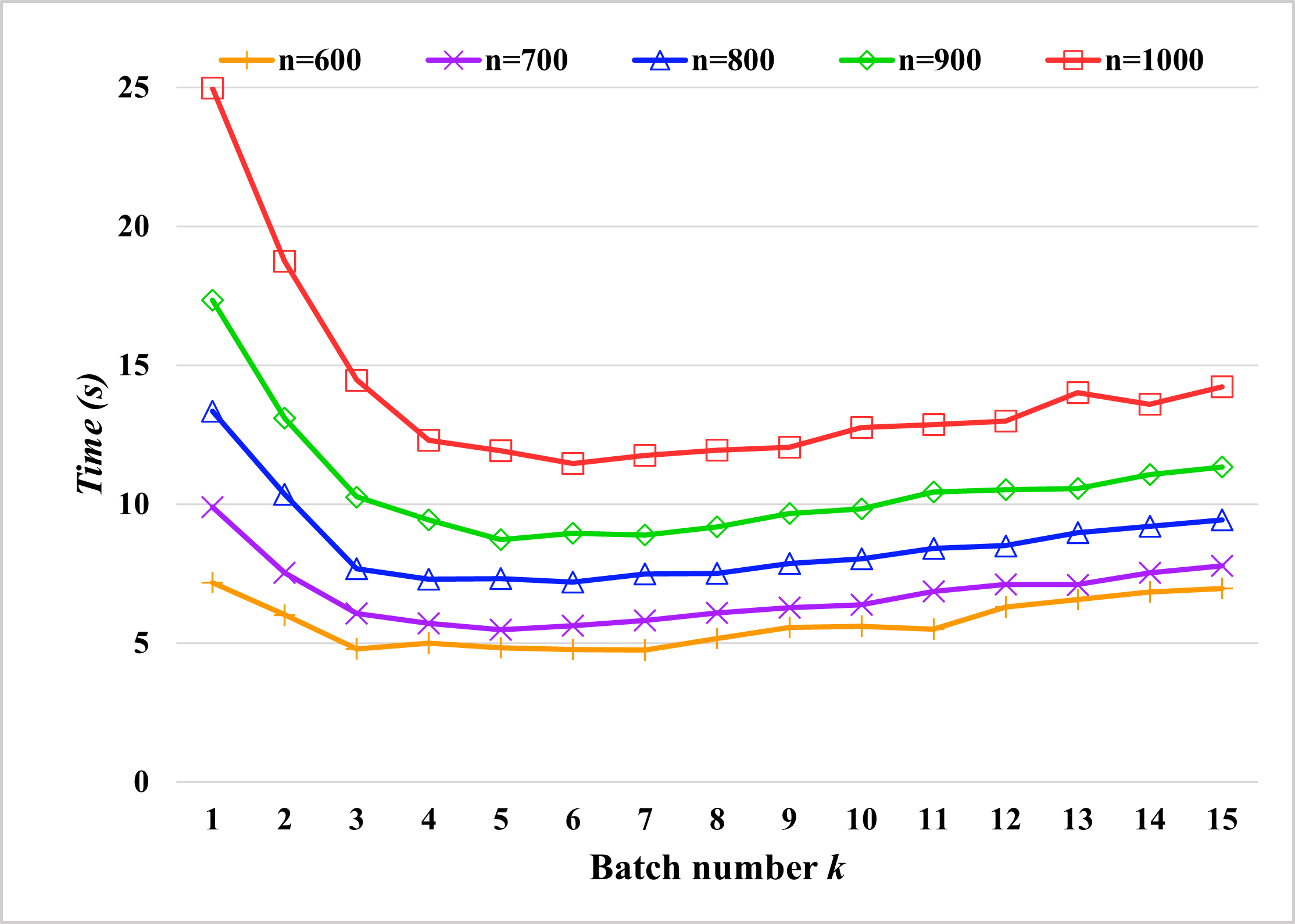} \label{batch-fig-scale-l}}
    \caption{Comparison between the different settings on the batch number of the \name method on different scale instances.}
    \label{batch-fig2}
\end{figure}

\begin{table}[]
\centering
\caption{The optimal batch number settings of \name for $n = 100, 200, ..., 1000$ compared with the non-batch setting.}
\label{tb-cmp-non-batch}

\scalebox{0.9}{
\begin{tabular}{lllllllllllllllllllll}
\toprule
$n$   &  & $T_{non.}~(s)$ &  & $opt.$ &  & $T_{opt.}~(s)$ &  & Ratio(\%) &  &  &  &  $n$   &  & $T_{non.}~(s)$ &  & $opt.$ &  & $T_{opt.}~(s)$ &  & Ratio(\%) \\ \midrule
100 &  & 0.09       &  & $k=1$  &  & 0.09        &  & 100.00\%  &  &  &  & 600  &  & 7.17       &  & $k=7$  &  & 4.74        &  & 66.16\%   \\
200 &  & 0.49       &  & $k=1$  &  & 0.49        &  & 100.00\%  &  &  &  & 700  &  & 9.90       &  & $k=5$  &  & 5.48        &  & 55.32\%   \\
300 &  & 1.23       &  & $k=3$  &  & 1.09        &  & 88.91\%   &  &  &  & 800  &  & 13.35      &  & $k=6$  &  & 7.19        &  & 53.81\%   \\
400 &  & 2.43       &  & $k=3$  &  & 1.89        &  & 77.78\%   &  &  &  & 900  &  & 17.35      &  & $k=5$  &  & 8.73        &  & 50.33\%   \\
500 &  & 4.04       &  & $k=5$  &  & 2.72        &  & 67.26\%   &  &  &  & 1000 &  & 24.97      &  & $k=6$  &  & 11.46       &  & 45.89\%  \\ \bottomrule
\end{tabular}
}
\end{table}

\textbf{On the batch number for various instance scales.}  
To evaluate the performance of \name on instances of different scales, we perform the \name method with sector partition on the scale $n = 100, 200, ..., 1000$ for the batch number from $k = 1$ to $k = 15$. We run each of the settings independently 1,000 times where each of the runs starts from a random initial layout and terminates at the energy being converged or the maximal iteration step is reached. The experimental results of the average time cost are shown in Figure~\ref{batch-fig2}. Note that the \name method degenerates to the classic BFGS method when the batch number $k$ is set to 1. 

Figures~\ref{batch-fig-scale-s} and ~\ref{batch-fig-scale-l} give the average time cost of the scale $n = 100, 200, ..., 500$ and the scale $n = 600, 700, ..., 1000$, respectively. The X-axis indicates the batch number $k$ and the Y-axis indicates the average time cost of 1,000 runs. Table~\ref{tb-cmp-non-batch} shows the comparison of the average time cost of the non-batch and optimal batch setting where the first column of the table gives $n$ of the instances, $T_{non.}$ and $T_{opt.}$ for the average time costs of the non-batch (i.e., $k = 1$) and the optimal batch setting, $opt.$ for the optimal setting of the batch number $k$ and Ratio (\%) $= \frac{T_{opt.}}{T_{non.}}$ for the average time cost ratio of the optimal batch to non-batch. 

From the two Figures~\ref{batch-fig-scale-s} and \ref{batch-fig-scale-l}, and Table~\ref{tb-cmp-non-batch}, we have the following observations:
\begin{itemize}
    \item [(1)] The curves of the scale for $n = 100$ and $n = 200$ show that \name does not work well on very small instances, applying the batch method will increase the convergence time, and \name needs more time to obtain a converged solution as the batch number increases. 
    \item [(2)] The curve of the scale for $n = 300$ shows that \name has a small advantage over the non-batch method on moderate scale instances. The average time costs of $k = 2, 3, 4$ and $5$ are slightly lower than non-batch (i.e., $k = 1$). The curves of the scale $400 \leq n \leq 1000$ show that \name can reduce the convergence time and boosts the convergence process significantly with a proper batch number setting. And the column of Ratio (\%) of the table shows that the ratio will be reduced with the increasing scale, which indicates that \name has more advantage of accelerating effect as the instance scale increases. The experimental results demonstrate that \name has excellent performance on large scale instances. 
    \item [(3)] The curves for the large scale instances, $500 \leq n \leq 1000$, show that the average time cost is increasing for batch number $k \geq 8$. It indicates that the batch number is not always better for larger values. And the optimal batch number $k$ of the large scale instances is located in interval $[5, 7]$.
\end{itemize}

According to the above discussion, we use the batch number $k = 3$ as the optimal setting for the moderate scale instances, $300 \leq n \leq 320$, and the batch number $k = 5$ as the trade-off setting for the large scale instances, $500 \leq n \leq 1000$. 


\begin{table}[t]
\centering
\caption{Computational results and comparison of the parameter $S_{iter}$ on the average result ($R_{avg}$) for 12 selected instances where the best results obtained among the tested parameter values are presented in bold.}
\label{tb-cmp-iteration}

\scalebox{0.8}{
\begin{tabular}{lllllllllllll}
\toprule
          &  & \multicolumn{11}{l}{$R_{avg}$}                                                                                                                     \\ \cline{3-13} 
$n$ \ / \ $S_{iter}$ &  & 100           &  & 200           &  & 300                    &  & 400                    &  & 500                    &  & 600                    \\ \midrule
305       &  & 19.0029499894 &  & 19.0022928525 &  & 19.0023365854          &  & \textbf{19.0022767126} &  & 19.0027346869          &  & 19.0026950563          \\
316       &  & 19.3350583426 &  & 19.3348928193 &  & 19.3352210818          &  & 19.3347673471          &  & \textbf{19.3345842813} &  & 19.3346305894          \\
513       &  & 24.4871818937 &  & 24.4850574870 &  & 24.4858380045          &  & 24.4843239209          &  & \textbf{24.4836476996} &  & 24.4850158014          \\
568       &  & 25.6753984748 &  & 25.6746996472 &  & 25.6746268914          &  & 25.6749397842          &  & 25.6747142118          &  & \textbf{25.6746092173} \\
608       &  & 26.5878240724 &  & 26.5866285288 &  & 26.5860789992          &  & 26.5865544926          &  & \textbf{26.5852665616} &  & 26.5867207111          \\
678       &  & 28.0509853708 &  & 28.0507532028 &  & 28.0503173118          &  & 28.0509759006          &  & \textbf{28.0486293696} &  & 28.0498132121          \\
740       &  & 29.2459673340 &  & 29.2469306141 &  & 29.2451838453          &  & 29.2451220192          &  & \textbf{29.2440743866} &  & 29.2458209932          \\
774       &  & 29.9177876254 &  & 29.9143968413 &  & \textbf{29.9117005742} &  & 29.9143441746          &  & 29.9157783799          &  & 29.9153122016          \\
846       &  & 31.2425443239 &  & 31.2333675583 &  & 31.2304459481          &  & 31.2320033828          &  & 31.2275469287          &  & \textbf{31.2261977327} \\
877       &  & 31.8025272145 &  & 31.7999463301 &  & 31.8016759748          &  & \textbf{31.7975804456} &  & 31.7991773790          &  & 31.7998976078          \\
923       &  & 32.5975528505 &  & 32.5942894739 &  & 32.5935176512          &  & 32.5949283036          &  & \textbf{32.5927036033} &  & 32.5947280012          \\
964       &  & 33.3335745251 &  & 33.3300879802 &  & 33.3314171856          &  & 33.3271953322          &  & 33.3287459497          &  & \textbf{33.3265773865} \\ \midrule
Average   &  & 27.6066126681 &  & 27.6044452780 &  & 27.6040300044          &  & 27.6037509847          &  & \textbf{27.6031336198} &  & 27.6035015425          \\ \bottomrule
\end{tabular}
}
\end{table}

\textbf{On iteration step of SED.} 
The rest of the parameters to be analyzed is the iteration step of SED. We perform SED (5-batch \name) with the several iteration steps $S_{iter} = 100, 200, ..., 600$ on the 12 instances, including 2, 6 and 4 randomly selected instances from the moderate scale, the large scale I and II respectively. The comparison of the iteration steps is shown in Table~\ref{tb-cmp-iteration}. $n$ is the number of items in the instances, and columns 2 to 7 show the average results, $R_{avg}$, of 10 or 20 runs (20 for the moderate scale and 10 for the large scale) for each tested iteration step $S_{iter}$, and the row of ``Average'' in the bottom shows the average value of the 12 instance results for each column.

Table~\ref{tb-cmp-iteration} shows that the algorithm with $S_{iter} = 500$ obtains the best performance in terms of $R_{avg}$ for 6 out of the tested 12 instances, a much higher number than the other 5 tested iteration steps. It has also obtained the best average value of the 12 instance results among the 6 tested iteration steps. As a result, we set the default value of $S_{iter}$ to $500$.

\section{Conclusions}
In this paper, we aim to address the most representative packing problem, the packing equal circles in a circle problem, on large scale. 
We propose a novel geometric batch optimization method that  
not only can significantly speed up the continuous optimization process but also can reduce the memory requirement for finding a local minimum packing configuration. 
We also propose a solution-space exploring and descent search heuristic accordingly for the search to find a global minimum for the optimization of the overall packing. 
Besides, we propose an adaptive neighbor object maintenance method, which handles some issues of the existing methods for maintaining the neighbor structure, and it is suitable for dynamic packing problems and online packing problems.
Extensive experiments on 21 moderate instances ($n = 300$ to 320) and 101 sampled large-scale instances ($n = 500$ to 1000) demonstrate the effectiveness and efficiency of our proposed methods. Our algorithm could often find new and better packing results than the current best records. In addition, our geometric batch optimization, heuristic search and adaptive maintenance methods are generic and can be used for other optimization problems. In future work, we will extend our methods for solving other packing problems.













\bibliographystyle{cas-model2-names}

\bibliography{mainfinal}


\end{document}